\documentclass[aps,prd,superscriptaddress,showpacs,nofootinbib,notitlepage,twocolumn]{revtex4-1}
\usepackage{graphicx,subfigure,amsmath,amsfonts,amssymb,multirow,mathrsfs,ulem}
\usepackage[colorlinks,linkcolor=blue,citecolor=blue,urlcolor=blue]{hyperref}

\def\nat{Nature\ }
\def\aap{Astron.\ Astrophys.\ }
\def\apj{Astrophys.\ J.\ }
\def\apjl{Astrophys.\ J.\ Lett.\ }
\def\apjs{Astrophys.\ J.\ Supp.\ }

\def\mnras{Mon.\ Not.\ Roy.\ Astron.\ Soc.\ }
\def\physrep{Phys.\ Rept.\ }
\def\prd{Phys.\ Rev.\ D\ }
\def\prl{Phys.\ Rev.\ Lett.\ }

\def\cjaa{Chin.\ J.\ Astron.\ Astrophys.\ }
\def\pasp{Publ.\ Astron.\ Soc.\ Pac.\ }

\newcommand{\ud}{\mathrm{d}}

\begin{document}
\newcommand{\PMO}{Key Laboratory of Dark Matter and Space Astronomy, Purple Mountain Observatory, Chinese Academy of Sciences, Nanjing 210023, People's Republic of China}
\newcommand{\USTC}{School of Astronomy and Space Science, University of Science and Technology of China, Hefei, Anhui 230026, People's Republic of China}

\title{Prospects of calibrating afterglow modeling of short GRBs with gravitational wave inclination angle measurements and resolving the Hubble tension with a GW-GRB association event}

\author{Yi-Ying Wang}
\author{Shao-Peng Tang}
\author{Xin-Yu Li}
\author{Zhi-Ping Jin}
\email[Corresponding author.~]{jin@pmo.ac.cn}
\author{Yi-Zhong Fan}
\email[Corresponding author.~]{yzfan@pmo.ac.cn}
\affiliation{\PMO}
\affiliation{\USTC}
\date{\today}

\begin{abstract}
In the past decades, an external forward shock model has been successfully developed to explain the main features of the afterglow emission of gamma-ray bursts (GRBs). In the numerical modeling of the GRB afterglow, some approximations have been made for simplicity, and different groups developed their codes. A robust test of these models/approaches is challenging because of the lack of directly measured physical parameters. Fortunately, the viewing angle inferred from the afterglow modeling is widely anticipated to be the same as the inclination angle of the binary neutron star (BNS) mergers that can be evaluated with the gravitational wave (GW) data. Therefore in the future, it is possible to calibrate the afterglow modeling with the GW inclination angle measurements. We take three methods, including both analytical estimations and direct simulations, to project the uncertainties of the inclination angle measurements. For some BNS mergers accompanied with electromagnetic counterparts detected in the O4/O5 runs of LIGO/Virgo/KAGRA/LIGO-India detectors, we show that the inclination angle can be determined within an uncertainty of $\leq 0.1$ rad, supposing that the Hubble constant is known with an accuracy of $\leq 3\%$ and the uncertainty of Hubble flow velocity is within $\sim 1 \%$. The off-axis GRB outflow will give rise to afterglow emission, and the most energetic ones may be detectable at the distance of $\sim$ 100-200 Mpc even for a viewing angle of $\geq 0.3$ rad.
Such events can thus serve as a robust test of the afterglow modeling approach. We have also evaluated the prospect of resolving the so-called Hubble tension with a single GW/GRB association event. We find out that a $\sim 3\%$ precision Hubble constant is obtainable if the uncertainty of the viewing angle can be constrained to be within $\sim 0.1$ rad, which is expected to be the case for some nearby ($\leq 250$ Mpc) bright/on-axis GRBs with a well-behaved afterglow light curve displaying a clear achromatic break at early times.
\end{abstract}
\maketitle

\section{introduction}
Thanks to the worldwide joint efforts in the last decades, a fireball model has been developed to interpret gamma-ray burst (GRB) observations (see, e.g., Refs.~\cite{Piran2004,Kumar+Zhang2015} for reviews). In such a framework, the black holes formed in the collapse of the massive stars (for the so-called long-duration bursts) or the mergers of binary neutron stars (BNSs) (mainly for the short-duration bursts) can launch ultrarelativistic narrowly beamed outflows. The collisions among the fast and slow material shells (or the dissipation of the magnetic fields if the outflows are highly magnetized) are widely believed to power the prompt gamma-ray flashes, which are observable if our line of sight is within the cone of the relativistic outflow. The GRB ejecta gets decelerated by the surrounding medium and drives an energetic blast wave (i.e., the forward shock). Fractions of the shock energy are assumed to be given to accelerate power-law distributed electrons and generate the turbulent magnetic fields. The synchrotron, as well as the synchrotron self-Compton radiation of the shock-accelerated electrons, give rise to the relatively long-lasting afterglow emission \citep[][]{katz1994ApJ,Sari+1998ApJL}. Later on, it is clear that the prolonged activity of the central engines can contribute to the afterglow emission via either the energy injection into the blast wave \citep[][]{Dai+Lu1998PRL,Zhang2001ApJL} or the prompt energy dissipation of the newly launched ejecta \citep[][]{Fan+Wei2005MNRAS,Burrows2005Sci,Fan2005ApJ,Gao+Fan2006,Zhang2006ApJ}. Anyhow, the prolonged activities of the GRB central engines are found to contribute at the early stages. In the late phase, the afterglow emissions are usually well behaved, which are widely believed to be from the forward shock radiation. The analytic expression of the forward shock emission is possible in the simplest scenario \citep[][]{Sari+1998ApJL,Chevalier1999ApJL,Granot1999ApJ,Rhoads1999ApJ}. The numerical calculations, of course, are more flexible \citep[][]{Panaitescu1999ApJ,Huang2000,Granot2002ApJL,Wei+Jin2003,Kumar+Granot2003ApJ,Fan2008MNRAS}. As a very complicated system, even the numerical calculations have to be done with some simplified approximation. For instance, the profiles of the jetted outflow are usually assumed to be top-hat or other simple functions \citep[][]{Meszaros1998ApJ,Salmonson2001ApJ,Dai&Gou2001ApJ,Rossi2002MNRAS,Berger2003Natur,Zhang2004ApJL,Wu2005MNRAS}, the shock parameters are assumed to be constant \citep[][]{Sari+1998ApJL,Chevalier1999ApJL}, the energy distribution of the shock-accelerated electrons is a single power law, the magnetic fields in the emitting region are assumed to be the same, and the sideways expansion velocities are assumed to be either zero or a simple function of the bulk Lorentz factor \citep[][]{Rhoads1999ApJ,Huang2000}. As for the relation between the radius ($R$) and the observer's time ($t$), in the case the bulk Lorentz factor of the GRB ejecta $\Gamma\gg 1$, some people used the relation of $dR \approx 4\Gamma^2 cdt/(1+z)$, while others adopted the function of $dR \approx 2 \Gamma^2 cdt/(1+z)$, where $c$ is the speed of the light and $z$ is the redshift of the burst \citep[][]{Sari+1998ApJL,Huang2000,Pe'er2012ApJL}. Different groups have also adopted different dynamical equations in calculating the afterglow emission \citep{Katz1997ApJ,Chiang1999ApJ,Huang1999MNRAS,Pe'er2012ApJL}.

Because of the adopted simplifications and the different treatments on some points, it is not surprising to find that different groups reported different fit parameters for many bursts. In most cases, it does not matter because the main purpose is to qualitatively/``reasonably" understand what happened in these GRB sources. However, this is unacceptable if we adopt the GRB data to infer some key parameters for further precision measurements. For instance, to obtain an accurate luminosity distance, we need a reliable inclination angle before breaking the degeneracy in the gravitational wave data analysis \citep{Schutz1986Natur}. The underlying assumption is that the viewing angle ($\theta_{\rm v}$) equals the inclination angle, which may be inferred from the afterglow modeling \footnote{Note that recently, the $\theta_{\rm v}$ is found to be degenerate with the half-opening angle of the energetic core ($\theta_{\rm c}$) of the ``structured" outflow if viewed off axis \citep{Nakar2021ApJ,Troja2022MNRAS}. One may need the complete afterglow data (i.e., including also the very late ones) or the dedicated numerical simulation rather than the simplified calculation of the afterglow emission of the structured outflow to break the degeneracy. The numerical simulations, however, are too time consuming, and some approximations are still needed to carry out the afterglow fit (see, e.g., Ref.~\cite{Wu2019APJ}). For the on-axis GRBs, the half-opening angle inferred from the afterglow modeling ($\theta_{\rm j}$) is a reasonable approximation of $\theta_{\rm c}$. For the most widely discussed top-hat jet model, we have $\theta_{\rm j}=\theta_{\rm c}$.} \citep{Panaitescu1999ApJ,Panaitescu2000ApJ,WH2021ApJ}. In this work we also adopt such an assumption (i.e., $\iota=\theta_{\rm v}$ and $\iota\leq \pi/2$). As the current unique BNS
merger event with detected multiwavelength counterparts, GW170817 \citep{Abbott2017PRL} has attracted very wide attention. Dedicated efforts have been made to fit the x-ray/optical/radio afterglow emission \citep{Abbott1708172017APJL}. The center of the ejecta is found to be significantly away from the line of our sight (i.e., it is an off-axis jet) and the ejecta is likely structured \citep[][]{Troja2017Nat,Lamb2018MNRAS,Lazzati2018PRL,Lyman2018NatAs,Mooley2018Nat2,Yue2018ApJ,Troja2019MNRAS,Ghirlanda2019Sci,Wu2019APJL,Hajela2019ApJL,Ioka2020MNRAS}. For GW170817, the inclination angle has been evaluated by superluminal motion of the GRB jet \cite{Mooley2018Nat2,Ghirlanda2019Sci,Hotokezaka2019NatAs}, kilonova modeling \cite{Coughlin2020NatCo,Heinzel2021MNRAS,Korobkin2021ApJ}, and the afterglow data \cite{Wu2019APJ,Lazzati2018PRL,Lamb2019ApJ,Hajela2019ApJL}. As pointed out in Ref.~\citep{Hotokezaka2019NatAs}, the hydrodynamics simulation of the jet superluminal motion gives $0.25\, {\rm rad}<\theta_{\rm v}({d_{\rm L}}/{41\,{\rm Mpc}})<0.5\,{\rm rad}$, and the joint fit with the afterglow light curve has improved the constraint to $0.30^{+0.04}_{-0.04} {\rm rad}$, where $d_{\rm L}$ is the luminosity distance of GW1708017/GRB 170817A. Likely, this constraint is dominated by the afterglow light curve data and might still suffer from the uncertainties involved in the modeling. For some future events with even better data, the superluminal motion alone may be able to yield high-precision measurement of $\theta_{\rm v}$ and can hence serve as an independent test of the afterglow light curve modeling. In this work, we will focus on whether the gravitational wave data can do such a job.

\begin{figure}[ht!]
	\centering
	\includegraphics[width=0.4\textwidth]{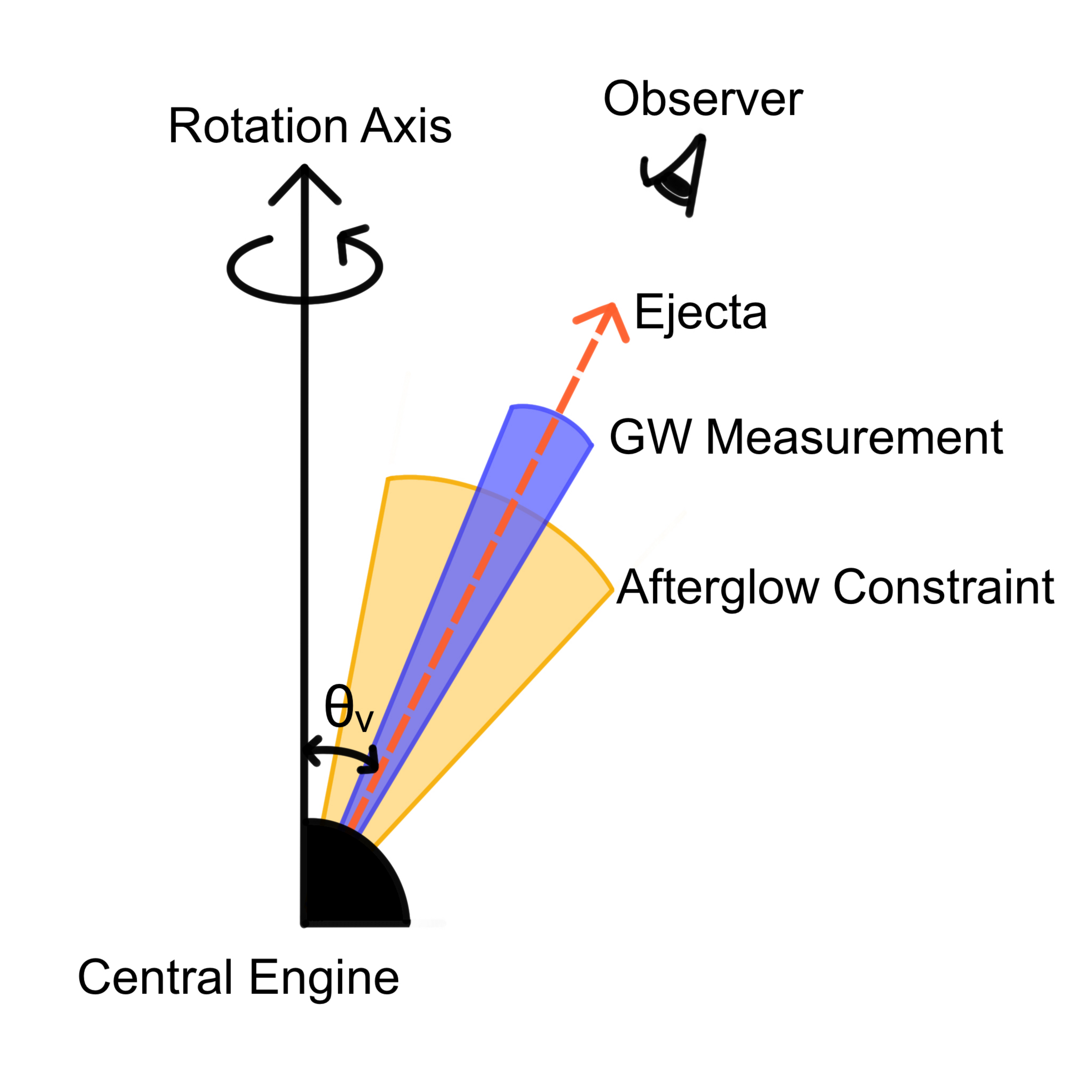}

	\caption{\small Schematic plot of the ejecta viewing geometries and the comparison of possible viewing angle $\theta_{\rm v}$ (which equals to $\iota$) constraints in two independent ways. The yellow range is the possible constraint on $\theta_{\rm v}$ by the afterglow data. The uncertainty has been suggested to arise from the degeneracy of $\theta_{\rm v}$ with the half-opening angle of the energetic core of the ``structured" outflow \citep{Nakar2021ApJ,Troja2022MNRAS}. The blue range is to show the possible bounds on $\iota$ in the gravitational wave data analysis.}
	\label{fig:cartoon}
\end{figure}

Being one of the most appealing approaches, the ``standard candle" has been studied in many works \cite{Betoule2014A&A, Scolnic2018ApJ, Riess2019ApJ, Dainotti2021ApJ, SHoES_2021}. With the sole gravitational wave (GW) data and the redshift information of GW170817/GRB 170817A/AT2017gfo, the Hubble constant has been measured to be $H_0=70.0^{+12.0}_{-8.0}\rm ~km\,s^{-1}\,Mpc^{-1}$ \citep{Abbott2017Nature}. 
After taking into account the inclination angle (i.e., the viewing angle $\theta_{\rm v}$ in the GRB afterglow modeling) constrained with the radio/optical/x-ray emission, \citet{Hotokezaka2019NatAs} have reported a new measurement $H_0 = 70.3^{+5.3}_{-5.0} \rm ~km\,s^{-1}\,Mpc^{-1}$. As demonstrated in Ref.~\cite{Chen2018Nature}, in a few years the gravitational wave ``standard siren" will likely be able to resolve the tension between the cosmic microwave background (CMB) measurements from the Planck Collaboration ($67.66 \pm 0.42 {\rm ~km\,s^{-1}\,Mpc^{-1}}$ \citep{Planck_2020}) and type Ia supernova measurement from the SHOES (Supernova, $H_0$, for the Equation of state of Dark Energy) team ($73.2 \pm 1.3 {\rm ~km\,s^{-1}\,Mpc^{-1}}$ \citep{SHoES_2021}). However, the afterglow modeling is likely unable to play an important role in achieving such a goal. This is because the inclination angles $\iota$ found in different afterglow modeling of GRB 170817A can differ from each other by a factor of 2 (i.e., ranging from $\sim 15^{\circ}$ to $\sim 30^{\circ}$ see Fig.~3 of Ref.~\cite{Nakar2021ApJ} for a summary) and may be due to the degeneracy of $\theta_{\rm c}$ and $\theta_{\rm v}$, which is one of the main sources of the uncertainty in measuring $H_0$. There is, fortunately, one exception. As demonstrated in this work, for some nearby ($\leq 250$ Mpc) bright on-axis\footnote{In this work we call the event with a viewing angle within the opening angle (top-hat case) or the energetic core (for instance, the Gaussian case) of the outflow as the on-axis GRB.} GRBs with a well-behaved afterglow light curve displaying a clear achromatic break at early times, the uncertainty of the estimated viewing angle is expected to be within $\sim 0.1$ rad (in accordance with the half-opening angle constraint), with which a $\sim$ 3\%-4\% precision Hubble constant is obtainable (see Sec.~\ref{sec:dl} for the extended discussion). Nevertheless, our main purpose is not to investigate how to resolve the Hubble tension. Instead, we will focus on the prospect of calibrating the afterglow light curve modeling approaches of short GRBs with the gravitational wave inclination angle measurements (see Fig.~\ref{fig:cartoon} for an illustration).

This work is organized as follows. In Sec.~\ref{sec:afterglow}, we discuss the distance range within which the afterglow emission of a highly off-axis short GRB can be bright enough to be used to measure the viewing angle of the ejecta. In Sec.~\ref{sec:method}, we describe three different methods to predict the estimated precision of $\iota$ and $d_{\rm L}$ for a BNS merger event accompanied by an afterglow. In parameter estimation, we use different prospective prior constraints for $\iota$ and $d_{\rm L}$ based on different assumptions. In Sec.~\ref{sec:result}, we first show the estimated result of $\iota$ and give the application to distinguish various jet profiles. The uncertainty of Hubble tension is also 
addressed for $\iota$ estimation. Then, we give prospective constraints for $d_{\rm L}$ with prior limited $\iota$ by electromagnetic counterparts and discuss the precision of $H_0$ detection in the future. In Sec.~\ref{sec:discussion}, we summarize our results with some discussions. 

\section{Prospect of evaluating the viewing angle of the highly off-axis ejecta with afterglow}\label{sec:afterglow}
The GRB afterglow emission depends upon the bulk Lorentz factor, the number density of the interstellar medium ($n$), and the viewing angle, as well as the physical parameters $\epsilon_{\rm e}$ and $\epsilon_{\rm B}$, the fractions of blast wave energy given to accelerate the electrons and generate the magnetic fields. Therefore, under the synchrotron radiation framework, people need the well-measured multiband (i.e., radio, optical, and X-ray) afterglow light curves to infer these parameters. 
To reasonably constrain the viewing angle of the GRB ejecta, the light curve before and after the temporal break should be well recorded. In this work we concentrate on the highly off-axis (i.e., $\theta_{\rm v}>0.3$ rad) ejecta scenario since for $\theta_{\rm v}\sim 0$ the gravitational wave data cannot set a reliably tight constraint on $\iota$ (equally, $\theta_{\rm v}$), as demonstrated in Sec.~\ref{sec:IVA}. While for $\iota\geq 0.3$ rad, the gravitational wave data in the O5 run of LIGO/Virgo detectors can yield a $\Delta \iota \leq 0.1$ rad, with which we can achieve the goal of calibrating the afterglow light curve modeling of short GRBs with the gravitational wave observations. However, for the highly off-axis events, the afterglow emission would be extremely suppressed until the blast waves driven by the ejecta have gottten decelerated to a bulk Lorentz factor of $\Gamma \approx 1/\theta_{\rm v}$. The corresponding time is $t_{\rm dec}\propto (E_{\rm k}/n)^{1/3}\theta_{\rm v}^{8/3}$, where $E_{\rm k}$ is the kinetic energy of outflow. For $t>t_{\rm dec}$, the afterglow emission is rather similar to the on-axis case of ejecta (e.g, \citep{Wei+Jin2003,Kumar+Granot2003ApJ}) and the flux drops with the time very quickly (i.e., $\propto t^{-p}$, where $p$ is the power-law distribution index of the shock-accelerated electrons). Therefore, the larger the $\theta_{\rm v}$, the dimmer the ``peak" of the forward shock emission. To well record the afterglow data of the highly off-axis events, the source cannot be too distant.

\begin{figure*}[ht!]
	\centering
	\subfigure{
	\includegraphics[width=0.45\textwidth]{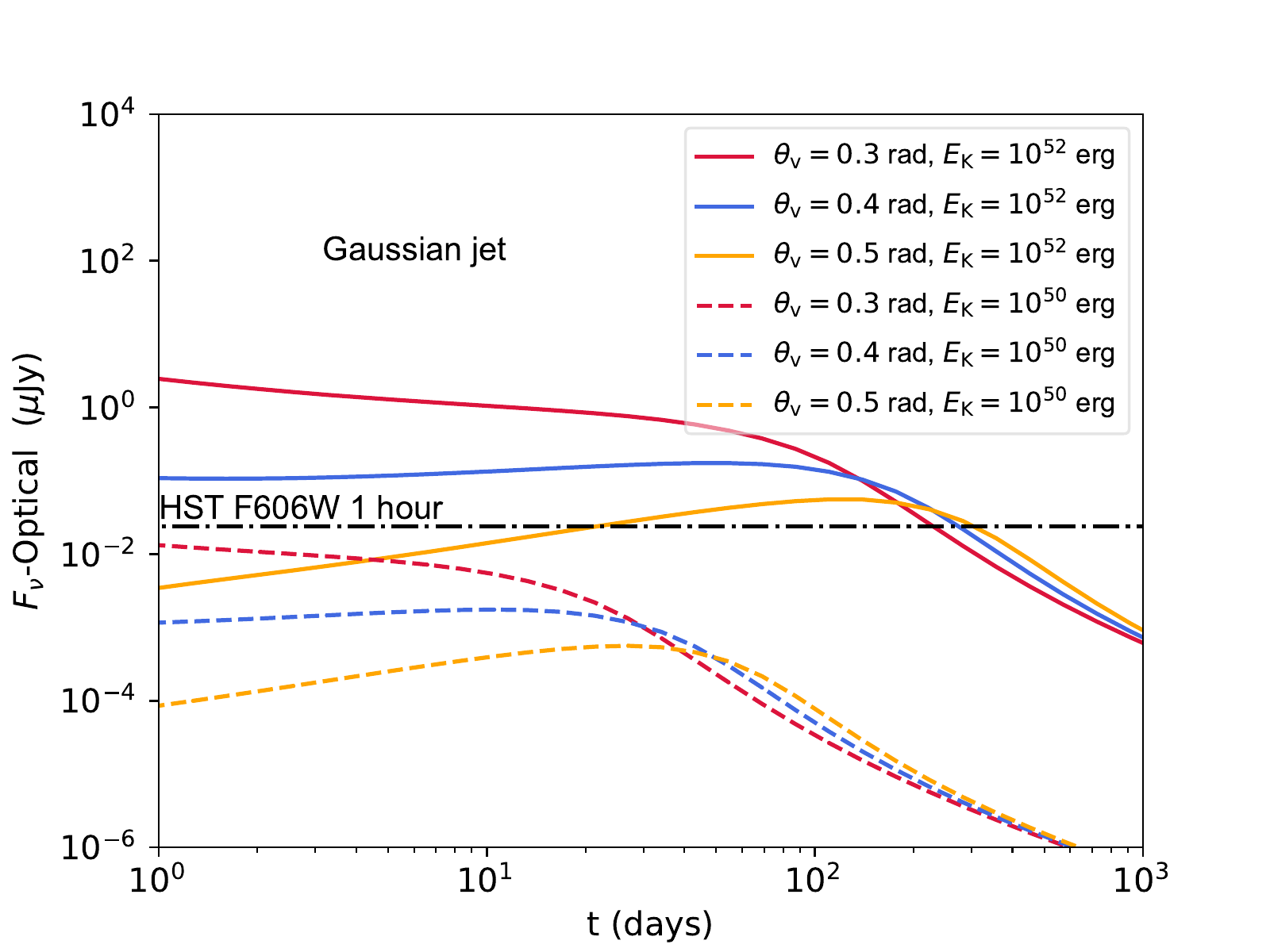}}
    \subfigure{
    \includegraphics[width=0.45\textwidth]{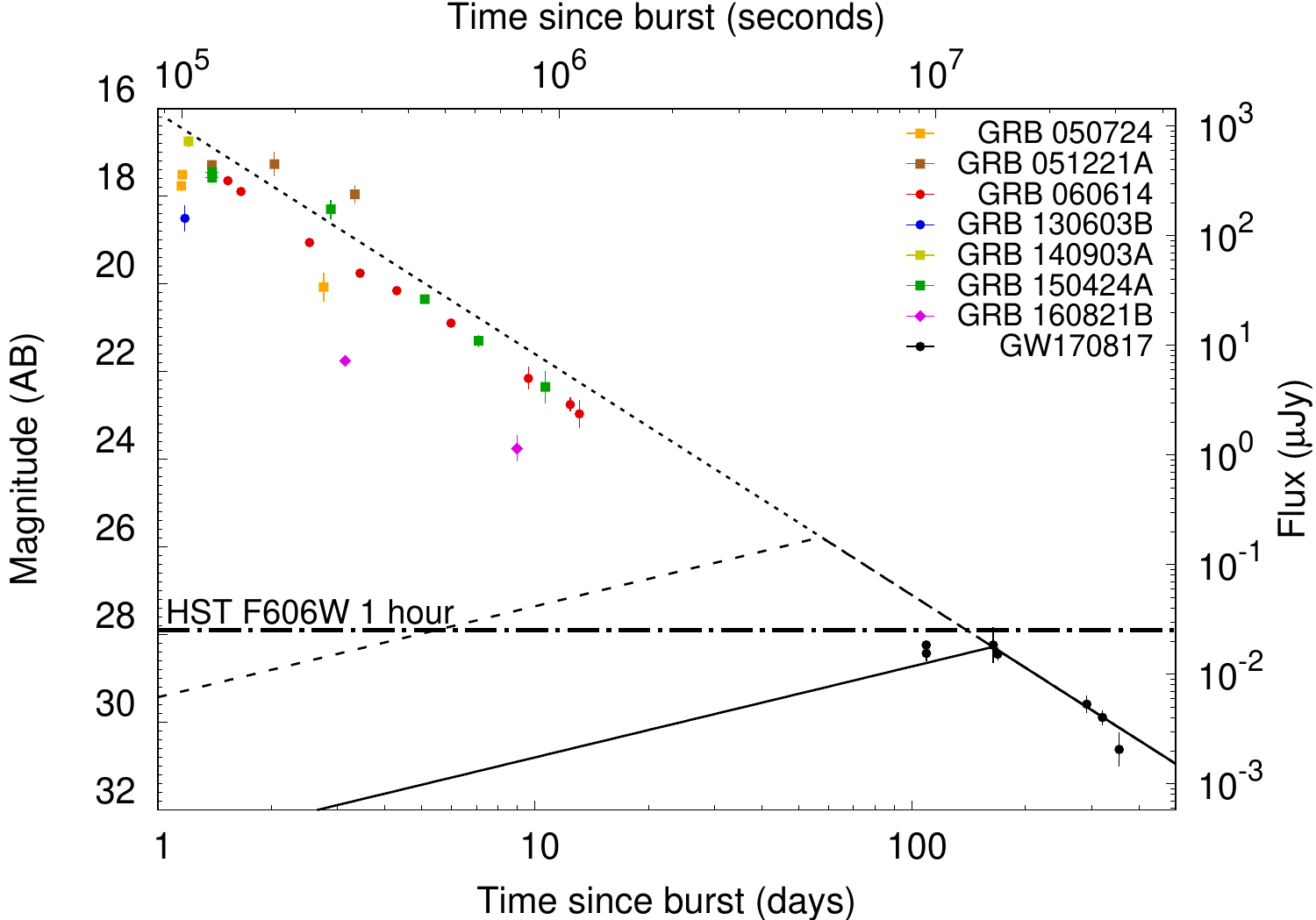}}
    \subfigure{
	\includegraphics[width=0.45\textwidth]{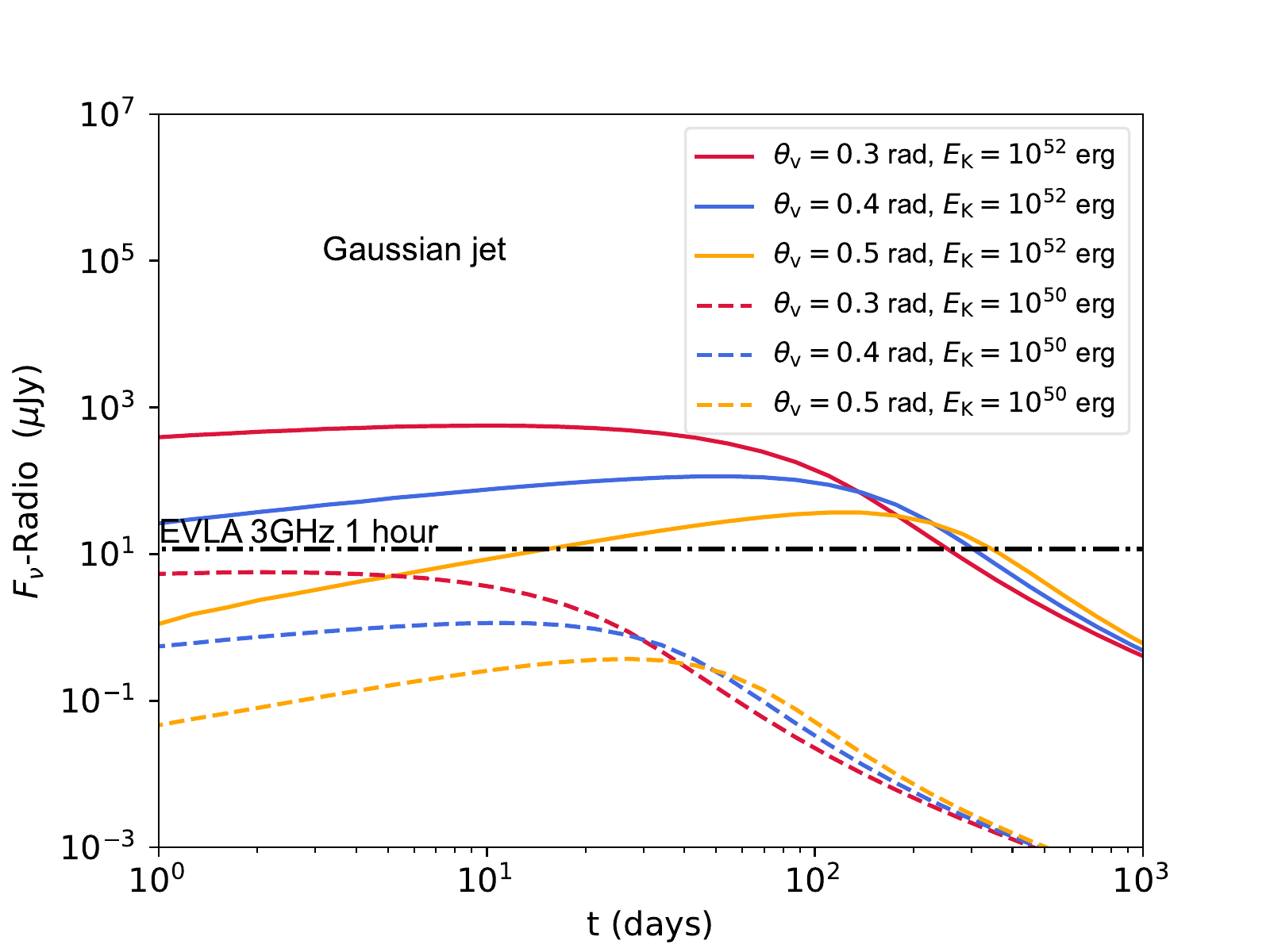}}
    \subfigure{
    \includegraphics[width=0.45\textwidth]{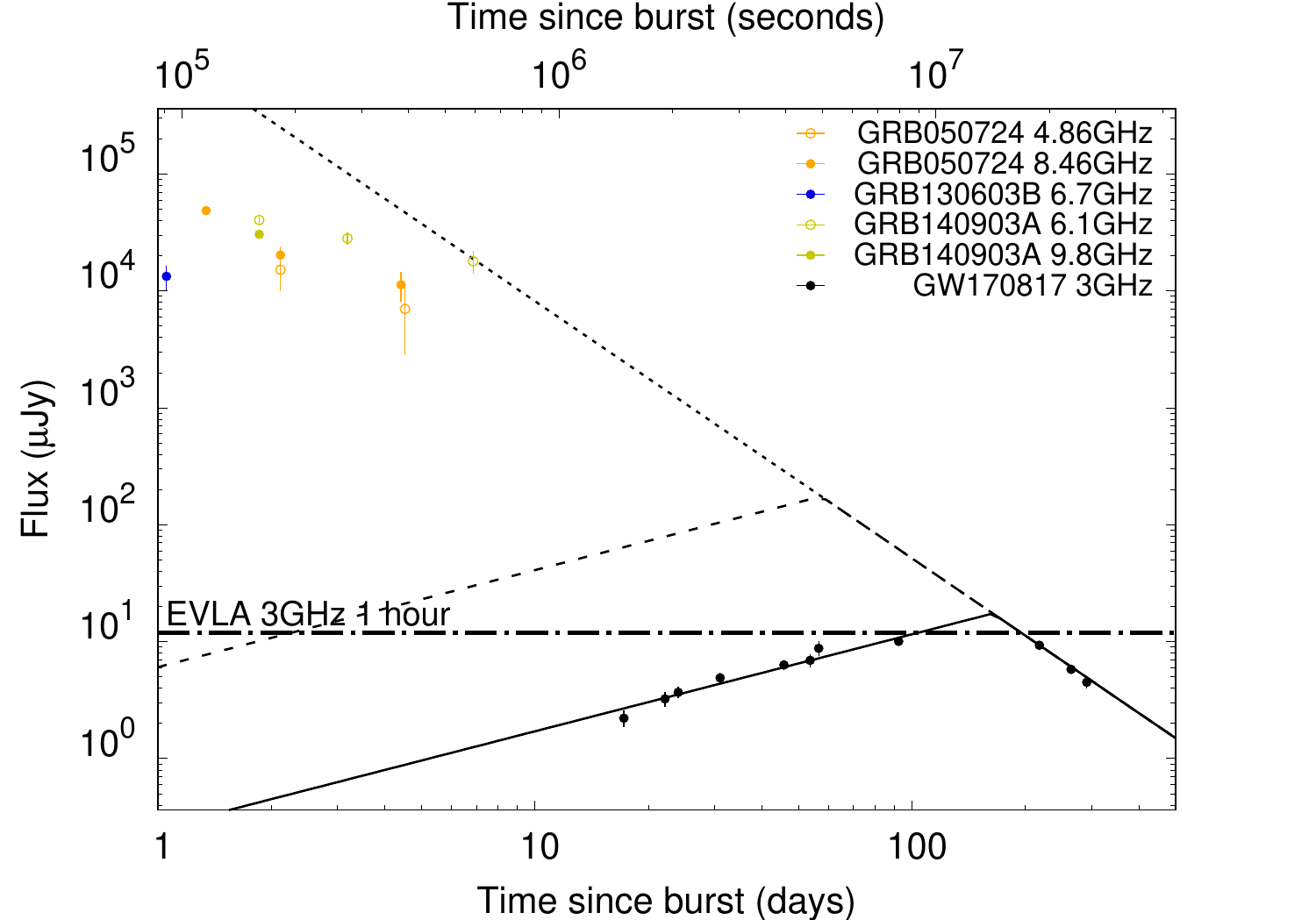}}
    \subfigure{
	\includegraphics[width=0.45\textwidth]{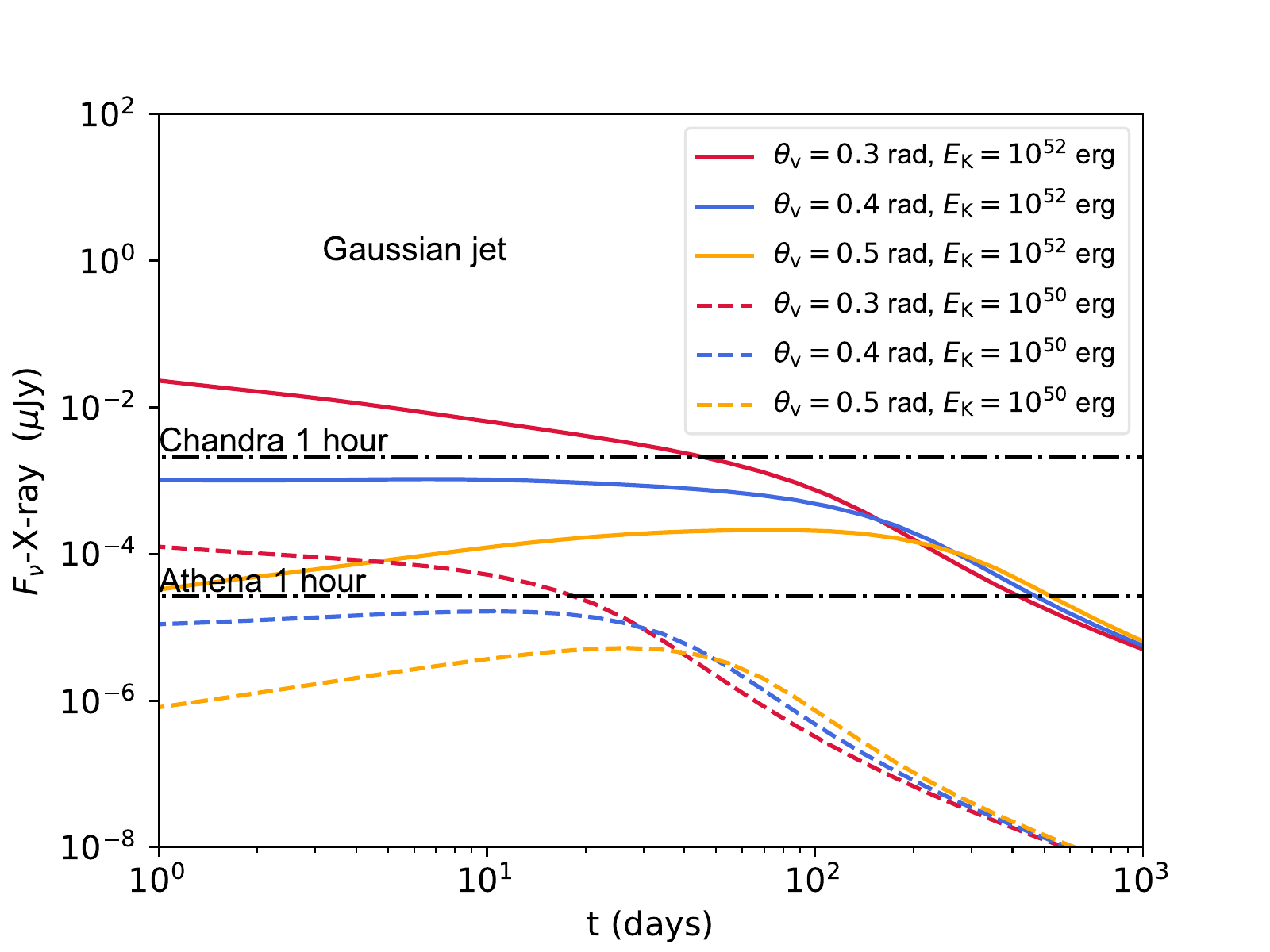}}
	\subfigure{
    \includegraphics[width=0.45\textwidth]{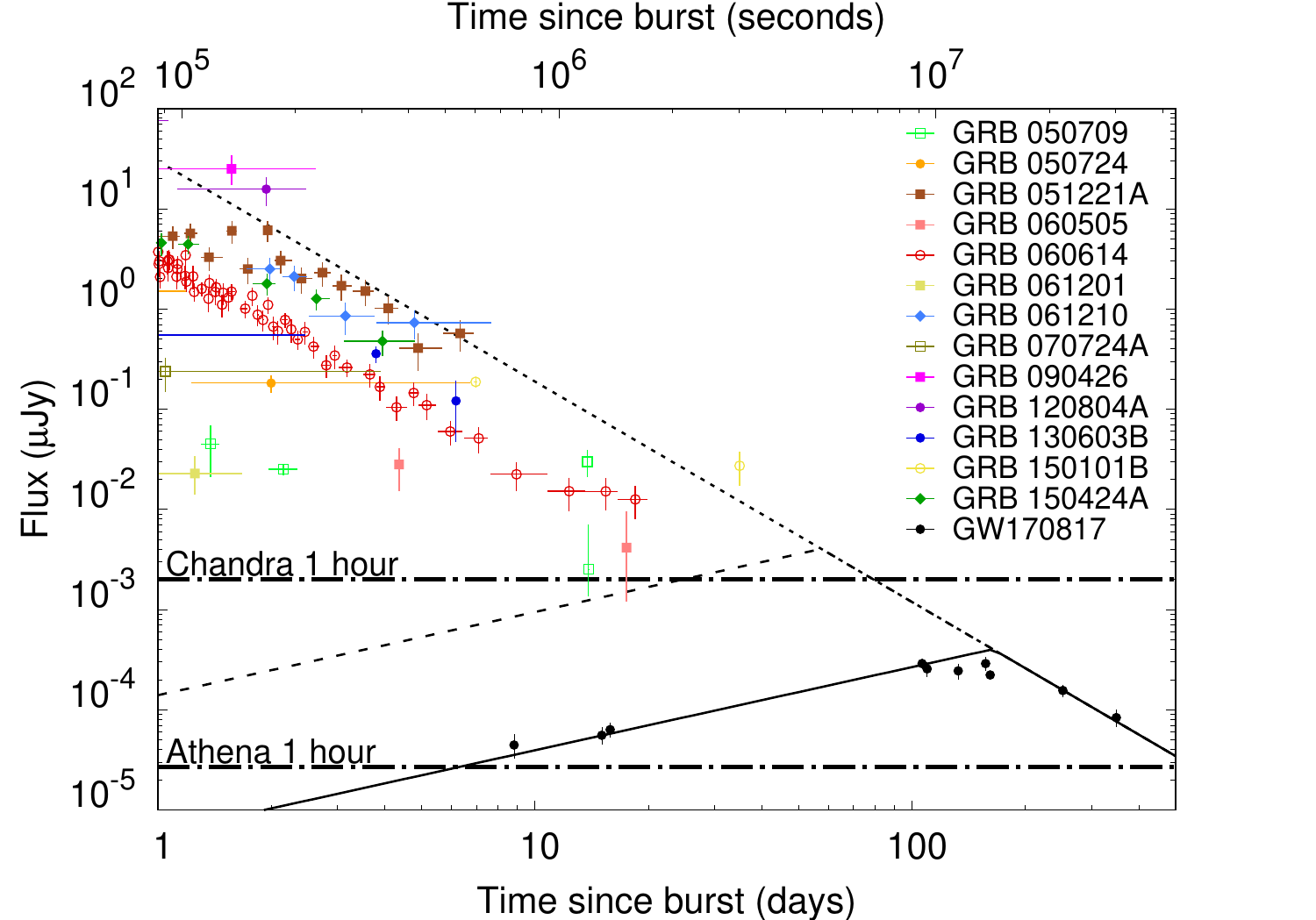}}
	
	\caption{\small The left column shows are the numerical results on the forward shock afterglow emission of off-axis Gaussian ejecta with different kinetic energies $E_{\rm k}$ and viewing angles $\theta_{\rm v}$. Other afterglow parameters are $n_0=10^{-3}~{\rm cm^{-3}}$, $\epsilon_{\rm e}=0.1$, $\epsilon_{\rm B}=0.01$, $p=2.2$, and $d_{\rm L}=120$ Mpc. The calculations are performed by the {\footnotesize AFTERGLOWPY} \citep{ryan2020gamma}. The sensitivity lines (i.e., the vertical dash-dotted lines) of \textit{Hubble Space Telescope} (HST), Expanded Very Large Array (EVLA), Chandra, and {\it Athena} observations are also shown. The right column shows the observed forward shock emission of some bright short GRBs and GRB 170817A/GW170817 (The datasets are adopted from \citet{Duan2019ApJL}). It is clear that if GRB 170817A was viewed on axis, the afterglow emission would be among the brightest ones (i.e., the gray dotted extrapolation line; see also \citet{Duan2019ApJL}). Even so, the shift of GRB 170817 to a distance of 120 Mpc would render the detection by both HST and EVLA challenging. However, if the viewing angle is reduced by $2/3$ (i.e., the dashed lines for the rising phase), the afterglow emission would be bright enough to be well recorded.}
	\label{fig:Gaussian_jet}
\end{figure*}

Here we estimate the detectability of the afterglow emission in two ways. One is to adopt {\footnotesize AFTERGLOWPY} \citep{ryan2020gamma}, which is a public {\footnotesize PYTHON} package for numerical modeling of structured jet afterglows, to calculate the afterglow emission. In the upper panels of Fig.~\ref{fig:Gaussian_jet}, the GRB ejecta is assumed to have a Gaussian profile, and the values of $E_{\rm k}$ and $\theta_{\rm v}$ have a wide distribution (other parameters are fixed). As one can see, at the luminosity distance of $120$ Mpc, only the very energetic outflow (i.e., $E_{\rm k}\geq 10^{52}$ erg) can generate the detectable afterglow emission for HST and EVLA-like telescopes. The other way is based on the statistical studies of the short GRB afterglow observations. Following \citet{Duan2019ApJL} we shift the ``long-lasting" afterglow emission of some bright short GRBs and GW170817/GRB 170817A to a luminosity distance of $d_{\rm L}=120$ Mpc. The forward shock emission of GW170817/GRB 170817A would be one of the bright afterglows of short GRBs if viewed on axis (see the gray dotted lines for the extrapolation to the ``on-axis" case in the bottom panels of Fig.~\ref{fig:Gaussian_jet}). Even so, the optical and radio forward shock emission can be only marginally detectable for HST and EVLA at a distance of $120$ Mpc. This is insufficient for a reliable measurement of the viewing angle. Then we may need an even brighter intrinsic afterglow emission or a smaller viewing angle. For instance, we can reduce $\theta_{\rm v}$ by a factor of 1.5. This would shorten the peak time of the forward shock emission by a factor of $\sim 3$. The flux would thus be enhanced by a factor of $\approx 3^{p} \sim 10$, which would be bright enough to be robustly measured in a reasonably long time range (see the gray dashed line in the bottom panels of Fig.~\ref{fig:Gaussian_jet}). The physical parameters can thus be reasonably inferred. The caution is that the intrinsic afterglow emission may be too dim for many BNS mergers. We, therefore, expect that our goal (i.e., the evaluation of the viewing angle with the afterglow data) can be achieved in a fraction of neutron star merger events that can generate bright afterglow emissions.

\section{Methods to evaluate the extrinsic GW parameters}\label{sec:method}
Here we describe the methods for estimating the extrinsic GW parameters. It consists of two major scenarios, including the constraints of inclination angle $\iota$ and luminosity distance $d_{\rm L}$ based on different prior assumptions, respectively. In Sec.~\ref{sec:iota}, we focus on future BNS merger events that have identified host galaxies with known redshifts. Then the luminosity distance $d_{\rm L}$ of each event can be constrained a priori by the redshift data of the host galaxy, which we assume to follow a Gaussian distribution. Taking advantage of such information for $d_{\rm L}$, the prospective estimation on $\iota \ (\theta_{\rm v})$ can be obtained by analyzing the GW signal, supposing $H_0$ is well determined. In Sec.~\ref{sec:dl}, we study a specific case in which the line of sight is within the energetic core of the ejecta (i.e., $\theta_{\rm v}\leq \theta_{\rm j}$, $\theta_{\rm j}$ represents the evaluated jet opening angle). The uncertainty of the $\theta_{\rm v}$ (equally, the $\iota$) can be estimated to be within $\sim 0.1$ rad. We show that the degeneracy between $\iota$ and $d_{\rm L}$ can be effectively broken, and a $\sim 3\%$ $(4\%)$ precision measurement of $H_0$ is reachable. In both scenarios, the constraints are obtained using similar methods (but with different assumptions/priors) introduced below. All methods estimate the required parameters from the analytical waveform function but use different ways to acquire the final constraints.
Specifically, we consider three different methods to estimate the inclination angle or luminosity distance. The first method approximates the likelihood of full GW parameter inference analytically, which is based on the Fisher information theories \citep{Finn_PRD_1993, Cutler_PRD_1994, sigma_epsilon_PRD_2019}. The second method further simplifies the approximated posterior distributions assuming they are Gaussian-like to estimate the uncertainties of parameters according to the theory of error. The third method uses the matched-filter technique under the Bayesian framework \citep{Allen:2005fk, Veitch:2014wba, Biwer:2018osg, Ashton:2018jfp, fll_Zackay_2018} to obtain the posterior distributions with simulated signals. The simulation method is reliable but rather time consuming, while the last two approximate methods are much faster, which can well complement the first method.

As anticipated in \citet{Abbott2018LRR}, during the O4 run, LIGO-Livingston/Hanford \cite{Abbott2016LRR, Martynov2016PRD} will reach their design sensitivity. Together with Virgo \cite{Acernese2015} and KAGRA \cite{Kagra2019NatAs}, a four detector network is expected to catch GW signals collaboratively. LIGO-India \cite{Abbott2018LRR, Saleem2021} will participate in the joint GW observation in 2025 to form the LIGO/Vrigo/KAGRA/LIGO-India (LHVKI) detector network. In this work, we adopt the prospective noise curves from \url{https://dcc.ligo.org/LIGO-T2000012/public} for O4 and O5 runs.

\subsection{Inclination angle estimation}\label{sec:iota}
\begin{figure}[ht!]
	\centering
	\includegraphics[width=0.5\textwidth]{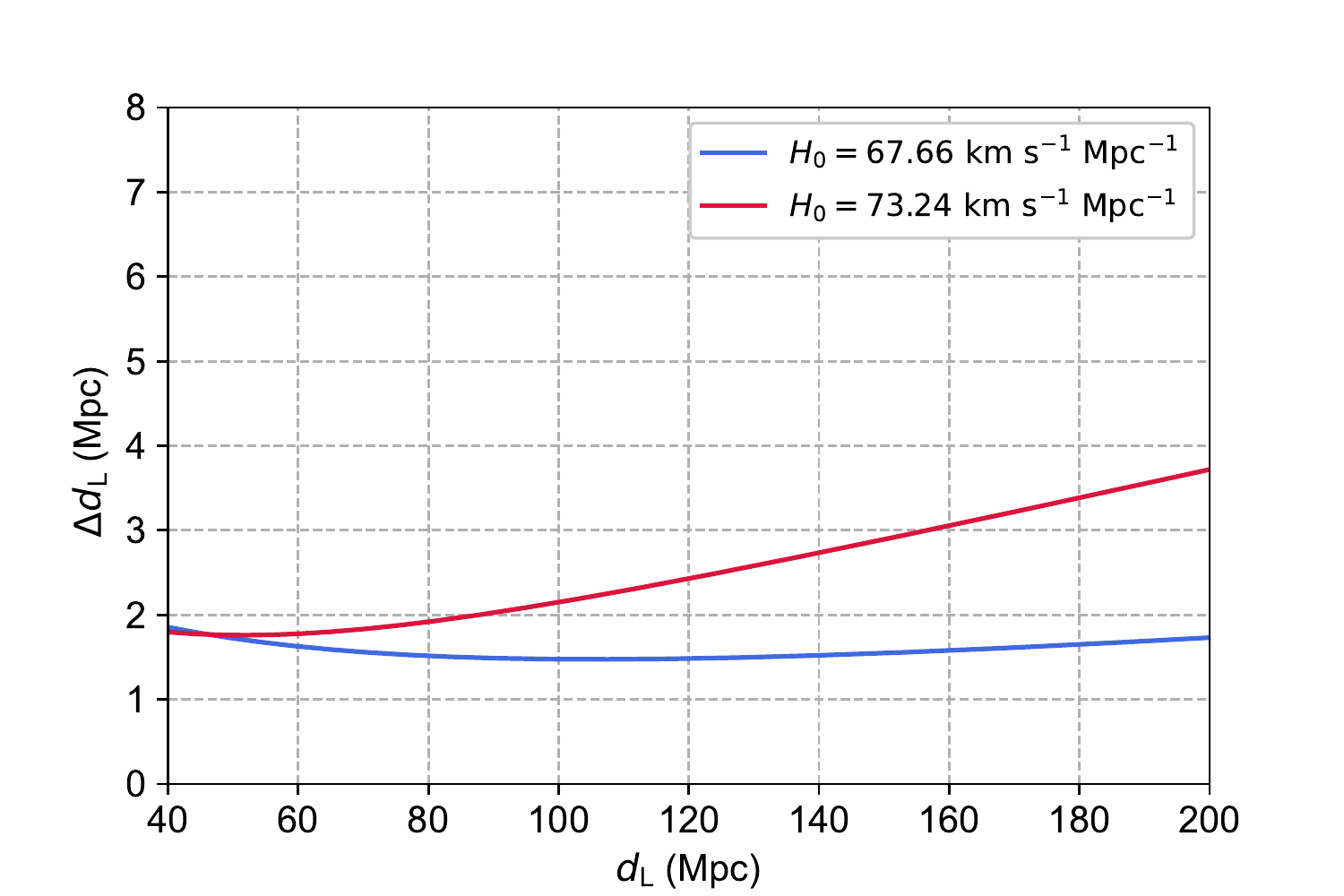}

	\caption{\small The distribution of standard deviation $\Delta d_{\rm L}$ with $d_{\rm L}$ increasing. The red and blue solid lines represent different $H_{0}$ from Planck and SHOES.}
	\label{fig:delta_dl}
\end{figure}

For a BNS merger event with a measured redshift, we can relate the system's luminosity distance and redshift by the Hubble's law, 
\begin{equation}\label{eq:H0}
	v_{\rm H}=v_{\rm r}-v_{\rm p}=H_0d_{\rm L},
\end{equation}
where $v_{\rm H}$ is the local Hubble flow velocity of the galaxy, $v_{\rm r}$ is the recession velocity of the galaxy relative to the CMB frame, and $v_{\rm p}$ is the peculiar velocity of the galaxy. Therefore, using the sole electromagnetic observations, the uncertainty of luminosity distance can be approximated to
\begin{equation}\label{eq:delta_d}
	\frac{\Delta^2 d_{\rm L}}{d^2_{\rm L}}=\frac{\Delta^2v_{\rm r} + \Delta^2v_{\rm p}}{H^2_0d^2_{\rm L}} + \frac{\Delta^2H_0}{H^2_0},
\end{equation}
where the uncertainty of $v_{\rm p}$ is estimated following \citet{Ma_Pan2014MNRAS}, in which the authors gave the median and $\pm 1\sigma$ variance of the bulk velocity magnitude as a function of the distance in their Fig.~4. Meanwhile, we choose a typical error of $50~\rm km\ s^{-1}$ \citep{Graziani2019MNRAS} for a spectroscopic redshift as the uncertainty of the recession velocity. For $H_0$ and $\Delta H_0$, two discrepant results from Planck \citep{Planck_2020} and SHOES \citep{SHoES_2021} are used. Thus, $\Delta d_{\rm L}$ calculated with Eq.~(\ref{eq:delta_d}) can be incorporated into the GW parameter estimation as a standard deviation of the Gaussian prior of $d_{\rm L}$. Figure.~\ref{fig:delta_dl} shows the values of $\Delta d_{\rm L}$ we used in the following calculations.

\subsubsection{The evaluation methods}
We neglect both the high-mode and precession effects, which is a reasonable approximation for BNS mergers. Besides, we do not consider the spin or tidal deformability terms in the waveform calculations and use the stationary phase approximation to the inspiral waveform in the frequency domain \citep{TaylorF2_PRD_2009} (except for the simulation method). Therefore, the parameters of the BNS merger events generally include the chirp mass $\mathcal{M}$, coalescence time $t_{\rm c}$, luminosity distance $d_{\rm L}$, inclination angle $\iota$, polarization angle $\psi$, phase angle $\phi$, and two sky position parameters, right ascension $\alpha$ and declination $\delta$. We first introduce method A and method B, which give analytic representations for $\Delta \iota$. We use two parameters, $\sigma_{\rm D}$ and $\epsilon_{\rm D}$ \citep{Cutler_PRD_1994}, to describe the response of a detector network, which can be related to the polarization angle of the source, the sensitivities of different detectors, and the relative position between detectors and the GW source. More specifically, $\sigma_{\rm D}$ reflects the detectability of a GW signal presented in the detectors. It will affect the signal to noise ratio (SNR), while $\epsilon_{\rm D}$ represents the response of a network to the two polarization modes. To check the robustness of the two analytical methods, some simulations (i.e., injection and recovery) of GW parameter inference for BNS mergers are carried out in method C. In this approach, the {\footnotesize BILBY} \citep{bilby2019ApJS,bilby2020MNRAS}, {\sc PyCBC} \citep{pycbc2019PASP} packages and {\footnotesize DYNESTY} \citep{dynesty2020MNRAS} sampler are applied to yield the posterior probability distribution for $\iota$. Below is a detailed description of these three methods.

\begin{itemize}
\item { \it {Method A: Cutler and Flanagan approximation.}} \label{sec:MA}
--For BNS merger with an identified electromagnetic counterpart, the sky position parameters $\alpha$ and $\delta$ and the intrinsic parameters $\mathcal{M}$ and $t_{\rm c}$ can be well determined. Only four other parameters are to be measured, and the posterior probability density can be written as
\begin{widetext}
\begin{equation}\label{eq:prob}
P({d_{\rm L}}^{\prime},{\cos \iota}^{\prime},{\phi_{\rm c}}^{\prime},{\psi}^{\prime})\propto \mathcal{L}(\boldsymbol{s}|{d_{\rm L}}^{\prime},{\cos \iota}^{\prime},{\phi_{\rm c}}^{\prime},{\psi}^{\prime},\alpha,\delta,\mathcal{M},t_{\rm c})\pi({d_{\rm L}}^{\prime},{\cos \iota}^{\prime},{\phi_{\rm c}}^{\prime},{\psi}^{\prime}),
\end{equation}
where $\mathcal{L}$ is the likelihood function and $\pi$ represents the prior of these parameters. \citet{sigma_epsilon_PRD_2019} gave an analytic likelihood for $\iota$ and $d_{\rm L}$ under \citet{Cutler_PRD_1994}'s approximation, i.e.,

\begin{equation}\label{eq:P_iota_dl}
\begin{split}
P(d_{\rm L}^{\prime}, \iota^{\prime}) \propto \exp \left( -\frac{\rho^2_0 \sigma_{\rm D}}{2} \left[ (1-\epsilon_{\rm D}\cos{4\bar{\psi}})\left(\cos{\iota} -\frac{\cos{\iota^{\prime}}\,d_{\rm L}}{d_{\rm L}^{\prime}}\right)^2 \right. \right.\\+ \left. \left. (1+\epsilon_{\rm D}\cos{4\bar{\psi}})\left(\frac{1+\cos^2\iota}{2} - \frac{(1+\cos^2 \iota^{\prime})\,d_{\rm L}}{2d_{\rm L}^{\prime}}\right)^2\right]\right).
\end{split}
\end{equation}

Then, $P(\iota)$ can be calculated by marginalizing over $d_{\rm L}$ with this posterior distribution,
\begin{equation}\label{eq:Pdv_cos}
\begin{split}
P(\iota^{\prime} ) \propto \sin(\iota^{\prime}) \int \ud d_{\rm L}^{\prime} \,\pi(d_{\rm L}^{\prime})\,P(d_{\rm L}^{\prime}, \iota^{\prime}).
\end{split}
\end{equation}
In the above two equations, $\pi (d^{\prime}_{\rm L})$ is the prior distribution of $d_{\rm L}$, $\bar{\psi} =\psi + \Delta \psi$ ( $\Delta \psi $ is an additional rotation angle for the detector network), and $\rho_0$ is the single-detector SNR for a face-on source located overhead, 
\begin{equation}\label{eq:rho0}
\rho_0^2 \equiv \frac{1}{d_{\rm L}^2}\biggl[\frac{5}{6\pi ^{4/3}} \frac{(G\mathcal{M})^{5/3}}{c^3} \int_{0}^{\infty} \frac {f^{-7/3} \ud f}{S_{\rm n,aver}}\biggr] 	
\end{equation}
where $S_{\rm n,aver}$ denotes the average noise spectral density over all detectors. And the total SNR ($\rho$) for a network can be expressed as \citep{sigma_epsilon_PRD_2019}

\begin{equation}\label{eq:snr}
\rho^2 =\frac{\sigma_{\rm D}\rho^2_0}{4} \biggl[ 1+\cos^4 \iota + 6\cos^2 \iota + \epsilon_{\rm D} (1-\cos^2\iota)^2\cos 4\bar{\psi} \biggr].
\end{equation}
\end{widetext}

\item {\it Method B: error synthesis theories.}\label{sec:MB}
 --We extend the Eqs.~(3.16) and (3.30) in Ref.~\cite{Finn_PRD_1993} to a multidetector case by Eq.~(\ref{eq:snr}), and the terms in square brackets of their (3.31) can be substituted with 
\begin{widetext}
\begin{equation}\label{eq:Theta_AD}
\begin{aligned}
\Theta^2 &\equiv \mathcal{A}^2 d^{2}_{\rm L}\\ 
&= 4 \sigma_{\rm D} [1+\cos^4 \iota + 6\cos^2 \iota + \epsilon_{\rm D} (1-\cos^2\iota)^2\cos4\bar{\psi}],
\end{aligned}
\end{equation}
which in turn can be expressed into
\begin{equation}\label{eq:cos_iota2}
\cos^2 \iota = \sqrt{\frac{\mathcal{A}^2d^2_{\rm L}}{4 \sigma_{\rm D} (1+\epsilon_{\rm D}\cos4\bar{\psi}) } +\frac{16}{(1+\epsilon_{\rm D}\cos4\bar{\psi})^2 } - \frac{8}{1+\epsilon_{\rm D}\cos4\bar{\psi}}} 
-\frac{4}{1+\epsilon_{\rm D}\cos4\bar{\psi}} + 1. 
\end{equation}

Thus, based on the contemporary theories of error synthesis, the uncertainty of inclination angle $\Delta \iota$ is estimated via
\begin{equation}\label{eq:delta_iota2}
\begin{aligned}
\Delta^2 \iota = & \bigg( \frac{1}{4\sin 2\iota \mathcal{Q}}\frac{\mathcal{A}^2 d_{\rm L}^2}{\sigma_{\rm D} (1+\epsilon_{\rm D} \cos4\bar{\psi})}\bigg) ^2\bigg(\frac{\Delta^2 \mathcal{A}^2}{\mathcal{A}^2} + \frac{\Delta^2 v_{\rm r}+\Delta^2 v_{\rm p}}{(v_{\rm r} -v_{\rm p})^2} + \frac{\Delta^2 H_0}{H_0^2 } \bigg) \\
&+ \biggl[ \frac{1}{2\sin2\iota \mathcal{Q}} \bigg( \frac{\mathcal{A}^2d_{\rm L}^2\epsilon_{\rm D}\sin4\bar{\psi}} {\sigma_{\rm D}(1+\epsilon_{\rm D}\cos4\bar{\psi})^2} + \frac{128\epsilon_{\rm D}\sin4\bar{\psi}}{(1+\epsilon_{\rm D}\cos4\bar{\psi})^3}- \frac{32\epsilon_{\rm D}\sin4\bar{\psi}}{(1+\epsilon_{\rm D}\cos4\bar{\psi})^2}\bigg) - \frac{16\epsilon_{\rm D}\sin4\bar{\psi}}{\sin2\iota(1+\epsilon_{\rm D}\cos4\bar{\psi})^2} \biggr]^2 \Delta^2 \bar{\psi}
\end{aligned}
\end{equation}
where
\begin{equation}\label{eq:Q}
\mathcal{Q}=\sqrt{\frac{\mathcal{A}^2d^2_{\rm L}}{4 \sigma_{\rm D} (1+\epsilon_{\rm D}\cos4\bar{\psi}) } +\frac{16}{(1+\epsilon_{\rm D}\cos4\bar{\psi})^2 } - \frac{8}{1+\epsilon_{\rm D}\cos4\bar{\psi}}},
\end{equation}
and $\Delta \mathcal{A}/\mathcal{A} \approx 1/\rho$.

The error of $\bar{\psi}$ has been estimated by the Fisher matrix approach \citep{Seto_iota_psi_PRD_2007}, which reads
\begin{equation}
\begin{aligned}
\langle \Delta \bar{\psi} \Delta \bar{\psi} \rangle \approx &\frac{d_{\rm L}^2[1+6\cos^2\iota + \cos^4\iota + (1-\cos^2 \iota)^2 \epsilon_{\rm D} \cos4\bar{\psi}]}{D_0^2 \sigma_{\rm D} (1-\epsilon^2_{\rm D})(1-\cos^2\iota)^4},
\end{aligned}
\end{equation}
\end{widetext}
where $D_0$ can be expressed by Eq.~(\ref{eq:rho0}) after comparing Eq.~(\ref{eq:snr}) with \citet{Seto_iota_psi_PRD_2007}'s Eq.~(18), i.e., 
\begin{equation}\label{eq:D0}
D^2_0 = \frac{5}{6\pi ^{4/3}} \frac{(G\mathcal{M})^{5/3}}{c^3} \int_{0}^{\infty} \frac {f^{-7/3} \ud f}{S_{\rm n}}.
\end{equation}
Thus, for the BNS merger accompanied by an electromagnetic counterpart, the uncertainty of the inclination angle can be directly estimated with the above equations.

\item {\it Method C: recovery of simulated signals.}\label{sec:MC}
--The previous two approaches are essentially analytic. As an independent check, we perform the full end-to-end Bayesian inference on synthetic data. We assume that both neutron stars have aligned and low spins, and the noises in the detectors are colored Gaussian with known power spectrum densities (PSDs). We first generate the simulated signals using the IMRPhenomD\_NRTidal \citep{IMRPhenomDNRTidal-2_PRD_2017, IMRPhenomDNRTidal-1_PRD_2019} waveform with the parameter configuration shown in Table.~\ref {tb:inject}. Then, these signals are injected into the detector network and recovered with the same approximant and PSDs used in generating the injection. Meanwhile, the ``relative binning" technique is applied for rapid parameter estimation \citep{fll_Zackay_2018,fll_Finstad_2020}, and the marginalized posterior distributions for each parameter can be obtained using Bayesian inference and nested sampling with the priors summarized in Table.~\ref {tb:inject}. For simplicity, five parameters $\alpha$, $\delta$, $t_{\rm c}$, and $\Lambda_{1,2}$ are fixed as the injection values. We assign uniform sine and Gaussian priors for $\iota$ and $d_{\rm L}$, respectively, and other parameters are uniform in their domains. 

\begin{table*}[ht!]
\begin{ruledtabular}
\centering
\caption{Parameters, injection configurations, and priors}
\label{tb:inject}
\begin{tabular}{lccc}
Names                   &Parameters                   &Injected value       &Priors of parameter inference         \\ \hline            
Chirp mass              &$\mathcal{M}/M_{\odot}$      &1.2                  &Uniform(0.4, 4.4)     \\
Mass ratio              &$q$                          &0.9                  &Uniform(0.125, 1.0) \\
Spin magnitude          &$\chi_{1,2}$                 &0.02, 0.03           &Uniform(0., 0.99)            \\
Coalescence phase       &$\phi_{\rm c}$               &1.57                 &Uniform(0, 2$\pi$)         \\
Polarization of GW      &$\psi$                       &1.57                 &Uniform(0, 2$\pi$)    \\	
Coalescence time        &$t_{\rm c}/\rm s$            &1187008882.42        &1187008882.42 \\
Right ascension         &$\alpha$                     &1.57                 &1.57       \\
Declination             &$\delta$                     &0                    &0      \\
Tidal deformability     &$\Lambda_{1,2}$              &412, 754             &412, 754\\
Inclination angle       &$\iota$                      &$\iota_0$            &Uniform sine and $\mathcal{N}$($\iota_0$, $\Delta \iota$)\textsuperscript{a}        \\
Luminosity distance     &$d_{\rm L}$/Mpc                  &$d_{\rm L,0}$         &$\mathcal{N}$($d_{\rm L,0}$, $\Delta d_{\rm L}$) and uniform comoving volume\textsuperscript{a}   \\
\end{tabular}
\tablenotetext{Note that, in the first scenario (i.e., the estimation of $\iota$), we use the former; otherwise, we use the latter. In the estimation of $\iota$, the values of $\Delta d_{\rm L}$ we use are presented in Fig.~\ref{fig:delta_dl}. In the estimation of the Hubble constant, the value of $\Delta \iota = 0.1$ rad is assumed.}
\end{ruledtabular}
\end{table*}

\end{itemize}

\subsubsection{The Hubble tension: An obstacle for precise inclination angle measurements}\label{sec:HubbleTension-caution}

\begin{figure}[ht!]
\centering
	\includegraphics[width=0.52\textwidth]{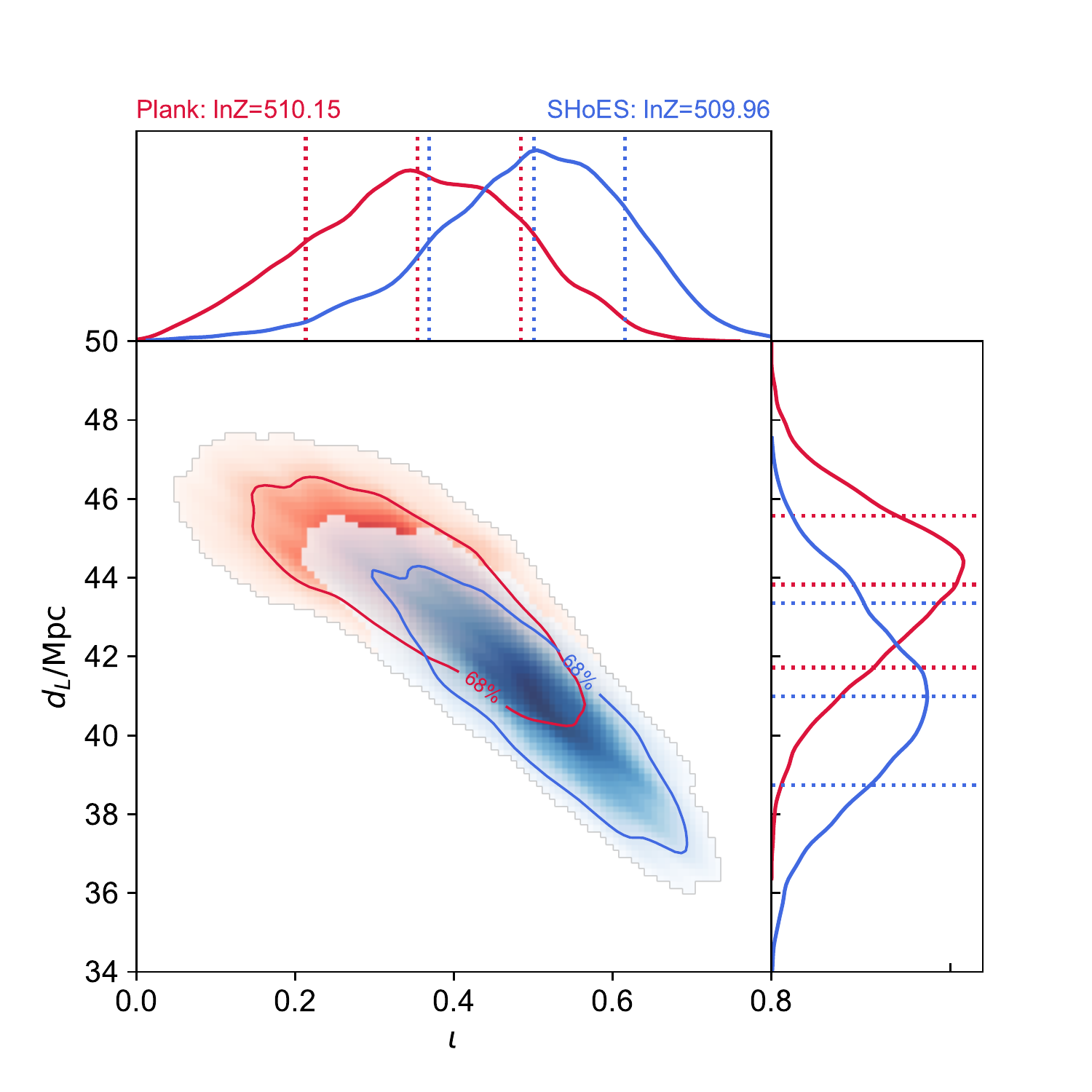}
	\caption{The estimation of luminosity distance and inclination angle for GW170817 with identified electromagnetic counterparts in the case of two different Hubble constants. The blue (red) line is for $H_0=$ 73.24 (67.66) $\rm km\,s^{-1}\,Mpc^{-1}$. The uncertainties of $H_0$ are taken from Refs.~\cite{Planck_2020,SHoES_2021}, respectively. }
	\label{fig:170817}
\end{figure}

Equation.~(\ref{eq:delta_d}) shows that the relative error of Hubble constant should impact the uncertainties of $d_{\rm L}$ and then $\Delta \iota$. If we consider two different results by the Plank Collaboration \cite{Planck_2020} and \citet{SHoES_2021} and take their systematic difference as $\Delta H_0$, the term $\Delta^2 H_0/H^2_0$ will be comparable to other terms or even become dominant. Thus, the influence of Hubble tension needs to be considered. 
As an example, we reanalyze the data of GW170817 with two different Hubble constants \citep{Planck_2020, SHoES_2021}. Except for the luminosity distance, the priors of other parameters are the same as those adopted by \citet{Abbott1708172017APJL}. Following Ref.~\cite{Abbott2017Nature}, the Hubble velocity is taken to be $v_{\rm H}=3017\pm 166 \rm \,km\,s^{-1}$ (the uncertainty $166 \rm \,km\,s^{-1}$ includes the uncertainties of recession velocity $72 \rm \,km\,s^{-1}$ and peculiar velocity $150 \rm \,km\,s^{-1}$). Therefore, the prior constraints on $d_{\rm L}$ follow Gaussian distribution with $\mu =44.59$ and $\sigma = 2.47 \rm \, Mpc$ for the Plank Collaboration \cite{Planck_2020} and $\mu=41.21$ and $\sigma =2.38 \rm \, Mpc$ for \citet{SHoES_2021}. Figure.~\ref{fig:170817} shows that these two different prior constraints of luminosity distance can affect the estimation of inclination angle. The similar values of the two logarithm of Bayes's evidences ${\rm ln}Z$ mean that the two discrepant measurements of the Hubble constant cannot be distinguished with GW170817, though the resulting $d_{\rm L}$ do show some difference.
Previously, \citet{Troja2019MNRAS} carried out a joint analysis of the GW data and the afterglow light curve of GRB 170817A to constrain $\theta_{\rm v}$ and $\theta_{\rm c}$. Here, we focus on analyzing the GW data and only take into account the redshift of the electromagnetic counterpart. Thus, our results do not suffer from the possible uncertainties involved in the afterglow modeling, which is important for one of our main purposes: to distinguish between different modeling approaches. Comparing with the case of Ref.~\citep{Planck_2020}, the larger $H_0$ reported in \citet{SHoES_2021} yields a higher inclination angle $\iota$. 
In conclusion, this result highlights that Hubble tension should significantly influence estimating $\iota$. 
The inclination angle could be robustly reconstructed only when the Hubble tension is solved (i.e., the uncertainty is within, for instance, $\sim 0.03H_0$); otherwise, the afterglow modeling cannot be calibrated with the unbroken intrinsic degeneracy.

\subsection{Hubble constant estimation}\label{sec:dl}
In this section, we try to measure the Hubble constant by extracting the luminosity distance from the GW data. In a specific case, the GRB is observed within the energetic core of the ejecta. Hence, we will have a small uncertainty for $\iota$ ($\theta_{\rm v}$) that follows a Gaussian distribution $\iota \sim \mathcal{N}(\iota_0, \Delta \iota)$. For a prospective estimate, only with $\Delta \iota \lesssim 10^{\circ}$ \citep{Chen2020PRL}, the Hubble tension might have a chance to be solved. Also, \citet{Jin2018ApJ} found a typical opening angle $\sim 0.1$ rad for the short GRB outflows, and such a value has been widely adopted in the multimessenger detection prospect projections (e.g., \cite{Mastrogiovanni2021AA}).
Therefore, if some bright on-axis afterglows are detected, the unambiguous detection of a jet break would yield an estimated $\theta_{\rm j}$, which sets an upper limit on $\theta_{\rm v}$ as well as its uncertainty (i.e., it is likely to be $\leq 0.1$ rad). Such events would play a crucial role in tightly constraining $H_0$ (see Sec.~\ref{sec:IVB} for further discussion).

With the help of this electromagnetic counterpart information, the degeneracy between $\iota$ and $d_{\rm L}$ can be effectively broken. We use very similar methods as Sec.~\ref{sec:iota} including the analytic calculations and GW simulations to obtain the posterior distribution of luminosity distance. Corresponding to method A, the posterior probability of $d_{\rm L}$ is given by adjusting Eq.~(\ref{eq:Pdv_cos}), i.e.,
\begin{equation}\label{eq:Pdl}
\begin{split}
P(d_{\rm L}^{\prime}) \propto \pi(d_{\rm L}^{\prime}) \int \ud \iota^{\prime} \pi(\iota^{\prime}) P(d_{\rm L}^{\prime}, \iota^{\prime}),
\end{split}
\end{equation}
where $\pi(\iota^{\prime})$ is the prior probability of $\iota$ following Gaussian distribution and $\pi(d_{\rm L}^{\prime})$ is the prior probability of $d_{\rm L}$ which is uniform in a comoving volume. For method B, the uncertainty of $d_{\rm L}$ can be estimated as

\begin{widetext}
\begin{equation}\label{eq:delta_dl}
\begin{aligned}
\frac{\Delta^2 d_{\rm L}}{d_{\rm L}^2} = 
\frac{\Delta^2 \mathcal{A}}{\mathcal{A}^2} + \bigg(\frac{8\sigma_{D}\epsilon_{D}\sin^4\iota \sin 4\bar{\psi}}{\mathcal{A}^2d_{\rm L}^2}\bigg)^2 \Delta^2 \bar{\psi}
+ \bigg[\frac{8\sigma_{D}\sin{\iota}\cos{\iota}\big(-\cos^2\iota -3 + \epsilon_{D} \sin^2\iota \cos 4\bar{\psi}\big)}{\mathcal{A}^2d_{\rm L}^2} \bigg]^2 \Delta^2 \iota .
\end{aligned}
\end{equation}
\end{widetext}
Both Eqs.~(\ref{eq:Pdl}) and (\ref{eq:delta_dl}) can give the uncertainty of $d_{\rm L}$. Thus, the uncertainty of $H_0$ for methods A and B can be obtained using Eq.~(\ref{eq:delta_d}).
As for method C here, we can use the same injection configurations as adopted in Sec.~\ref{sec:iota}, but with different settings of priors (see Table.~\ref{tb:inject} for details). Once we have got the posterior samples for $d_{\rm L}$, according to Hubble's law, it is pretty convenient to obtain the probability distribution for $H_0$ by the Monte Carlo way.

\section{Prospective Estimation and Analysis}\label{sec:result}
With the methods introduced in the previous section, we present the estimation results and discuss their applications. First, we give the uncertainties of inclination angle estimation in different cases and ``identify" the most optimistic constraint on $\Delta \iota$. Then, we discuss the prospect of calibrating the afterglow modeling with such constraints. Second, we give the prospective constraints on luminosity distance in the case of on-axis observation of BNS mergers and predict the precision for measuring the Hubble constant in the future.

\subsection{The uncertainties of inclination angle}\label{sec:IVA}

\begin{figure*}[ht!]
\centering	
\subfigure[]{
\includegraphics[width=0.45\textwidth]{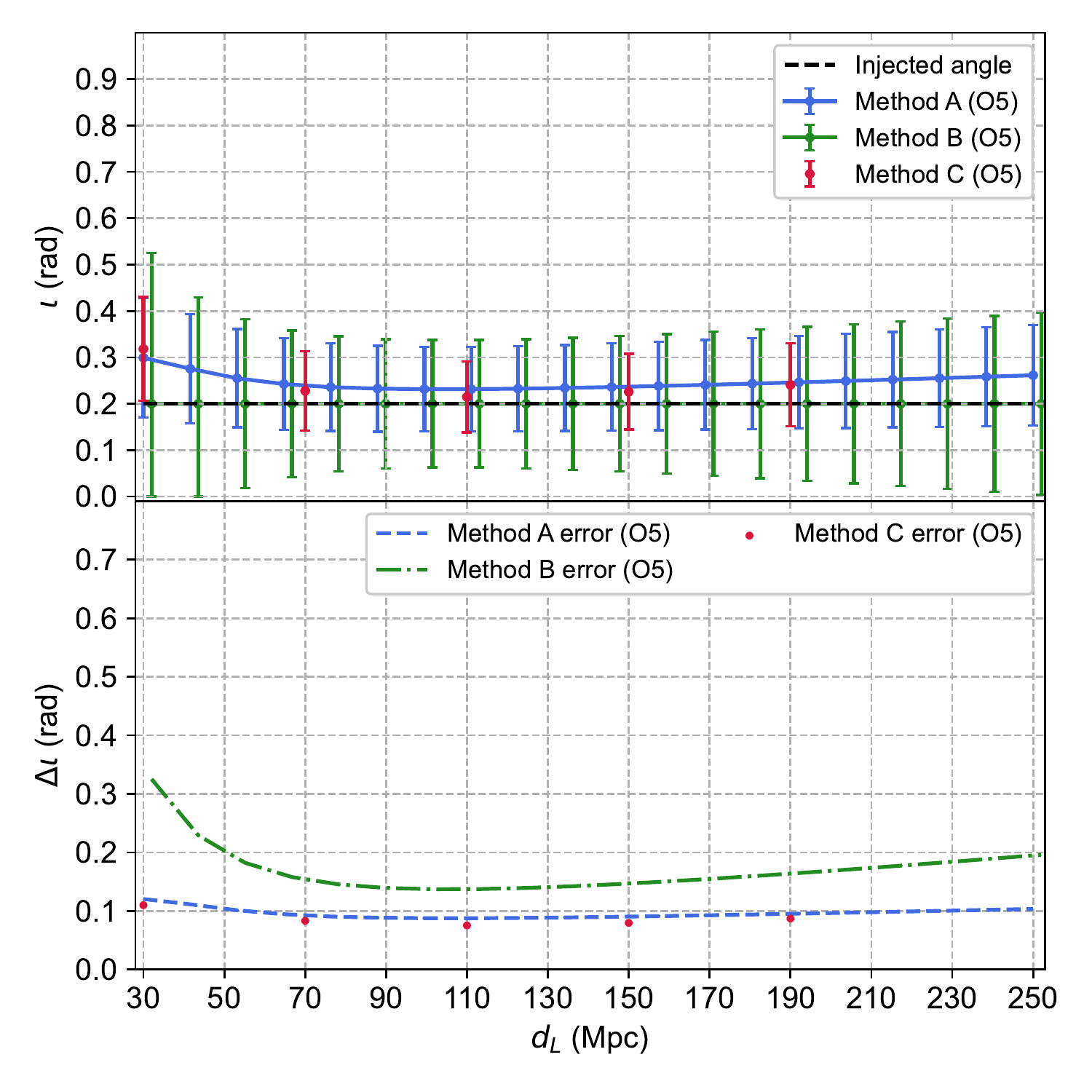}
\label{subfig:DLa}}
\subfigure[]{
\includegraphics[width=0.45\textwidth]{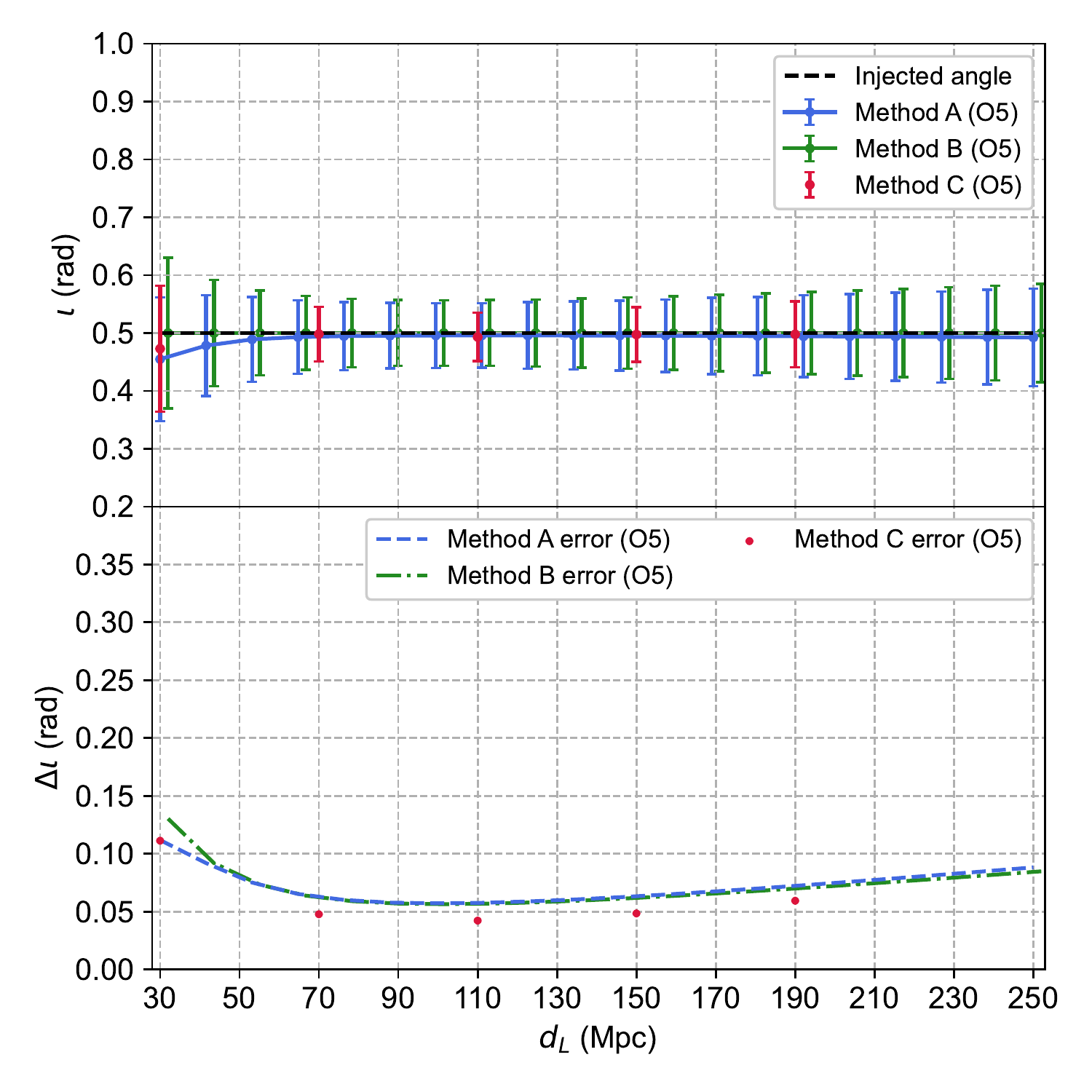}
\label{subfig:DLb}}
\caption{\small The projected distribution of $\Delta \iota$ as a function of the luminosity distance during the O5 run. (a),(b) The inclination angle estimation with the injection values of $(0.2, 0.5)$ rad. The top part in each panel shows the median values with error bars for $68\%$ credible intervals. The bottom part presents $\Delta \iota$ in each method. The black dashed line represents the injected $\iota_0$. The blue, green, and red represent method A, B, and C, respectively. In methods A and B, we show $\Delta \iota$ in dashed lines and compare them with $\Delta \iota$ (red dot) in method C.}
\label{fig:DL}
\end{figure*}

We first examine how the injected $d_{\rm L}$ affects the $\iota$'s uncertainty at a certain inclination angle. The results of two cases of $\iota=(0.2,~0.5)$ rad with varying luminosity distance of the GW source from 30 to 250 Mpc are, respectively, shown in Figs.~\ref{subfig:DLa} and \ref{subfig:DLb}. Here, we only present the figures for the results of O5 run sensitivities, and the results expected in the O4 run are displayed in the Appendix. The prior of each $d_{\rm L}$ follows the Gaussian distribution (see Sec.~\ref{sec:iota} and Fig.~\ref{fig:delta_dl}) and the Hubble constant takes the value of $67.66~{\rm km\,s^{-1}\,Mpc^{-1}}$. Since methods A and C can yield posterior distribution for $\iota$, we use the definition that is similar to Eq.~(34) of Ref.~\cite{sigma_epsilon_PRD_2019} to represent the measurement accuracy for $\iota$ (see also Ref.~\cite{Cutler_PRD_1994}), i.e.,
\begin{equation}\label{eq:deltaiota}
\Delta \iota \equiv \sqrt{\int \ud \iota P(\iota)(\iota - \bar{\iota})^2},
\end{equation}
where $P(\iota)$ is the posterior distribution in Eq.~(\ref{eq:Pdv_cos}), and $\bar{\iota}$ is the expected value of inclination angle.

Moreover, these error bars are defined as symmetric $68\%$ confidence intervals, and the central value is the median value in the posterior probability distribution of $\iota$. When the posterior distribution follows Gaussian distribution, the symmetric $68\%$ confidence intervals will be in accordance with Eq.~(\ref{eq:deltaiota}). In method B, we use Eq.~(\ref{eq:delta_iota2}) as the value of the half error bar.

We find that all methods give rather similar results. Please notice that the SNRs predicted by the three methods stay almost the same (with a relative error of $\sim 2.4\%$). Method B tends to overestimate $\Delta \iota$ compared with the other two methods, which is, in particular, the case for small $\iota$ or $d_{\rm L}$, while method C yields the tightest constraints of $\iota$ if $d_{\rm L}$ is less than about 130 Mpc. The results obtained with the three methods show that $\Delta \iota$ has a decreasing trend until it reaches the minimum value at about 130 (90) Mpc for the O5 (O4) run. The differences between the estimated median values and the injections are significant at small inclination angles. This phenomenon, i.e., decreases first and then increases for $\Delta \iota$, is mainly caused by two competing effects. For BNS mergers at a small luminosity distance (though they will have higher SNR), the uncertainty of peculiar velocity dominates the width of the $d_{\rm L}$ prior and hence influences the uncertainty of inclination angle. Whereas for distant luminosity distance events, the uncertainty contributed from peculiar velocity becomes less important, and the effect of SNR takes over and governs $\Delta \iota$.

\begin{figure*}[ht!]
\subfigure[]{
	\includegraphics[width=0.45\textwidth]{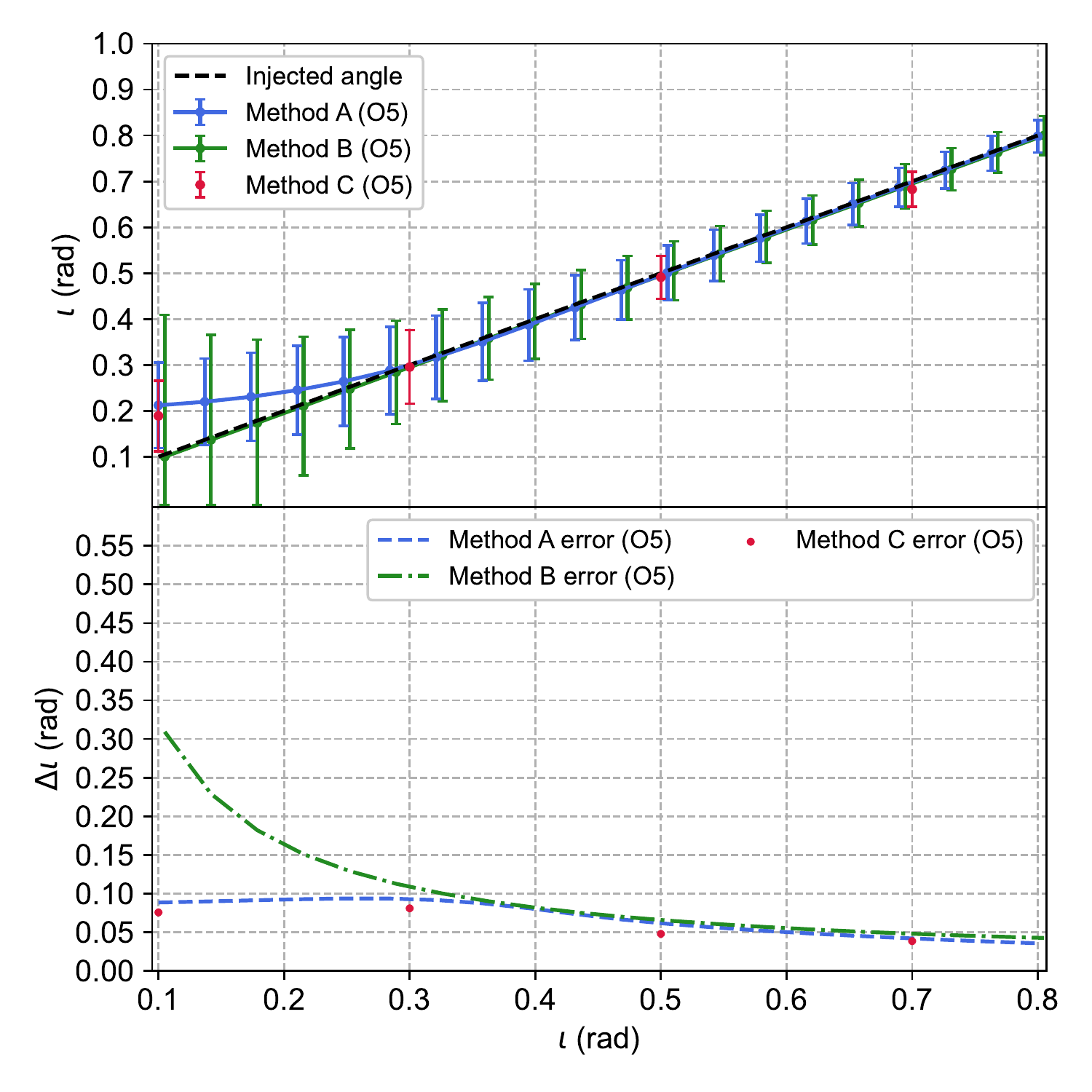}}
\subfigure[]{
	\includegraphics[width=0.45\textwidth]{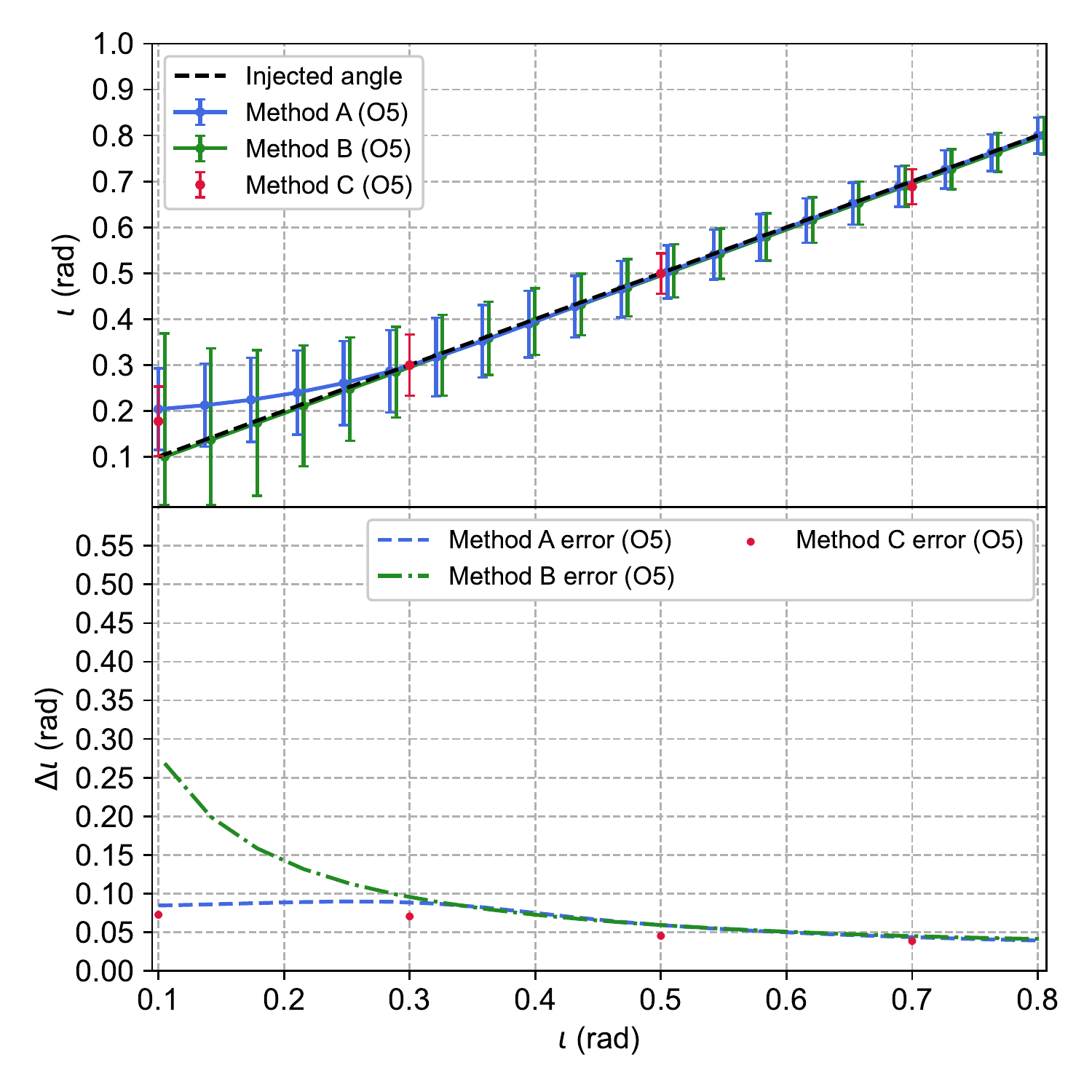}}
	\caption{\small The projected distributions of $\Delta \iota$ as a function of inclination angle during O5 run. (a),(b) The luminosity distances in 70 and 130 Mpc, respectively. The top part in each panel shows the median value with error bars for $68\%$ credible intervals. The bottom part presents the size of $\Delta \iota$. The black dashed line shows the injected $d_{\rm L0}$. The blue, green, and red represent method A, B, and C, respectively. In methods A and B, we represent $\Delta \iota$ in dashed lines and compare them with $\Delta \iota$ (red dot) in method C.} 
    \label{fig:iota}
\end{figure*}

Next, we fix the luminosity distance to 70 and 130 Mpc based on the previous discussion about the optimal $d_{\rm L}$ and then examine how the inclination angle impacts its uncertainty. In Fig.~\ref{fig:iota}, we present the estimation of $\iota$ during O5. Also, the expected results in the O4 run can be found in the Appendix. Similar to Fig.~\ref{fig:DL}, the three methods yield consistent results except that at small inclination angles, method B overestimates the uncertainties, while method C yields the tightest constraints. We find that the errors of $\iota$ decrease with an increasing $\iota$, and an appropriate luminosity distance can reduce the holistic error. For $\iota\leq 0.2$ rad, the uncertainties of the inclination angle reconstructed with the gravitational wave data are very large (with $\Delta \iota \geq 0.1$ rad) and are not suitable for further calibration unless $\Delta \iota \leq 0.1$ rad, while for $\iota\geq 0.3$ rad, a $\Delta \iota\sim$ 0.05-0.1 rad is possible. (Again, we would like to remind the readers that such high accuracy is only possible when the Hubble tension has been satisfactorily solved.) Therefore, it is sufficiently good to be used to calibrate the afterglow modeling of some GW-associated off-axis relativistic ejecta. References.~\cite{Nakar2021ApJ,Troja2022MNRAS} pointed out that there was an intrinsic degeneracy between $\theta_{\rm v}$ and $\theta_{\rm c}$ and only the ratio $\theta_{\rm v}/\theta_{\rm c}$ can be constrained by light curves. While our results do not rely on afterglow models, it might be a good chance to break such degeneracy.

\begin{figure*}[ht!]
	\centering\vspace{-6mm}\hspace{19mm}
	\includegraphics[width=0.6\textwidth]{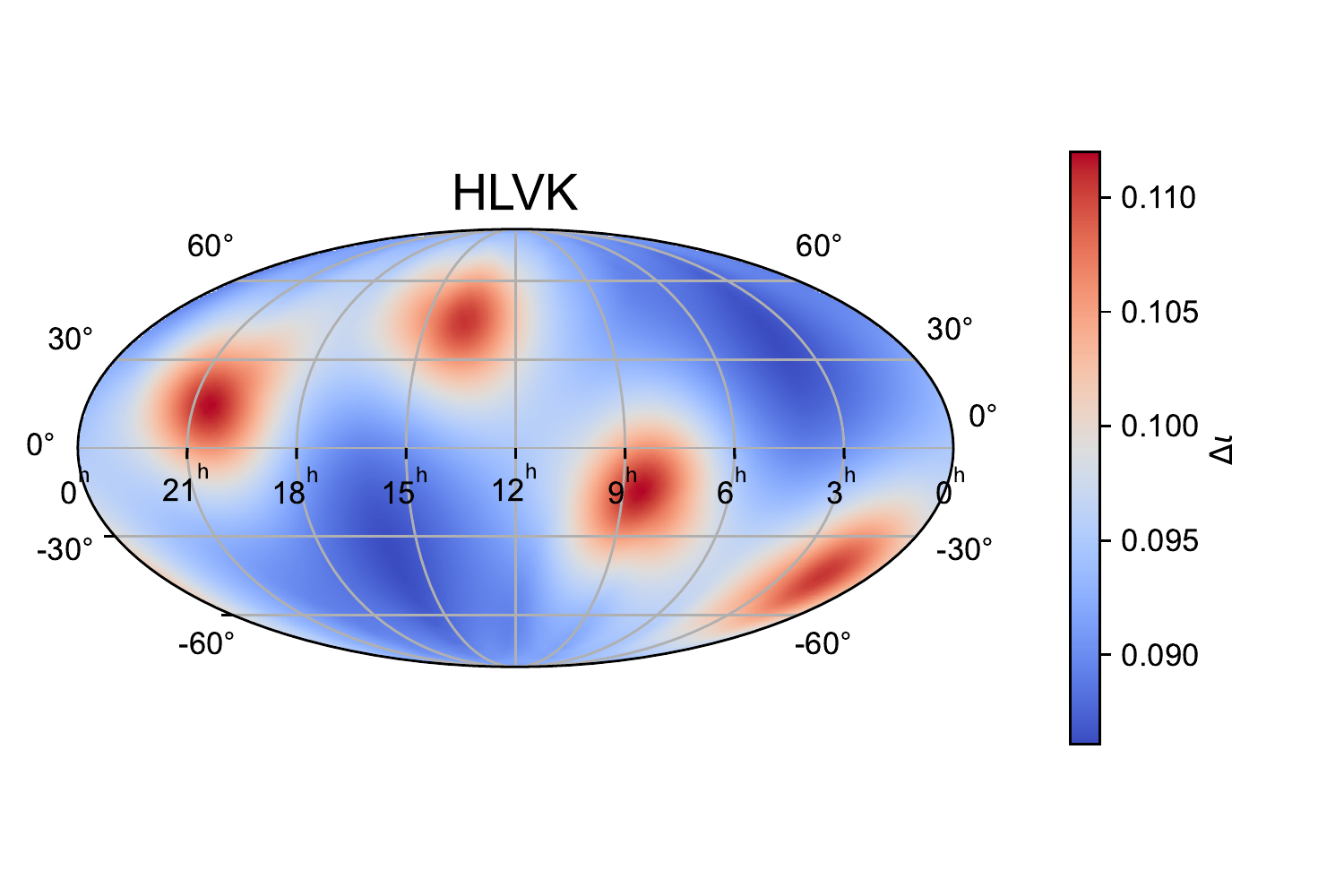}\vspace{-10mm}\\
	\includegraphics[width=0.6\textwidth]{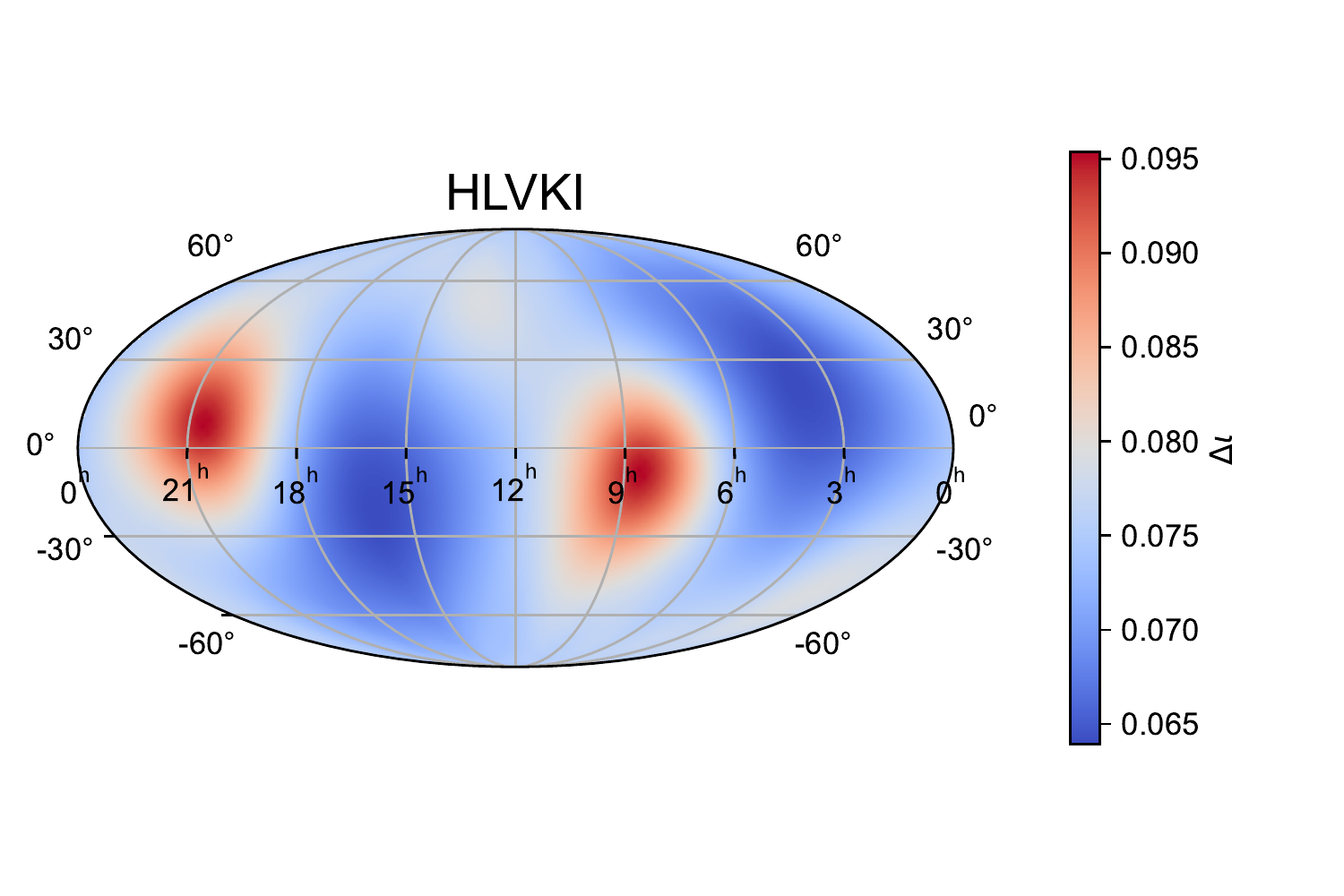}\vspace{-10mm}\hspace{-20mm}
   \caption{\small The spatial distributions of $\Delta \iota$ in O4 (upper) and O5 (bottom) period. The luminosity distances of the source are fixed as 90 Mpc (upper) and 130 Mpc (bottom), respectively.This sky map is given in equatorial coordinates at a fiducial epoch arbitrarily fixed to $t_{\rm GPS} =1187008882.42 \,\rm s $.}
	\label{fig:sky}
	\hfill
\end{figure*}

In our analysis, it is also found that the polarization angle $\psi$ has a minor effect on the inclination angle estimate (one can see this directly with method B). For Eq.~(\ref{eq:delta_iota2}), using the parameters in Table.~\ref {tb:inject}, the sum value of the terms in square brackets approximates to $10^{-5}$ ($10^{-3}$) when $\iota = 0.1$ (0.7) rad. In that case, $\sigma_{\rm D}=1.055$ and $\epsilon_{\rm D}=0.313$. Even if there were no constraints for $\Delta \bar{\psi}$, the error term caused by $\bar{\psi}$ should have the same order of magnitude from $10^{-5}$ to $10^{-3}$. Thus, the first line in Eq.~(\ref{eq:delta_iota2}) will be the main contribution to $\Delta \iota$. We also investigate the spatial influence on evaluating $\iota$ and give the projected sky distribution of the uncertainties of inclination angle with the fixed $\iota$ and $d_{\rm L}$ in O4 and O5 runs by method A (shown in Fig.~\ref{fig:sky}). We take $d_{\rm L} =130$ (90) Mpc for O5 (O4) run, and set the Global Positioning System (GPS) time to be 1187008882.42s. Other parameters are identical to the injection parameters in Table.~\ref{tb:inject}. It is evident that the measurement uncertainties of $\iota$ will significantly decrease in the O5 run due to the increase of the detectors' sensitivities and the participation of the LIGO-India. In the optimistic case (i.e., all five detectors have detected the same gravitational wave event with high SNRs, we have $\Delta \iota \leq 0.1$ rad in the O5 run. We also find that the distribution of $\Delta \iota$ has a correlation with the distribution of $\sigma_{\rm D}$ \cite{sigma_epsilon_PRD_2019}, this is because larger $\sigma_{\rm D}$ causes higher SNR [as indicated in Eq.~(\ref{eq:snr})] and hence reduces the uncertainties of inclination angle.

\subsection{The uncertainties of luminosity distance and Hubble constant}\label{sec:IVB}
Though the main purpose of this work is to investigate the prospect of calibrating the afterglow modeling with the gravitational-wave-based inclination angle measurements, it is also interesting to investigate whether it is possible to get a robust evaluation of the Hubble constant (i.e., the influence of $\iota$ can be minimized). The answer is yes. With Eqs.~(\ref{eq:delta_d}) and (\ref{eq:delta_dl}) , it is straightforward to see that for a fixed/small $\Delta^2\iota$ we have a tightly bounded $\Delta H_0/H_0$.

Interestingly, a prior of $\Delta \iota\leq 0.1$ rad is achievable in the following scenario: as long as our line of sight is within the energetic core of the structured ejecta, the afterglow emission will be similar to that viewed on axis and the evaluated jet opening angle ($\theta_{\rm j}$) should be a robust upper limit on the viewing angle $\theta_{\rm v}$, as found in the numerical calculations (e.g., \citep{Wei+Jin2003,Kumar+Granot2003ApJ}). Whereas for the nearby bright short GRBs, we have a typical $\theta_{\rm j}\approx 0.1$ rad (see Table 3 of \citep{Jin2018ApJ}), with which it is very reasonable to assume $\Delta \iota\leq 0.1$ rad. Then, we analyze GW signal combined with the prior bound of $\iota$ to constrain the uncertainty of $d_{\rm L}$ which has similar definition to Eq.~(\ref{eq:deltaiota}). In such a case, the prospective precision of Hubble constant can be directly evaluated via 
\begin{equation}\label{eq:delta_H0}
	\frac{\Delta^2H_0}{H^2_0}=\frac{\Delta^2v_{\rm r} + \Delta^2v_{\rm p}}{H^2_0d^2_{\rm L}} + \frac{\Delta^2 d_{\rm L}}{d^2_{\rm L}},
\end{equation}
where the values of $v_{\rm r}$ and $v_{\rm p}$ are the same as those used in Sec.~\ref{sec:iota}. One thing that should be specified is that Eq.~(\ref{eq:delta_H0}) is only valid in a simplified situation; i.e., the posterior distribution of $d_{\rm L}$ follows Gaussian distribution.

\begin{figure*}[ht!]
	\centering
	\subfigure[]{
		\includegraphics[width=0.45\textwidth]{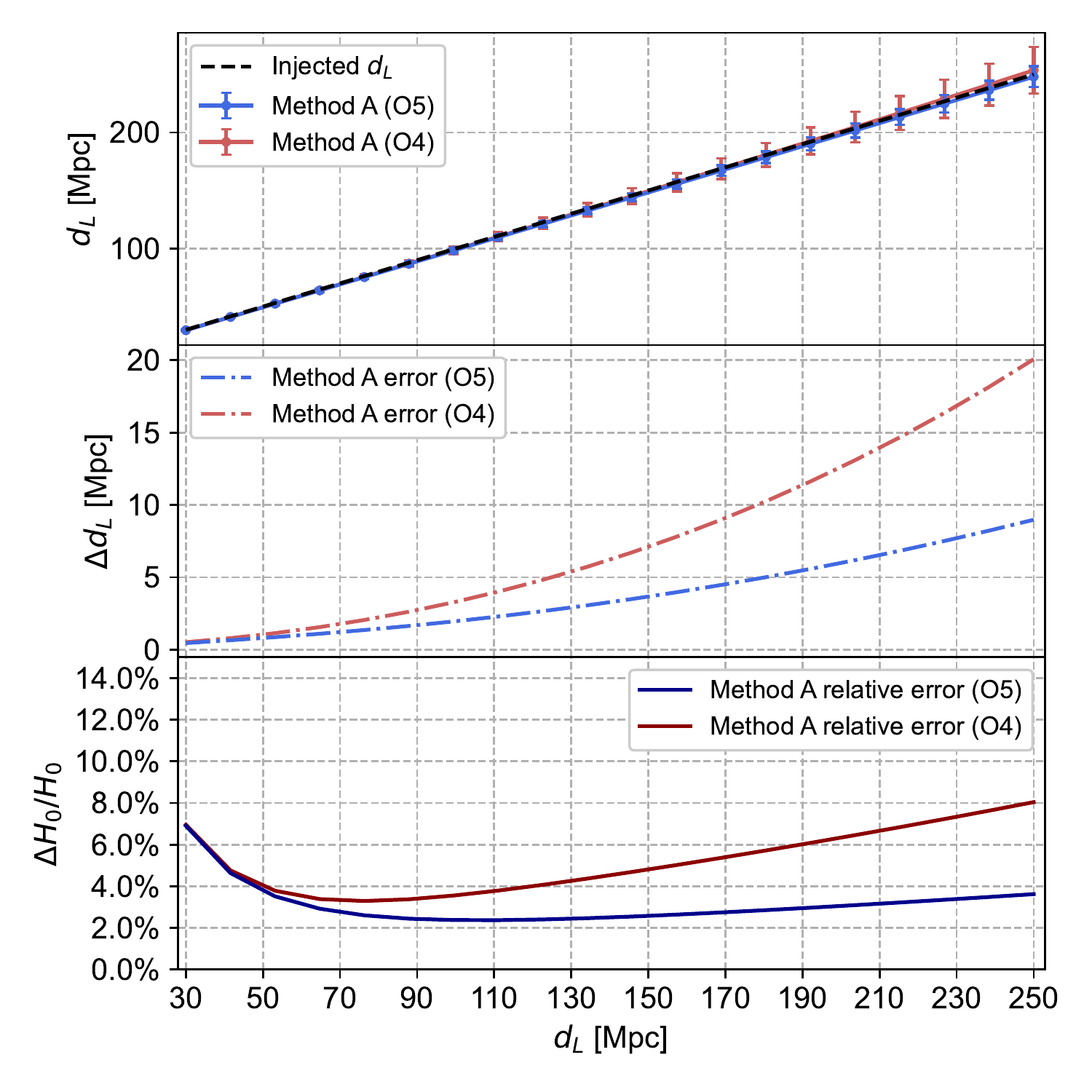}
		\label{fig:dl_estimationa}}
	\subfigure[]{
		\includegraphics[width=0.45\textwidth]{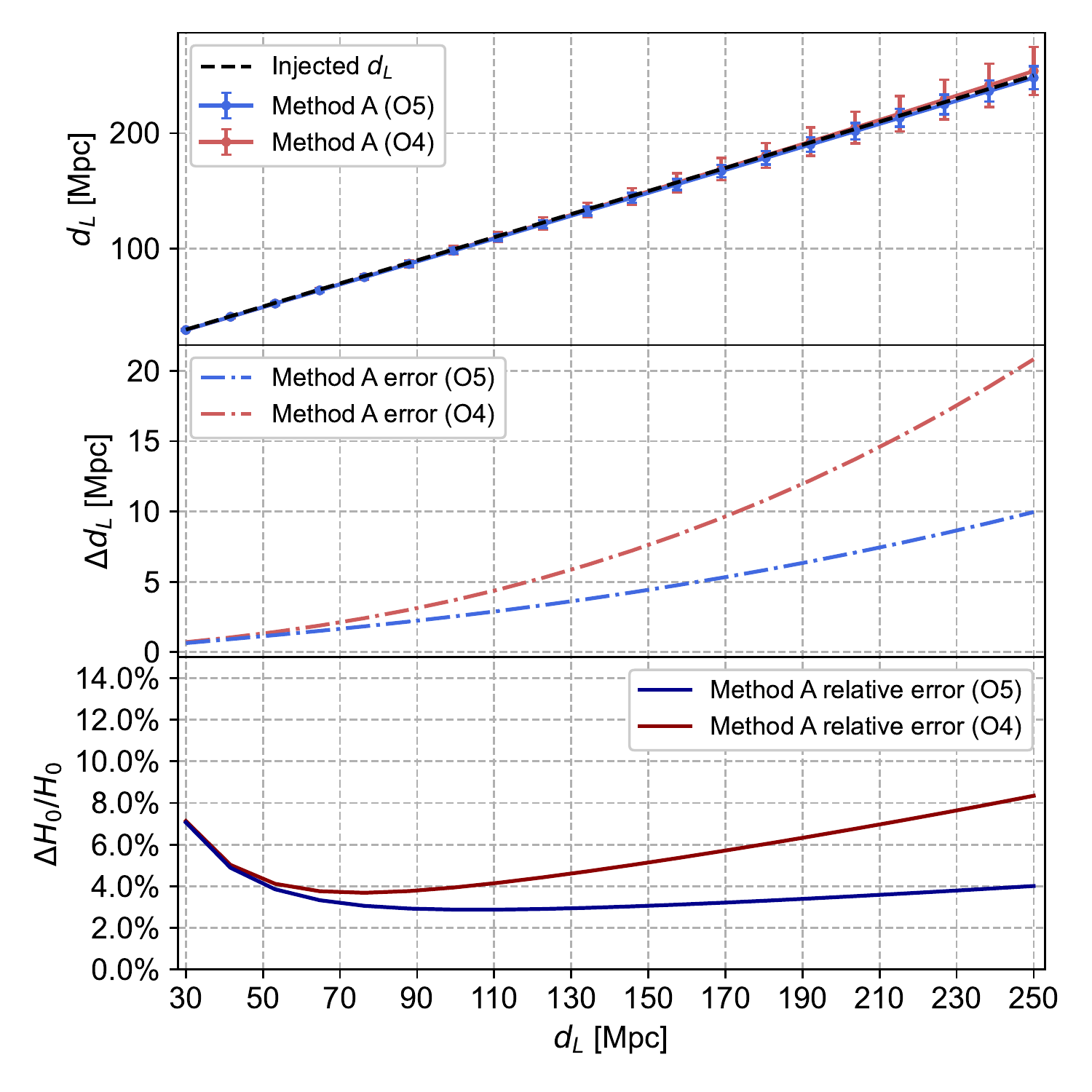}
		\label{fig:dl_estimationb}}

	\caption{\small The distribution of luminosity distance and Hubble constant estimation with variable luminosity distance using method A. (a),(b) The estimated results with fixed inclination angles (0.1, 0.2) rad, respectively. The uncertainty of $\iota$ is assumed to be 0.1 rad, which is obtainable for some bright/on-axis GRBs. In the top sections, we show the posterior distribution of $d_{\rm L}$ with error bars for $68\%$ credible intervals. The black dashed lines show the injected $d_{\rm L}$. The blue and red solid lines represent O5 and O4 separately. In the medium sections, the red and blue dashed lines represent the errors in O5 and O4, respectively. The bottom sections show the relative error of the Hubble constant.}
	\label{fig:dl_estimation}
	\hfill
\end{figure*}

For method A, we take the same sets of injection parameters in Table.~\ref{tb:inject} to estimate the uncertainties of luminosity distance. Since there are many complicated factors such as the wide variety of telescopes, observing strategies, observing conditions, and so on, we do not consider these selection effects in our analysis, which may introduce a bias for $H_0$ estimation \citep{Chen2020PRL}. Currently, we only give approximate results based on some reasonable simplifications. Our results are shown in the Figs.~\ref{fig:dl_estimationa} and \ref{fig:dl_estimationb} with $\iota_0 = 0.1~(0.2)$ rad. The luminosity distance $d_{\rm L,0}$ ranges from 30 to 250 Mpc, and the prior uncertainty of $\iota$ is set as 0.1 rad. 
We find that the higher the sensitivity, the less the inferred $\Delta d_{\rm L}$. It is apparent that $\Delta H_0/H_0$ has a similar tendency to $\Delta \iota$ in Fig.~\ref{fig:iota}. The high relative error of estimated $H_0$ at a small distance is mainly caused by the peculiar velocity uncertainty. At a large distance, the uncertainties of $d_{\rm L}$ become the dominant influence on $H_0$'s uncertainty. 
The other general trend is that $\Delta d_{\rm L}$ increases with $d_{\rm L}$, and $\Delta d_{\rm L}$ ($\Delta H_0/H_0$) is also found to be higher for a larger $\iota$. Encouraging, in the most optimistic case, the $H_0$ can be measured to a precision of $3\%$ ($4\%$ in the O4 run) for a GW/GRB association event, supposing our line of sight is within the energetic core of the GRB ejecta and the jet break in the afterglow light curve can be well measured. 

There would be another case that some on-axis afterglows are bright enough to constrain $
\theta_{\rm v}/\theta_{\rm j}$ with high-resolution imaging. For example, \citet{Ryan2015ApJ} reported very good constraints on $\theta_{\rm j}$ and the ratio $\theta_{\rm v}/\theta_{\rm j}$ ($\theta_{\rm v}=0.0733^{+0.011}_{-0.0098} \, \rm rad, \,\theta_{\rm v}/\theta_{\rm j}=0.676^{+0.035}_{-0.050}$) for GRB 110422A. Therefore, we suppose $\theta_{\rm v}=0.05$ and $\Delta \theta_{\rm v}=0.015\,\rm rad$ to predict if the uncertainty of Hubble constant can be constrained tighter. We find that method A gives a precision of $2\%$ for the Hubble constant. Though such bright events 
are rare, the high-precision estimation with a single ``lucky" event is still a possible solution to Hubble tension in the future.

\subsection{The probability of detecting BNS mergers with detectable afterglow}\label{sec:IVC}
\begin{figure}[ht!]
	\centering
	\includegraphics[width=0.5\textwidth]{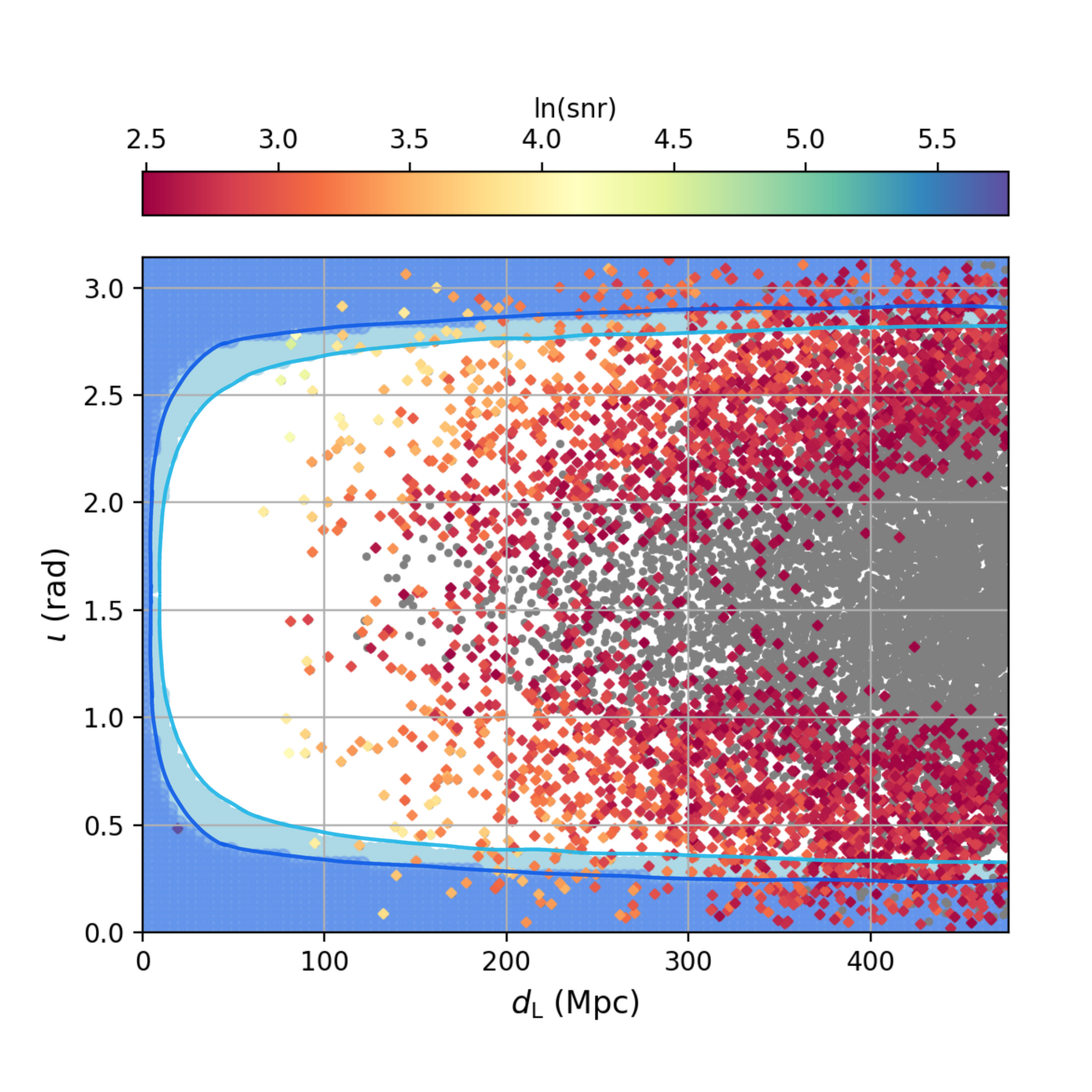}\vspace{-5mm}
	\caption{\small Relation of SNR between luminosity distance and inclination angle for BNS mergers. This figure contains 10000 simulated BNS events. Only the events satisfying SNRs $\ge 12$ are shown in colored points and others are in gray. The points falling into the blue (light blue) region represent that their light curves satisfy the criterion of peak flux $F_{\rm peak} > 10\times F_{\rm lim}$ ($10\times F_{\rm lim} > F_{\rm peak} > F_{\rm lim}$).}
	\label{fig:BNS_simulation}
\end{figure}

Although previous discussions give positive prospects for constraining viewing angle or Hubble constant, it is crucial to estimate the number of BNS detections with or without electromagnetic counterparts. Please note that below we assume the localization of the GW/GRB association event is well determined\footnote{Here, we do not consider the effect of different fields of view and the sky localization requirements. Detailed discussions can be found in Ref.~\cite{2022PatricelliMNRAS}}. Wide field of view gamma-ray and x-ray observatories will become more and more crucial to work in synergy with GW detectors \citep{Mura2021MNRAS, 2022Ronchini}. This is reasonable since, in the space, there are some dedicated gamma-ray detectors with a very wide field of view to hunt for the sub-MeV flash from the neutron star merger events (one example is the Gravitational Wave High-Energy Electromagnetic Counterpart All-sky Monitor \citep{GeCAM} launched at the end of 2020). In the near future, even more detectors (such as the All-sky Medium Energy Gamma-ray Observatory \citep{AMEGO}, Gamma-Ray Monitor equipped in the Space Variable Objects Monitor \citep{SVOM}, and the Southern Wide Field-of-View Gamma-ray Observatory \citep{SGWO,SGWO2}) for the same purpose will be launched. The next generation of GW detectors will also combine with many monitors of multimessengers, such as the High Energy Rapid Modular Ensemble \citep{HERMES}, the Gravitational-Wave Optical Transient Observer \citep{GOTO}, the Transient Astrophysics Probe \citep{TAP}, and so on. Although usually, a single sub-MeV detector is unable to yield a very accurate localization. A small error region can be triangulated with the data from quite a few observatories. The follow-up observations of big telescopes with high sensitivity but a small field of view can be carried out. Following \citet{Mastrogiovanni2021AA}, we simulate 10000 BNS merger events (within the range of $z\le0.1$) and set SNR $\ge$ 12 as the threshold of GW detection. About $40.7\%$ ($32.8\%$) events that exceed the GW SNR threshold can be afterglow candidates during O5 (O4), as shown in the upper section of Fig.~\ref{fig:BNS_simulation}. To simulate the flux densities of these candidate events, we take the same ejecta model and parameters in generating Fig.~\ref{fig:Gaussian_jet} (the kinetic energy is taken to be $E_{\rm k}=1.9 \times 10^{51} \rm erg$ \citep{Fong2015ApJ}). We take Eq.~(4) of Ref.~\cite{Mastrogiovanni2021AA} (i.e., $F_{\rm peak} > 10\times F_{\rm lim}$) as a criterion for $\iota$ measurement with the afterglow light curve; this is because the weaker candidates cannot well constrain $\Delta \iota$. Setting the sensitivity lines of HST, EVLA, and Chandra as $F_{\rm lim}$, these BNS simulations can be divided into three parts, without afterglow detection, with detected afterglow, but $\theta_{\rm v}$ cannot be well constrained, and with bright afterglow and reliably constrained $\theta_{\rm v}$. In the last two cases, about $2.3\%$ events can be used to constrain the viewing angle by combining optical, radio, and x-ray bands (as we cautioned before, such constraints likely suffer from the uncertainties from the degeneracy between $\theta_{\rm v}$ and $\theta_{\rm c}$ \cite{Nakar2021ApJ,Troja2022MNRAS}). And the detected events (only exceeding the peak flux slightly) are about twice over. Therefore, considering the limited sensitivity of these three bands, we plot the ranges of these two cases in blue and light blue in Fig.~\ref{fig:BNS_simulation}. The gravitational wave data yield a BNS merger rate of $\rho_{\rm BNS}\sim 320^{+490}_{-240} \, \rm Gpc^{-3}$ \citep{Abbott2021ApJL}, while the low redshift short GRB observations suggest a rate of $\rho_{\rm BNS} \sim 10^{3}~{\rm Gpc^{-3}}$ \citep{Jin2018ApJ}. We find that there will be $\sim 360~{\rm yr^{-1}}~(\rho_{\rm BNS}/10^{3}~{\rm Gpc^{-3}})$ BNS merger events taking place within the range of $z\le0.1$. So we predict that there will be ${\cal O}(10)$ [${\cal O}(1)$] BNS mergers accompanied by bright afterglow during O5 (O4). Moreover, with the {\it James Webb} Space Telescope (JWST) \cite{JWST} that will be in formal performance in 2022 and the {\it Athena} that is expected to be available in the 2030s \cite{Athena}, the detection prospect of the afterglow emission would be further enhanced. 
Indeed, a good fraction of the BNS mergers simulated in Fig.~\ref{fig:BNS_simulation} is detectable for JWST with a sensitivity of $1.0\times10^{-17} \rm erg \, s ^{-1} \rm cm^{-2}$ at 9.2 $\mu$m for an observation time of $10^{4}$ s. 
Finally, we would remind the reader that, for the nearly on-axis, very-nearby short GRBs that are almost certainly able to yield detectable afterglow emission and can hence play an important role in constraining $H_0$, we can estimate their detection rate as $\sim 2\pi \theta_{\rm j}^{2}\rho_{\rm BNS}\times (150~{\rm Mpc})^3/3 \sim 0.08~{\rm yr}^{-1}~(\theta_{\rm j}/0.1)^{2}(\rho_{\rm BNS}/10^{3}~{\rm Gpc^{-3}})$; here a distance range of $\sim 150$ Mpc is adopted to yield a $\sim 3\%$ accuracy of $H_0$ (see Fig.~\ref{fig:dl_estimation}). Such detection is possible in the future, though the chance is not high. For an accuracy of $\sim 4\%$, the distance can be extended to 250 Mpc for an O5 run, which would further enhance the detection chance by a factor of $\sim 4.7$.

\section{Summary}\label{sec:discussion}
Thanks to dedicated observational and theoretical efforts made in the last decades, an external forward shock model has been successfully developed to interpret the main features of the afterglow radiation of GRBs. Because of the simplifications and approximations involved in the modeling and partly because of the incomplete dataset, the fit to the data usually gives different physical parameters. It is challenging to distinguish the real one in these results with the sole electromagnetic radiation data. In this work,
motivated by the fact that the gravitational wave can directly measure the inclination angle and the afterglow modeling can infer the viewing angle, we have examined the possibility of calibrating the afterglow modeling with the gravitational wave measurements. The basic assumption for such an approach is that $\theta_{\rm v}=\iota$, which is reasonable for the BNS mergers since usually such objects are rotating very slowly and the angular momentum of the formed remnants are perpendicular to the merger plan (note that the ejecta is widely believed to be launched along the rotation axis of the massive remnant). 
We have taken three different methods, including both analytical estimations and direct simulations, to predict the prospective uncertainties of the inclination angle. For some neutron star mergers accompanied with electromagnetic counterparts detected in the O4/O5 and later runs of LIGO/Virgo/KAGRA/LIGO-India detectors, we show that the inclination angle can be determined within an uncertainty of $\leq 0.1$ rad, supposing the Hubble constant has been well determined (i.e., within an uncertainty of $3\%$). We also find that LIGO-India's participation will significantly decrease the proportion of worse-detected positions. At least for some neutron stars, the off-axis relativistic outflow will be launched, giving rise to afterglow emission. The most energetic ones may be detectable at the distance of $\sim$ 100-200 Mpc even for a viewing angle of $\geq 0.3$ rad. Such events can thus serve as a robust test of the afterglow modelings. One thing that should be noticed is that these tight constraints are based on an accurate determination of the Hubble constant. Thus, we discuss the implication of Hubble tension to the reconstruction of $\iota$. Because the luminosity distance is negatively correlated with the Hubble constant, higher $H_0$ will trend to estimate higher $d_{\rm L}$ and then impact the value of $\iota$. If the Hubble tension remains, our initial purpose is not achievable.

We have also evaluated the prospect of resolving the Hubble tension with a single GW/GRB association event. A $\sim 3\%$ ($4\%$) precision Hubble constant is obtainable in the O5 (O4) run if the uncertainty of the viewing angle can be constrained to be within $\sim 0.1$ rad, which is achievable for some nearby ($\leq 250$ Mpc) bright/on-axis GRBs with a well-behaved afterglow light curve displaying a clear achromatic break at early times. Though with a single GW/GRB association event (even in the optimistic case), it seems hard to resolve the Hubble tension completely, the statistical studies of a group of such events, anyhow, are expected to play a significant role. As \citet{Nissanke2010ApJ} pointed out, the precision of $H_0$ can be improved to $\Delta H_0 / \sqrt{N}$, where $N$ is the number of BNS observations. Consequently, four such events will yield a $\sim 1.5\%$ accuracy Hubble constant measurement, which is sufficiently accurate to resolve the Hubble tension. Therefore, a large neutron star merger sample is crucial. Given a BNS merger rate of $\sim 10^{3}~{\rm Gpc^{-3}~yr^{-1}}$ \citep{LIGOScientific:2020ibl,Jin2018ApJ}, we would expect a bright afterglow combined detection rate of $\sim ${\cal O}(10) [{\cal O}(1)] ${\rm yr^{-1}}$ in the O5 (O4) run. Consequently, detecting an almost on-axis GRB/GW association event, though with a chance much lower than the off-axis ones, is still possible. Indeed, previously, the nearest candidate of an almost on-axis merger-driven burst was GRB 060505 at a redshift of 0.089 \citep{JinZP2021}. There could be some on-axis events taking place even closer because the field of view of {\it Swift}, which has an angular resolution of $\sim$ a few arc minutes and hence enables the successful detection of the afterglow as well as the redshift, is just $\approx 2$ rad \citep{2004ApJ...611.1005G} and many more events were missed. In the upcoming O4 and O5 runs, the situation will be significantly improved because several gamma-ray burst monitors with wide fields of view are in performance, with which the nearby bright on-axis merger-driven GRBs are expected to be well recorded.

\section*{ACKNOWLEDGMENTS}
{We thank the anonymous referees for helpful comments and suggestions.} This work was supported in part by NSFC under Grants No. 11921003, No. 11773078, and No. 11525313, the Funds for Distinguished Young Scholars of Jiangsu Province (No. BK20180050), the Chinese Academy of Sciences via the Strategic Priority Research Program (Grant No. XDB23040000), and Key Research Program of Frontier Sciences (No. QYZDJ-SSW-SYS024). This research has made use of data and software obtained from the Gravitational Wave Open Science Center \url{https://www.gw-openscience.org}, a service of LIGO Laboratory, the LIGO Scientific Collaboration, and the Virgo Collaboration. LIGO is funded by the U.S. National Science Foundation. Virgo is funded by the French Centre National de Recherche Scientifique (CNRS), the Italian Istituto Nazionale della Fisica Nucleare (INFN), and the Dutch Nikhef, with contributions by Polish and Hungarian institutes.\\

\appendix
\section{THE ESTIMATION OF INCLINATION ANGLE DURING O4}
\begin{figure*}[ht!]
	\centering
	\vspace{-5mm}
	\subfigure[]{
		\includegraphics[width=0.45\textwidth]{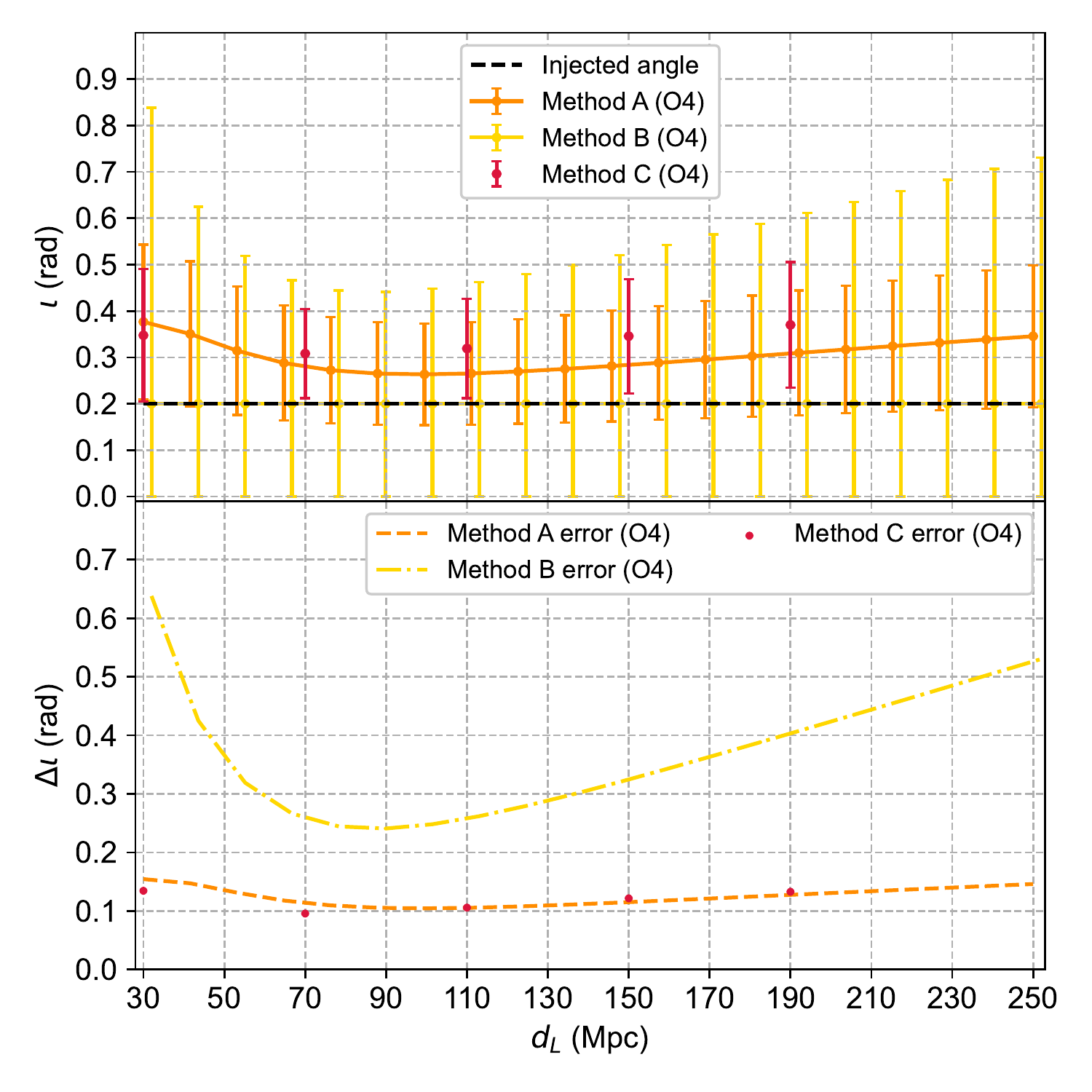}}
	\subfigure[]{
		\includegraphics[width=0.45\textwidth]{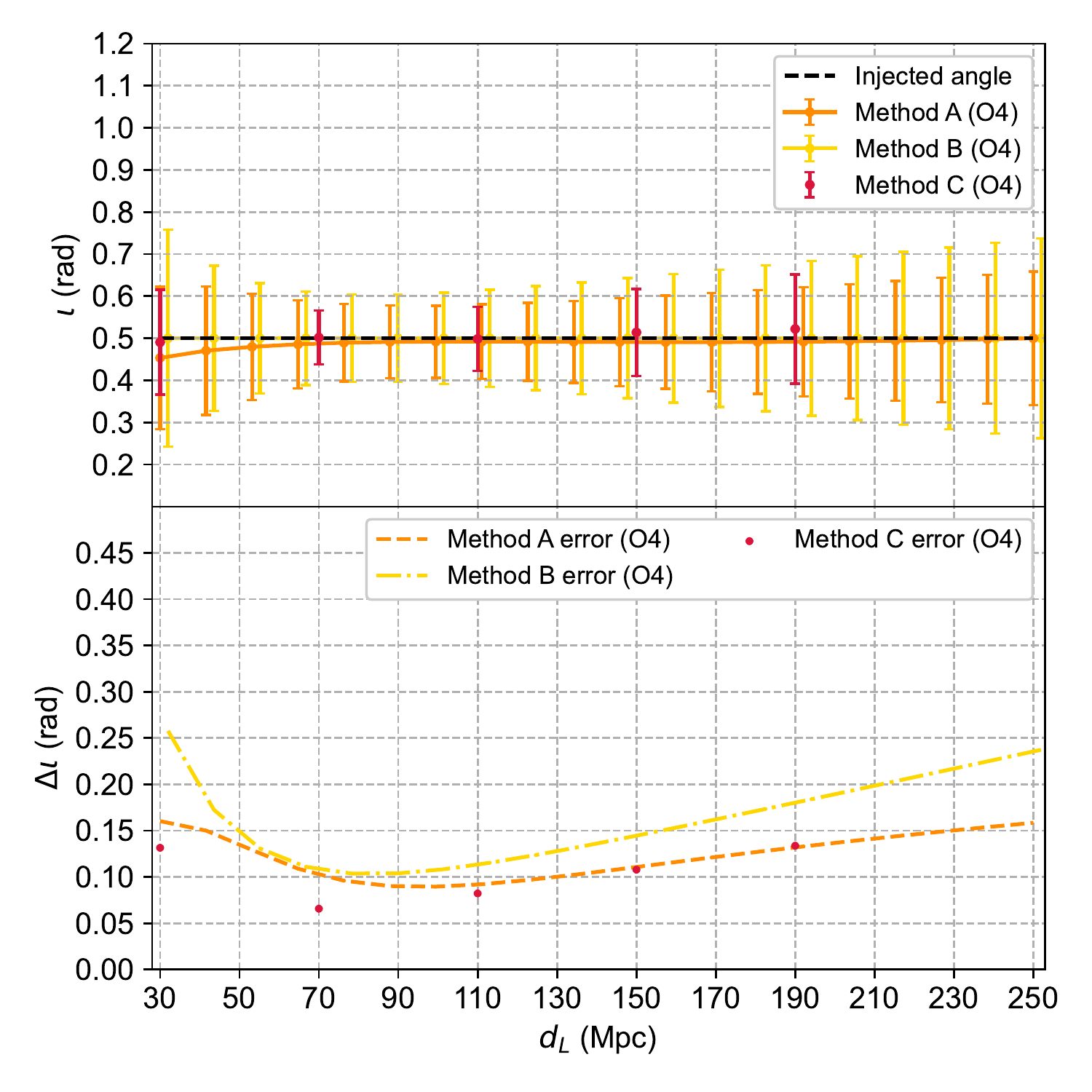}}
	\caption{\small The prospective distribution of $\Delta \iota$ with a changed luminosity distance during O4. (a),(b) The inclination angle estimation at (0.2, 0.5) rad including two parts respectively. The top part shows the estimated median value with error bars for 68$\%$ credible intervals. The bottom part shows the errors $\Delta \iota$ for three methods. The black dashed line shows the injected $\iota_0$. The orange, yellow, and red represent method A, B, and C respectively.} 
	\label{fig:DL_O4}
\end{figure*}

\begin{figure*}[ht!]
	\centering\subfigure[]{
		\includegraphics[width=0.45\textwidth]{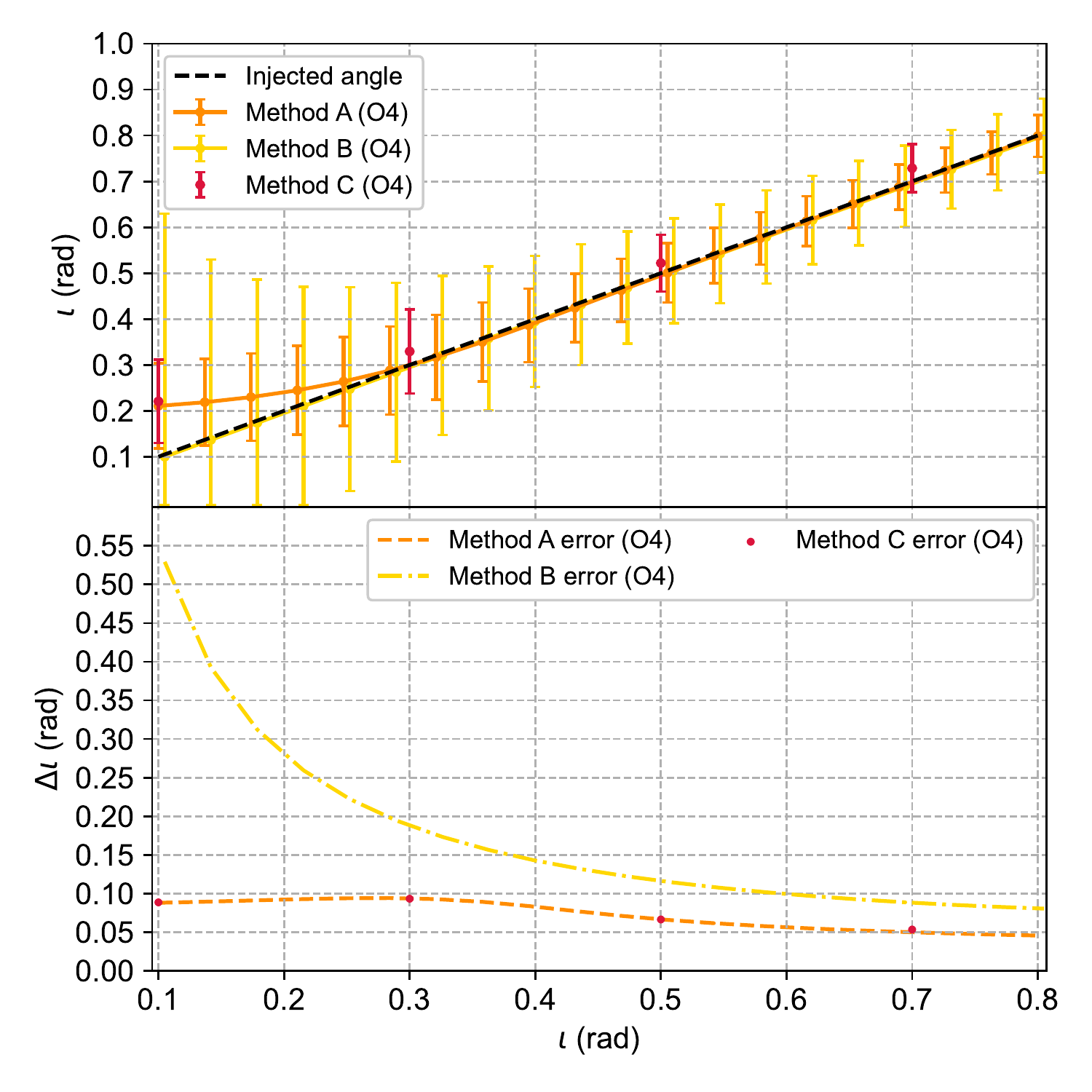}}
	\subfigure[]{
		\includegraphics[width=0.45\textwidth]{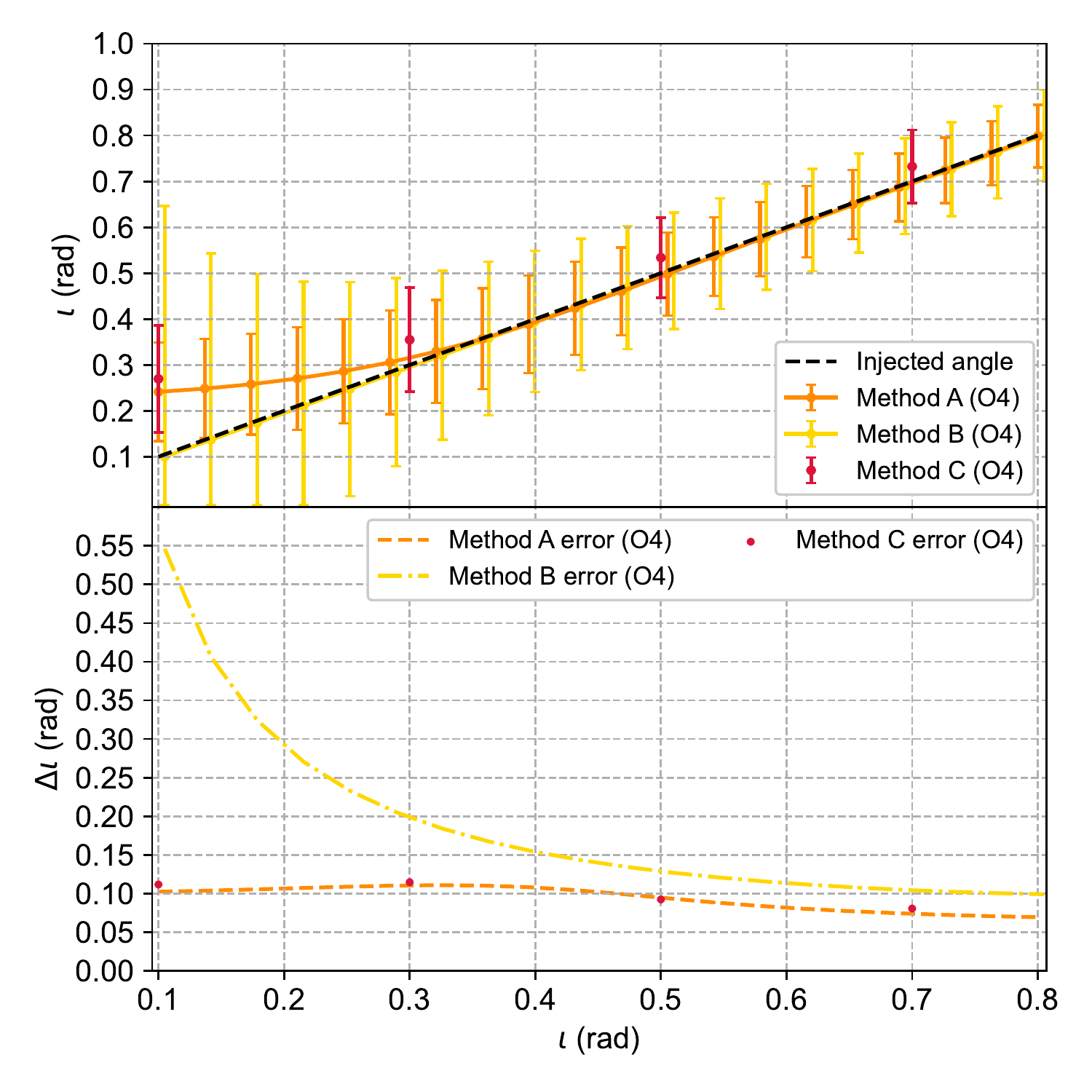}}
	\caption{\small The prospect distributions of $\Delta \iota$ with a changed inclination angle during O4. (a),(b) The inclination angle estimation at 70 and 130 Mpc including two parts respectively. The top part shows the estimated median value with error bar for 68$\%$ credible intervals. The bottom part shows the error $\Delta \iota$ for three methods. The black dashed line shows the injected $d_{L0}$. The orange, yellow, and red represent method A, B, and C respectively.}
	\label{fig:iota_O4}
\end{figure*}

In Figs.~\ref{fig:DL_O4} and \ref{fig:iota_O4}, we show the uncertainties of $\iota$ with varying luminosity distance or $\iota$ itself during the O4 run, respectively. The parameter configurations are same as Figs.~\ref{fig:DL} and \ref{fig:iota}. The injected parameters and the priors of parameter inference follow Table~\ref{tb:inject}. 

\clearpage
\bibliography{ref}

\begin{thebibliography}{113}%
\makeatletter
\providecommand \@ifxundefined [1]{%
 \@ifx{#1\undefined}
}%
\providecommand \@ifnum [1]{%
 \ifnum #1\expandafter \@firstoftwo
 \else \expandafter \@secondoftwo
 \fi
}%
\providecommand \@ifx [1]{%
 \ifx #1\expandafter \@firstoftwo
 \else \expandafter \@secondoftwo
 \fi
}%
\providecommand \natexlab [1]{#1}%
\providecommand \enquote  [1]{``#1''}%
\providecommand \bibnamefont  [1]{#1}%
\providecommand \bibfnamefont [1]{#1}%
\providecommand \citenamefont [1]{#1}%
\providecommand \href@noop [0]{\@secondoftwo}%
\providecommand \href [0]{\begingroup \@sanitize@url \@href}%
\providecommand \@href[1]{\@@startlink{#1}\@@href}%
\providecommand \@@href[1]{\endgroup#1\@@endlink}%
\providecommand \@sanitize@url [0]{\catcode `\\12\catcode `\$12\catcode
  `\&12\catcode `\#12\catcode `\^12\catcode `\_12\catcode `\%12\relax}%
\providecommand \@@startlink[1]{}%
\providecommand \@@endlink[0]{}%
\providecommand \url  [0]{\begingroup\@sanitize@url \@url }%
\providecommand \@url [1]{\endgroup\@href {#1}{\urlprefix }}%
\providecommand \urlprefix  [0]{URL }%
\providecommand \Eprint [0]{\href }%
\providecommand \doibase [0]{http://dx.doi.org/}%
\providecommand \selectlanguage [0]{\@gobble}%
\providecommand \bibinfo  [0]{\@secondoftwo}%
\providecommand \bibfield  [0]{\@secondoftwo}%
\providecommand \translation [1]{[#1]}%
\providecommand \BibitemOpen [0]{}%
\providecommand \bibitemStop [0]{}%
\providecommand \bibitemNoStop [0]{.\EOS\space}%
\providecommand \EOS [0]{\spacefactor3000\relax}%
\providecommand \BibitemShut  [1]{\csname bibitem#1\endcsname}%
\let\auto@bib@innerbib\@empty
\bibitem [{\citenamefont {{Piran}}(2005)}]{Piran2004}%
  \BibitemOpen
  \bibfield  {author} {\bibinfo {author} {\bibfnamefont {T.}~\bibnamefont
  {{Piran}}},\ }\href {\doibase 10.1103/RevModPhys.76.1143} {\bibfield
  {journal} {\bibinfo  {journal} {Reviews of Modern Physics}\ }\textbf
  {\bibinfo {volume} {76}},\ \bibinfo {pages} {1143} (\bibinfo {year}
  {2005})},\ \Eprint {http://arxiv.org/abs/astro-ph/0405503}
  {arXiv:astro-ph/0405503 [astro-ph]} \BibitemShut {NoStop}%
\bibitem [{\citenamefont {{Kumar}}\ and\ \citenamefont
  {{Zhang}}(2015)}]{Kumar+Zhang2015}%
  \BibitemOpen
  \bibfield  {author} {\bibinfo {author} {\bibfnamefont {P.}~\bibnamefont
  {{Kumar}}}\ and\ \bibinfo {author} {\bibfnamefont {B.}~\bibnamefont
  {{Zhang}}},\ }\href {\doibase 10.1016/j.physrep.2014.09.008} {\bibfield
  {journal} {\bibinfo  {journal} {\physrep}\ }\textbf {\bibinfo {volume}
  {561}},\ \bibinfo {pages} {1} (\bibinfo {year} {2015})},\ \Eprint
  {http://arxiv.org/abs/1410.0679} {arXiv:1410.0679 [astro-ph.HE]} \BibitemShut
  {NoStop}%
\bibitem [{\citenamefont {{Katz}}(1994)}]{katz1994ApJ}%
  \BibitemOpen
  \bibfield  {author} {\bibinfo {author} {\bibfnamefont {J.~I.}\ \bibnamefont
  {{Katz}}},\ }\href {\doibase 10.1086/173723} {\bibfield  {journal} {\bibinfo
  {journal} {\apj}\ }\textbf {\bibinfo {volume} {422}},\ \bibinfo {pages} {248}
  (\bibinfo {year} {1994})},\ \Eprint {http://arxiv.org/abs/astro-ph/9212006}
  {arXiv:astro-ph/9212006 [astro-ph]} \BibitemShut {NoStop}%
\bibitem [{\citenamefont {{Sari}}\ \emph {et~al.}(1998)\citenamefont {{Sari}},
  \citenamefont {{Piran}},\ and\ \citenamefont {{Narayan}}}]{Sari+1998ApJL}%
  \BibitemOpen
  \bibfield  {author} {\bibinfo {author} {\bibfnamefont {R.}~\bibnamefont
  {{Sari}}}, \bibinfo {author} {\bibfnamefont {T.}~\bibnamefont {{Piran}}}, \
  and\ \bibinfo {author} {\bibfnamefont {R.}~\bibnamefont {{Narayan}}},\ }\href
  {\doibase 10.1086/311269} {\bibfield  {journal} {\bibinfo  {journal} {\apjl}\
  }\textbf {\bibinfo {volume} {497}},\ \bibinfo {pages} {L17} (\bibinfo {year}
  {1998})},\ \Eprint {http://arxiv.org/abs/astro-ph/9712005}
  {arXiv:astro-ph/9712005 [astro-ph]} \BibitemShut {NoStop}%
\bibitem [{\citenamefont {{Dai}}\ and\ \citenamefont
  {{Lu}}(1998)}]{Dai+Lu1998PRL}%
  \BibitemOpen
  \bibfield  {author} {\bibinfo {author} {\bibfnamefont {Z.~G.}\ \bibnamefont
  {{Dai}}}\ and\ \bibinfo {author} {\bibfnamefont {T.}~\bibnamefont {{Lu}}},\
  }\href {\doibase 10.1103/PhysRevLett.81.4301} {\bibfield  {journal} {\bibinfo
   {journal} {\prl}\ }\textbf {\bibinfo {volume} {81}},\ \bibinfo {pages}
  {4301} (\bibinfo {year} {1998})},\ \Eprint
  {http://arxiv.org/abs/astro-ph/9810332} {arXiv:astro-ph/9810332 [astro-ph]}
  \BibitemShut {NoStop}%
\bibitem [{\citenamefont {{Zhang}}\ and\ \citenamefont
  {{M{\'e}sz{\'a}ros}}(2001)}]{Zhang2001ApJL}%
  \BibitemOpen
  \bibfield  {author} {\bibinfo {author} {\bibfnamefont {B.}~\bibnamefont
  {{Zhang}}}\ and\ \bibinfo {author} {\bibfnamefont {P.}~\bibnamefont
  {{M{\'e}sz{\'a}ros}}},\ }\href {\doibase 10.1086/320255} {\bibfield
  {journal} {\bibinfo  {journal} {\apjl}\ }\textbf {\bibinfo {volume} {552}},\
  \bibinfo {pages} {L35} (\bibinfo {year} {2001})},\ \Eprint
  {http://arxiv.org/abs/astro-ph/0011133} {arXiv:astro-ph/0011133 [astro-ph]}
  \BibitemShut {NoStop}%
\bibitem [{\citenamefont {{Fan}}\ and\ \citenamefont
  {{Wei}}(2005)}]{Fan+Wei2005MNRAS}%
  \BibitemOpen
  \bibfield  {author} {\bibinfo {author} {\bibfnamefont {Y.~Z.}\ \bibnamefont
  {{Fan}}}\ and\ \bibinfo {author} {\bibfnamefont {D.~M.}\ \bibnamefont
  {{Wei}}},\ }\href {\doibase 10.1111/j.1745-3933.2005.00102.x} {\bibfield
  {journal} {\bibinfo  {journal} {\mnras}\ }\textbf {\bibinfo {volume} {364}},\
  \bibinfo {pages} {L42} (\bibinfo {year} {2005})},\ \Eprint
  {http://arxiv.org/abs/astro-ph/0506155} {arXiv:astro-ph/0506155 [astro-ph]}
  \BibitemShut {NoStop}%
\bibitem [{\citenamefont {{Burrows}}\ \emph {et~al.}(2005)\citenamefont
  {{Burrows}}, \citenamefont {{Romano}}, \citenamefont {{Falcone}},
  \citenamefont {{Kobayashi}}, \citenamefont {{Zhang}} \emph
  {et~al.}}]{Burrows2005Sci}%
  \BibitemOpen
  \bibfield  {author} {\bibinfo {author} {\bibfnamefont {D.~N.}\ \bibnamefont
  {{Burrows}}}, \bibinfo {author} {\bibfnamefont {P.}~\bibnamefont {{Romano}}},
  \bibinfo {author} {\bibfnamefont {A.}~\bibnamefont {{Falcone}}}, \bibinfo
  {author} {\bibfnamefont {S.}~\bibnamefont {{Kobayashi}}}, \bibinfo {author}
  {\bibfnamefont {B.}~\bibnamefont {{Zhang}}},  \emph {et~al.},\ }\href
  {\doibase 10.1126/science.1116168} {\bibfield  {journal} {\bibinfo  {journal}
  {Science}\ }\textbf {\bibinfo {volume} {309}},\ \bibinfo {pages} {1833}
  (\bibinfo {year} {2005})},\ \Eprint {http://arxiv.org/abs/astro-ph/0506130}
  {arXiv:astro-ph/0506130 [astro-ph]} \BibitemShut {NoStop}%
\bibitem [{\citenamefont {{Fan}}\ \emph {et~al.}(2005)\citenamefont {{Fan}},
  \citenamefont {{Zhang}},\ and\ \citenamefont {{Proga}}}]{Fan2005ApJ}%
  \BibitemOpen
  \bibfield  {author} {\bibinfo {author} {\bibfnamefont {Y.~Z.}\ \bibnamefont
  {{Fan}}}, \bibinfo {author} {\bibfnamefont {B.}~\bibnamefont {{Zhang}}}, \
  and\ \bibinfo {author} {\bibfnamefont {D.}~\bibnamefont {{Proga}}},\ }\href
  {\doibase 10.1086/499489} {\bibfield  {journal} {\bibinfo  {journal} {\apjl}\
  }\textbf {\bibinfo {volume} {635}},\ \bibinfo {pages} {L129} (\bibinfo {year}
  {2005})},\ \Eprint {http://arxiv.org/abs/astro-ph/0509019}
  {arXiv:astro-ph/0509019 [astro-ph]} \BibitemShut {NoStop}%
\bibitem [{\citenamefont {{Gao}}\ and\ \citenamefont
  {{Fan}}(2006)}]{Gao+Fan2006}%
  \BibitemOpen
  \bibfield  {author} {\bibinfo {author} {\bibfnamefont {W.-H.}\ \bibnamefont
  {{Gao}}}\ and\ \bibinfo {author} {\bibfnamefont {Y.-Z.}\ \bibnamefont
  {{Fan}}},\ }\href {\doibase 10.1088/1009-9271/6/5/01} {\bibfield  {journal}
  {\bibinfo  {journal} {\cjaa}\ }\textbf {\bibinfo {volume} {6}},\ \bibinfo
  {pages} {513} (\bibinfo {year} {2006})},\ \Eprint
  {http://arxiv.org/abs/astro-ph/0512646} {arXiv:astro-ph/0512646 [astro-ph]}
  \BibitemShut {NoStop}%
\bibitem [{\citenamefont {{Zhang}}\ \emph {et~al.}(2006)\citenamefont
  {{Zhang}}, \citenamefont {{Fan}}, \citenamefont {{Dyks}}, \citenamefont
  {{Kobayashi}}, \citenamefont {{M{\'e}sz{\'a}ros}}, \citenamefont {{Burrows}},
  \citenamefont {{Nousek}},\ and\ \citenamefont {{Gehrels}}}]{Zhang2006ApJ}%
  \BibitemOpen
  \bibfield  {author} {\bibinfo {author} {\bibfnamefont {B.}~\bibnamefont
  {{Zhang}}}, \bibinfo {author} {\bibfnamefont {Y.~Z.}\ \bibnamefont {{Fan}}},
  \bibinfo {author} {\bibfnamefont {J.}~\bibnamefont {{Dyks}}}, \bibinfo
  {author} {\bibfnamefont {S.}~\bibnamefont {{Kobayashi}}}, \bibinfo {author}
  {\bibfnamefont {P.}~\bibnamefont {{M{\'e}sz{\'a}ros}}}, \bibinfo {author}
  {\bibfnamefont {D.~N.}\ \bibnamefont {{Burrows}}}, \bibinfo {author}
  {\bibfnamefont {J.~A.}\ \bibnamefont {{Nousek}}}, \ and\ \bibinfo {author}
  {\bibfnamefont {N.}~\bibnamefont {{Gehrels}}},\ }\href {\doibase
  10.1086/500723} {\bibfield  {journal} {\bibinfo  {journal} {\apj}\ }\textbf
  {\bibinfo {volume} {642}},\ \bibinfo {pages} {354} (\bibinfo {year}
  {2006})},\ \Eprint {http://arxiv.org/abs/astro-ph/0508321}
  {arXiv:astro-ph/0508321 [astro-ph]} \BibitemShut {NoStop}%
\bibitem [{\citenamefont {{Chevalier}}\ and\ \citenamefont
  {{Li}}(1999)}]{Chevalier1999ApJL}%
  \BibitemOpen
  \bibfield  {author} {\bibinfo {author} {\bibfnamefont {R.~A.}\ \bibnamefont
  {{Chevalier}}}\ and\ \bibinfo {author} {\bibfnamefont {Z.-Y.}\ \bibnamefont
  {{Li}}},\ }\href {\doibase 10.1086/312147} {\bibfield  {journal} {\bibinfo
  {journal} {\apjl}\ }\textbf {\bibinfo {volume} {520}},\ \bibinfo {pages}
  {L29} (\bibinfo {year} {1999})},\ \Eprint
  {http://arxiv.org/abs/astro-ph/9904417} {arXiv:astro-ph/9904417 [astro-ph]}
  \BibitemShut {NoStop}%
\bibitem [{\citenamefont {{Granot}}\ \emph {et~al.}(1999)\citenamefont
  {{Granot}}, \citenamefont {{Piran}},\ and\ \citenamefont
  {{Sari}}}]{Granot1999ApJ}%
  \BibitemOpen
  \bibfield  {author} {\bibinfo {author} {\bibfnamefont {J.}~\bibnamefont
  {{Granot}}}, \bibinfo {author} {\bibfnamefont {T.}~\bibnamefont {{Piran}}}, \
  and\ \bibinfo {author} {\bibfnamefont {R.}~\bibnamefont {{Sari}}},\ }\href
  {\doibase 10.1086/306884} {\bibfield  {journal} {\bibinfo  {journal} {\apj}\
  }\textbf {\bibinfo {volume} {513}},\ \bibinfo {pages} {679} (\bibinfo {year}
  {1999})},\ \Eprint {http://arxiv.org/abs/astro-ph/9806192}
  {arXiv:astro-ph/9806192 [astro-ph]} \BibitemShut {NoStop}%
\bibitem [{\citenamefont {{Rhoads}}(1999)}]{Rhoads1999ApJ}%
  \BibitemOpen
  \bibfield  {author} {\bibinfo {author} {\bibfnamefont {J.~E.}\ \bibnamefont
  {{Rhoads}}},\ }\href {\doibase 10.1086/307907} {\bibfield  {journal}
  {\bibinfo  {journal} {\apj}\ }\textbf {\bibinfo {volume} {525}},\ \bibinfo
  {pages} {737} (\bibinfo {year} {1999})},\ \Eprint
  {http://arxiv.org/abs/astro-ph/9903399} {arXiv:astro-ph/9903399 [astro-ph]}
  \BibitemShut {NoStop}%
\bibitem [{\citenamefont {{Panaitescu}}\ and\ \citenamefont
  {{M{\'e}sz{\'a}ros}}(1999)}]{Panaitescu1999ApJ}%
  \BibitemOpen
  \bibfield  {author} {\bibinfo {author} {\bibfnamefont {A.}~\bibnamefont
  {{Panaitescu}}}\ and\ \bibinfo {author} {\bibfnamefont {P.}~\bibnamefont
  {{M{\'e}sz{\'a}ros}}},\ }\href {\doibase 10.1086/308005} {\bibfield
  {journal} {\bibinfo  {journal} {\apj}\ }\textbf {\bibinfo {volume} {526}},\
  \bibinfo {pages} {707} (\bibinfo {year} {1999})},\ \Eprint
  {http://arxiv.org/abs/astro-ph/9806016} {arXiv:astro-ph/9806016 [astro-ph]}
  \BibitemShut {NoStop}%
\bibitem [{\citenamefont {{Huang}}\ \emph {et~al.}(2000)\citenamefont
  {{Huang}}, \citenamefont {{Dai}},\ and\ \citenamefont {{Lu}}}]{Huang2000}%
  \BibitemOpen
  \bibfield  {author} {\bibinfo {author} {\bibfnamefont {Y.~F.}\ \bibnamefont
  {{Huang}}}, \bibinfo {author} {\bibfnamefont {Z.~G.}\ \bibnamefont {{Dai}}},
  \ and\ \bibinfo {author} {\bibfnamefont {T.}~\bibnamefont {{Lu}}},\
  }\href@noop {} {\bibfield  {journal} {\bibinfo  {journal} {\aap}\ }\textbf
  {\bibinfo {volume} {355}},\ \bibinfo {pages} {L43} (\bibinfo {year}
  {2000})},\ \Eprint {http://arxiv.org/abs/astro-ph/0002433}
  {arXiv:astro-ph/0002433 [astro-ph]} \BibitemShut {NoStop}%
\bibitem [{\citenamefont {{Granot}}\ \emph {et~al.}(2002)\citenamefont
  {{Granot}}, \citenamefont {{Panaitescu}}, \citenamefont {{Kumar}},\ and\
  \citenamefont {{Woosley}}}]{Granot2002ApJL}%
  \BibitemOpen
  \bibfield  {author} {\bibinfo {author} {\bibfnamefont {J.}~\bibnamefont
  {{Granot}}}, \bibinfo {author} {\bibfnamefont {A.}~\bibnamefont
  {{Panaitescu}}}, \bibinfo {author} {\bibfnamefont {P.}~\bibnamefont
  {{Kumar}}}, \ and\ \bibinfo {author} {\bibfnamefont {S.~E.}\ \bibnamefont
  {{Woosley}}},\ }\href {\doibase 10.1086/340991} {\bibfield  {journal}
  {\bibinfo  {journal} {\apjl}\ }\textbf {\bibinfo {volume} {570}},\ \bibinfo
  {pages} {L61} (\bibinfo {year} {2002})},\ \Eprint
  {http://arxiv.org/abs/astro-ph/0201322} {arXiv:astro-ph/0201322 [astro-ph]}
  \BibitemShut {NoStop}%
\bibitem [{\citenamefont {{Wei}}\ and\ \citenamefont
  {{Jin}}(2003)}]{Wei+Jin2003}%
  \BibitemOpen
  \bibfield  {author} {\bibinfo {author} {\bibfnamefont {D.~M.}\ \bibnamefont
  {{Wei}}}\ and\ \bibinfo {author} {\bibfnamefont {Z.~P.}\ \bibnamefont
  {{Jin}}},\ }\href {\doibase 10.1051/0004-6361:20030007} {\bibfield  {journal}
  {\bibinfo  {journal} {\aap}\ }\textbf {\bibinfo {volume} {400}},\ \bibinfo
  {pages} {415} (\bibinfo {year} {2003})},\ \Eprint
  {http://arxiv.org/abs/astro-ph/0212514} {arXiv:astro-ph/0212514 [astro-ph]}
  \BibitemShut {NoStop}%
\bibitem [{\citenamefont {{Kumar}}\ and\ \citenamefont
  {{Granot}}(2003)}]{Kumar+Granot2003ApJ}%
  \BibitemOpen
  \bibfield  {author} {\bibinfo {author} {\bibfnamefont {P.}~\bibnamefont
  {{Kumar}}}\ and\ \bibinfo {author} {\bibfnamefont {J.}~\bibnamefont
  {{Granot}}},\ }\href {\doibase 10.1086/375186} {\bibfield  {journal}
  {\bibinfo  {journal} {\apj}\ }\textbf {\bibinfo {volume} {591}},\ \bibinfo
  {pages} {1075} (\bibinfo {year} {2003})},\ \Eprint
  {http://arxiv.org/abs/astro-ph/0303174} {arXiv:astro-ph/0303174 [astro-ph]}
  \BibitemShut {NoStop}%
\bibitem [{\citenamefont {{Fan}}\ \emph {et~al.}(2008)\citenamefont {{Fan}},
  \citenamefont {{Piran}}, \citenamefont {{Narayan}},\ and\ \citenamefont
  {{Wei}}}]{Fan2008MNRAS}%
  \BibitemOpen
  \bibfield  {author} {\bibinfo {author} {\bibfnamefont {Y.-Z.}\ \bibnamefont
  {{Fan}}}, \bibinfo {author} {\bibfnamefont {T.}~\bibnamefont {{Piran}}},
  \bibinfo {author} {\bibfnamefont {R.}~\bibnamefont {{Narayan}}}, \ and\
  \bibinfo {author} {\bibfnamefont {D.-M.}\ \bibnamefont {{Wei}}},\ }\href
  {\doibase 10.1111/j.1365-2966.2007.12765.x} {\bibfield  {journal} {\bibinfo
  {journal} {\mnras}\ }\textbf {\bibinfo {volume} {384}},\ \bibinfo {pages}
  {1483} (\bibinfo {year} {2008})},\ \Eprint {http://arxiv.org/abs/0704.2063}
  {arXiv:0704.2063 [astro-ph]} \BibitemShut {NoStop}%
\bibitem [{\citenamefont {{M{\'e}sz{\'a}ros}}\ \emph
  {et~al.}(1998)\citenamefont {{M{\'e}sz{\'a}ros}}, \citenamefont {{Rees}},\
  and\ \citenamefont {{Wijers}}}]{Meszaros1998ApJ}%
  \BibitemOpen
  \bibfield  {author} {\bibinfo {author} {\bibfnamefont {P.}~\bibnamefont
  {{M{\'e}sz{\'a}ros}}}, \bibinfo {author} {\bibfnamefont {M.~J.}\ \bibnamefont
  {{Rees}}}, \ and\ \bibinfo {author} {\bibfnamefont {R.~A.~M.~J.}\
  \bibnamefont {{Wijers}}},\ }\href {\doibase 10.1086/305635} {\bibfield
  {journal} {\bibinfo  {journal} {\apj}\ }\textbf {\bibinfo {volume} {499}},\
  \bibinfo {pages} {301} (\bibinfo {year} {1998})},\ \Eprint
  {http://arxiv.org/abs/astro-ph/9709273} {arXiv:astro-ph/9709273 [astro-ph]}
  \BibitemShut {NoStop}%
\bibitem [{\citenamefont {{Salmonson}}(2001)}]{Salmonson2001ApJ}%
  \BibitemOpen
  \bibfield  {author} {\bibinfo {author} {\bibfnamefont {J.~D.}\ \bibnamefont
  {{Salmonson}}},\ }\href {\doibase 10.1086/318063} {\bibfield  {journal}
  {\bibinfo  {journal} {\apjl}\ }\textbf {\bibinfo {volume} {546}},\ \bibinfo
  {pages} {L29} (\bibinfo {year} {2001})},\ \Eprint
  {http://arxiv.org/abs/astro-ph/0010123} {arXiv:astro-ph/0010123 [astro-ph]}
  \BibitemShut {NoStop}%
\bibitem [{\citenamefont {{Dai}}\ and\ \citenamefont
  {{Gou}}(2001)}]{Dai&Gou2001ApJ}%
  \BibitemOpen
  \bibfield  {author} {\bibinfo {author} {\bibfnamefont {Z.~G.}\ \bibnamefont
  {{Dai}}}\ and\ \bibinfo {author} {\bibfnamefont {L.~J.}\ \bibnamefont
  {{Gou}}},\ }\href {\doibase 10.1086/320463} {\bibfield  {journal} {\bibinfo
  {journal} {\apj}\ }\textbf {\bibinfo {volume} {552}},\ \bibinfo {pages} {72}
  (\bibinfo {year} {2001})},\ \Eprint {http://arxiv.org/abs/astro-ph/0010261}
  {arXiv:astro-ph/0010261 [astro-ph]} \BibitemShut {NoStop}%
\bibitem [{\citenamefont {{Rossi}}\ \emph {et~al.}(2002)\citenamefont
  {{Rossi}}, \citenamefont {{Lazzati}},\ and\ \citenamefont
  {{Rees}}}]{Rossi2002MNRAS}%
  \BibitemOpen
  \bibfield  {author} {\bibinfo {author} {\bibfnamefont {E.}~\bibnamefont
  {{Rossi}}}, \bibinfo {author} {\bibfnamefont {D.}~\bibnamefont {{Lazzati}}},
  \ and\ \bibinfo {author} {\bibfnamefont {M.~J.}\ \bibnamefont {{Rees}}},\
  }\href {\doibase 10.1046/j.1365-8711.2002.05363.x} {\bibfield  {journal}
  {\bibinfo  {journal} {\mnras}\ }\textbf {\bibinfo {volume} {332}},\ \bibinfo
  {pages} {945} (\bibinfo {year} {2002})},\ \Eprint
  {http://arxiv.org/abs/astro-ph/0112083} {arXiv:astro-ph/0112083 [astro-ph]}
  \BibitemShut {NoStop}%
\bibitem [{\citenamefont {{Berger}}\ \emph {et~al.}(2003)\citenamefont
  {{Berger}}, \citenamefont {{Kulkarni}}, \citenamefont {{Pooley}},
  \citenamefont {{Frail}}, \citenamefont {{McIntyre}}, \citenamefont {{Wark}},
  \citenamefont {{Sari}}, \citenamefont {{Soderberg}}, \citenamefont {{Fox}},
  \citenamefont {{Yost}},\ and\ \citenamefont {{Price}}}]{Berger2003Natur}%
  \BibitemOpen
  \bibfield  {author} {\bibinfo {author} {\bibfnamefont {E.}~\bibnamefont
  {{Berger}}}, \bibinfo {author} {\bibfnamefont {S.~R.}\ \bibnamefont
  {{Kulkarni}}}, \bibinfo {author} {\bibfnamefont {G.}~\bibnamefont
  {{Pooley}}}, \bibinfo {author} {\bibfnamefont {D.~A.}\ \bibnamefont
  {{Frail}}}, \bibinfo {author} {\bibfnamefont {V.}~\bibnamefont {{McIntyre}}},
  \bibinfo {author} {\bibfnamefont {R.~M.}\ \bibnamefont {{Wark}}}, \bibinfo
  {author} {\bibfnamefont {R.}~\bibnamefont {{Sari}}}, \bibinfo {author}
  {\bibfnamefont {A.~M.}\ \bibnamefont {{Soderberg}}}, \bibinfo {author}
  {\bibfnamefont {D.~W.}\ \bibnamefont {{Fox}}}, \bibinfo {author}
  {\bibfnamefont {S.}~\bibnamefont {{Yost}}}, \ and\ \bibinfo {author}
  {\bibfnamefont {P.~A.}\ \bibnamefont {{Price}}},\ }\href {\doibase
  10.1038/nature01998} {\bibfield  {journal} {\bibinfo  {journal} {\nat}\
  }\textbf {\bibinfo {volume} {426}},\ \bibinfo {pages} {154} (\bibinfo {year}
  {2003})},\ \Eprint {http://arxiv.org/abs/astro-ph/0308187}
  {arXiv:astro-ph/0308187 [astro-ph]} \BibitemShut {NoStop}%
\bibitem [{\citenamefont {{Zhang}}\ \emph {et~al.}(2004)\citenamefont
  {{Zhang}}, \citenamefont {{Dai}}, \citenamefont {{Lloyd-Ronning}},\ and\
  \citenamefont {{M{\'e}sz{\'a}ros}}}]{Zhang2004ApJL}%
  \BibitemOpen
  \bibfield  {author} {\bibinfo {author} {\bibfnamefont {B.}~\bibnamefont
  {{Zhang}}}, \bibinfo {author} {\bibfnamefont {X.}~\bibnamefont {{Dai}}},
  \bibinfo {author} {\bibfnamefont {N.~M.}\ \bibnamefont {{Lloyd-Ronning}}}, \
  and\ \bibinfo {author} {\bibfnamefont {P.}~\bibnamefont
  {{M{\'e}sz{\'a}ros}}},\ }\href {\doibase 10.1086/382132} {\bibfield
  {journal} {\bibinfo  {journal} {\apjl}\ }\textbf {\bibinfo {volume} {601}},\
  \bibinfo {pages} {L119} (\bibinfo {year} {2004})},\ \Eprint
  {http://arxiv.org/abs/astro-ph/0311190} {arXiv:astro-ph/0311190 [astro-ph]}
  \BibitemShut {NoStop}%
\bibitem [{\citenamefont {{Wu}}\ \emph {et~al.}(2005)\citenamefont {{Wu}},
  \citenamefont {{Dai}}, \citenamefont {{Huang}},\ and\ \citenamefont
  {{Lu}}}]{Wu2005MNRAS}%
  \BibitemOpen
  \bibfield  {author} {\bibinfo {author} {\bibfnamefont {X.~F.}\ \bibnamefont
  {{Wu}}}, \bibinfo {author} {\bibfnamefont {Z.~G.}\ \bibnamefont {{Dai}}},
  \bibinfo {author} {\bibfnamefont {Y.~F.}\ \bibnamefont {{Huang}}}, \ and\
  \bibinfo {author} {\bibfnamefont {T.}~\bibnamefont {{Lu}}},\ }\href {\doibase
  10.1111/j.1365-2966.2005.08685.x} {\bibfield  {journal} {\bibinfo  {journal}
  {\mnras}\ }\textbf {\bibinfo {volume} {357}},\ \bibinfo {pages} {1197}
  (\bibinfo {year} {2005})},\ \Eprint {http://arxiv.org/abs/astro-ph/0412011}
  {arXiv:astro-ph/0412011 [astro-ph]} \BibitemShut {NoStop}%
\bibitem [{\citenamefont {{Pe'er}}(2012)}]{Pe'er2012ApJL}%
  \BibitemOpen
  \bibfield  {author} {\bibinfo {author} {\bibfnamefont {A.}~\bibnamefont
  {{Pe'er}}},\ }\href {\doibase 10.1088/2041-8205/752/1/L8} {\bibfield
  {journal} {\bibinfo  {journal} {\apjl}\ }\textbf {\bibinfo {volume} {752}},\
  \bibinfo {eid} {L8} (\bibinfo {year} {2012})},\ \Eprint
  {http://arxiv.org/abs/1203.5797} {arXiv:1203.5797 [astro-ph.HE]} \BibitemShut
  {NoStop}%
\bibitem [{\citenamefont {{Katz}}\ and\ \citenamefont
  {{Piran}}(1997)}]{Katz1997ApJ}%
  \BibitemOpen
  \bibfield  {author} {\bibinfo {author} {\bibfnamefont {J.~I.}\ \bibnamefont
  {{Katz}}}\ and\ \bibinfo {author} {\bibfnamefont {T.}~\bibnamefont
  {{Piran}}},\ }\href {\doibase 10.1086/304913} {\bibfield  {journal} {\bibinfo
   {journal} {\apj}\ }\textbf {\bibinfo {volume} {490}},\ \bibinfo {pages}
  {772} (\bibinfo {year} {1997})}\BibitemShut {NoStop}%
\bibitem [{\citenamefont {{Chiang}}\ and\ \citenamefont
  {{Dermer}}(1999)}]{Chiang1999ApJ}%
  \BibitemOpen
  \bibfield  {author} {\bibinfo {author} {\bibfnamefont {J.}~\bibnamefont
  {{Chiang}}}\ and\ \bibinfo {author} {\bibfnamefont {C.~D.}\ \bibnamefont
  {{Dermer}}},\ }\href {\doibase 10.1086/306789} {\bibfield  {journal}
  {\bibinfo  {journal} {\apj}\ }\textbf {\bibinfo {volume} {512}},\ \bibinfo
  {pages} {699} (\bibinfo {year} {1999})},\ \Eprint
  {http://arxiv.org/abs/astro-ph/9803339} {arXiv:astro-ph/9803339 [astro-ph]}
  \BibitemShut {NoStop}%
\bibitem [{\citenamefont {{Huang}}\ \emph {et~al.}(1999)\citenamefont
  {{Huang}}, \citenamefont {{Dai}},\ and\ \citenamefont
  {{Lu}}}]{Huang1999MNRAS}%
  \BibitemOpen
  \bibfield  {author} {\bibinfo {author} {\bibfnamefont {Y.~F.}\ \bibnamefont
  {{Huang}}}, \bibinfo {author} {\bibfnamefont {Z.~G.}\ \bibnamefont {{Dai}}},
  \ and\ \bibinfo {author} {\bibfnamefont {T.}~\bibnamefont {{Lu}}},\ }\href
  {\doibase 10.1046/j.1365-8711.1999.02887.x} {\bibfield  {journal} {\bibinfo
  {journal} {\mnras}\ }\textbf {\bibinfo {volume} {309}},\ \bibinfo {pages}
  {513} (\bibinfo {year} {1999})},\ \Eprint
  {http://arxiv.org/abs/astro-ph/9906370} {arXiv:astro-ph/9906370 [astro-ph]}
  \BibitemShut {NoStop}%
\bibitem [{\citenamefont {{Schutz}}(1986)}]{Schutz1986Natur}%
  \BibitemOpen
  \bibfield  {author} {\bibinfo {author} {\bibfnamefont {B.~F.}\ \bibnamefont
  {{Schutz}}},\ }\href {\doibase 10.1038/323310a0} {\bibfield  {journal}
  {\bibinfo  {journal} {\nat}\ }\textbf {\bibinfo {volume} {323}},\ \bibinfo
  {pages} {310} (\bibinfo {year} {1986})}\BibitemShut {NoStop}%
\bibitem [{\citenamefont {{Nakar}}\ and\ \citenamefont
  {{Piran}}(2021)}]{Nakar2021ApJ}%
  \BibitemOpen
  \bibfield  {author} {\bibinfo {author} {\bibfnamefont {E.}~\bibnamefont
  {{Nakar}}}\ and\ \bibinfo {author} {\bibfnamefont {T.}~\bibnamefont
  {{Piran}}},\ }\href {\doibase 10.3847/1538-4357/abd6cd} {\bibfield  {journal}
  {\bibinfo  {journal} {\apj}\ }\textbf {\bibinfo {volume} {909}},\ \bibinfo
  {eid} {114} (\bibinfo {year} {2021})},\ \Eprint
  {http://arxiv.org/abs/2005.01754} {arXiv:2005.01754 [astro-ph.HE]}
  \BibitemShut {NoStop}%
\bibitem [{\citenamefont {{Troja}}\ \emph {et~al.}(2021)\citenamefont
  {{Troja}}, \citenamefont {{O'Connor}}, \citenamefont {{Ryan}}, \citenamefont
  {{Piro}}, \citenamefont {{Ricci}}, \citenamefont {{Zhang}} \emph
  {et~al.}}]{Troja2022MNRAS}%
  \BibitemOpen
  \bibfield  {author} {\bibinfo {author} {\bibfnamefont {E.}~\bibnamefont
  {{Troja}}}, \bibinfo {author} {\bibfnamefont {B.}~\bibnamefont {{O'Connor}}},
  \bibinfo {author} {\bibfnamefont {G.}~\bibnamefont {{Ryan}}}, \bibinfo
  {author} {\bibfnamefont {L.}~\bibnamefont {{Piro}}}, \bibinfo {author}
  {\bibfnamefont {R.}~\bibnamefont {{Ricci}}}, \bibinfo {author} {\bibfnamefont
  {B.}~\bibnamefont {{Zhang}}},  \emph {et~al.},\ }\href {\doibase
  10.1093/mnras/stab3533} {\bibfield  {journal} {\bibinfo  {journal} {\mnras}\
  }\textbf {\bibinfo {volume} {510}},\ \bibinfo {pages} {1902} (\bibinfo {year}
  {2021})},\ \Eprint {http://arxiv.org/abs/2104.13378} {arXiv:2104.13378
  [astro-ph.HE]} \BibitemShut {NoStop}%
\bibitem [{\citenamefont {{Wu}}\ and\ \citenamefont
  {{MacFadyen}}(2019{\natexlab{a}})}]{Wu2019APJ}%
  \BibitemOpen
  \bibfield  {author} {\bibinfo {author} {\bibfnamefont {Y.}~\bibnamefont
  {{Wu}}}\ and\ \bibinfo {author} {\bibfnamefont {A.}~\bibnamefont
  {{MacFadyen}}},\ }\href {\doibase 10.3847/2041-8213/ab2fd4} {\bibfield
  {journal} {\bibinfo  {journal} {\apjl}\ }\textbf {\bibinfo {volume} {880}},\
  \bibinfo {eid} {L23} (\bibinfo {year} {2019}{\natexlab{a}})},\ \Eprint
  {http://arxiv.org/abs/1905.02665} {arXiv:1905.02665 [astro-ph.HE]}
  \BibitemShut {NoStop}%
\bibitem [{\citenamefont {{Panaitescu}}\ and\ \citenamefont
  {{Kumar}}(2000)}]{Panaitescu2000ApJ}%
  \BibitemOpen
  \bibfield  {author} {\bibinfo {author} {\bibfnamefont {A.}~\bibnamefont
  {{Panaitescu}}}\ and\ \bibinfo {author} {\bibfnamefont {P.}~\bibnamefont
  {{Kumar}}},\ }\href {\doibase 10.1086/317090} {\bibfield  {journal} {\bibinfo
   {journal} {\apj}\ }\textbf {\bibinfo {volume} {543}},\ \bibinfo {pages} {66}
  (\bibinfo {year} {2000})},\ \Eprint {http://arxiv.org/abs/astro-ph/0003246}
  {arXiv:astro-ph/0003246 [astro-ph]} \BibitemShut {NoStop}%
\bibitem [{\citenamefont {{Wang}}\ and\ \citenamefont
  {{Giannios}}(2021)}]{WH2021ApJ}%
  \BibitemOpen
  \bibfield  {author} {\bibinfo {author} {\bibfnamefont {H.}~\bibnamefont
  {{Wang}}}\ and\ \bibinfo {author} {\bibfnamefont {D.}~\bibnamefont
  {{Giannios}}},\ }\href {\doibase 10.3847/1538-4357/abd39c} {\bibfield
  {journal} {\bibinfo  {journal} {\apj}\ }\textbf {\bibinfo {volume} {908}},\
  \bibinfo {eid} {200} (\bibinfo {year} {2021})},\ \Eprint
  {http://arxiv.org/abs/2009.04427} {arXiv:2009.04427 [astro-ph.HE]}
  \BibitemShut {NoStop}%
\bibitem [{\citenamefont {{Abbott}}\ \emph
  {et~al.}(2017{\natexlab{a}})\citenamefont {{Abbott}}, \citenamefont
  {{Abbott}}, \citenamefont {{Abbott}}, \citenamefont {{Acernese}},
  \citenamefont {{Ackley}} \emph {et~al.}}]{Abbott2017PRL}%
  \BibitemOpen
  \bibfield  {author} {\bibinfo {author} {\bibfnamefont {B.~P.}\ \bibnamefont
  {{Abbott}}}, \bibinfo {author} {\bibfnamefont {R.}~\bibnamefont {{Abbott}}},
  \bibinfo {author} {\bibfnamefont {T.~D.}\ \bibnamefont {{Abbott}}}, \bibinfo
  {author} {\bibfnamefont {F.}~\bibnamefont {{Acernese}}}, \bibinfo {author}
  {\bibfnamefont {K.}~\bibnamefont {{Ackley}}},  \emph {et~al.},\ }\href
  {\doibase 10.1103/PhysRevLett.119.161101} {\bibfield  {journal} {\bibinfo
  {journal} {\prl}\ }\textbf {\bibinfo {volume} {119}},\ \bibinfo {eid}
  {161101} (\bibinfo {year} {2017}{\natexlab{a}})},\ \Eprint
  {http://arxiv.org/abs/1710.05832} {arXiv:1710.05832 [gr-qc]} \BibitemShut
  {NoStop}%
\bibitem [{\citenamefont {{Abbott}}\ \emph
  {et~al.}(2017{\natexlab{b}})\citenamefont {{Abbott}}, \citenamefont
  {{Abbott}}, \citenamefont {{Abbott}}, \citenamefont {{Acernese}},
  \citenamefont {{Ackley}} \emph {et~al.}}]{Abbott1708172017APJL}%
  \BibitemOpen
  \bibfield  {author} {\bibinfo {author} {\bibfnamefont {B.~P.}\ \bibnamefont
  {{Abbott}}}, \bibinfo {author} {\bibfnamefont {R.}~\bibnamefont {{Abbott}}},
  \bibinfo {author} {\bibfnamefont {T.~D.}\ \bibnamefont {{Abbott}}}, \bibinfo
  {author} {\bibfnamefont {F.}~\bibnamefont {{Acernese}}}, \bibinfo {author}
  {\bibfnamefont {K.}~\bibnamefont {{Ackley}}},  \emph {et~al.},\ }\href
  {\doibase 10.3847/2041-8213/aa91c9} {\bibfield  {journal} {\bibinfo
  {journal} {\apjl}\ }\textbf {\bibinfo {volume} {848}},\ \bibinfo {eid} {L12}
  (\bibinfo {year} {2017}{\natexlab{b}})},\ \Eprint
  {http://arxiv.org/abs/1710.05833} {arXiv:1710.05833 [astro-ph.HE]}
  \BibitemShut {NoStop}%
\bibitem [{\citenamefont {{Troja}}\ \emph {et~al.}(2017)\citenamefont
  {{Troja}}, \citenamefont {{Piro}}, \citenamefont {{van Eerten}},
  \citenamefont {{Wollaeger}}, \citenamefont {{Im}}, \citenamefont {{Fox}}
  \emph {et~al.}}]{Troja2017Nat}%
  \BibitemOpen
  \bibfield  {author} {\bibinfo {author} {\bibfnamefont {E.}~\bibnamefont
  {{Troja}}}, \bibinfo {author} {\bibfnamefont {L.}~\bibnamefont {{Piro}}},
  \bibinfo {author} {\bibfnamefont {H.}~\bibnamefont {{van Eerten}}}, \bibinfo
  {author} {\bibfnamefont {R.~T.}\ \bibnamefont {{Wollaeger}}}, \bibinfo
  {author} {\bibfnamefont {M.}~\bibnamefont {{Im}}}, \bibinfo {author}
  {\bibfnamefont {O.~D.}\ \bibnamefont {{Fox}}},  \emph {et~al.},\ }\href
  {\doibase 10.1038/nature24290} {\bibfield  {journal} {\bibinfo  {journal}
  {\nat}\ }\textbf {\bibinfo {volume} {551}},\ \bibinfo {pages} {71} (\bibinfo
  {year} {2017})},\ \Eprint {http://arxiv.org/abs/1710.05433} {arXiv:1710.05433
  [astro-ph.HE]} \BibitemShut {NoStop}%
\bibitem [{\citenamefont {{Lamb}}\ and\ \citenamefont
  {{Kobayashi}}(2018)}]{Lamb2018MNRAS}%
  \BibitemOpen
  \bibfield  {author} {\bibinfo {author} {\bibfnamefont {G.~P.}\ \bibnamefont
  {{Lamb}}}\ and\ \bibinfo {author} {\bibfnamefont {S.}~\bibnamefont
  {{Kobayashi}}},\ }\href {\doibase 10.1093/mnras/sty1108} {\bibfield
  {journal} {\bibinfo  {journal} {\mnras}\ }\textbf {\bibinfo {volume} {478}},\
  \bibinfo {pages} {733} (\bibinfo {year} {2018})},\ \Eprint
  {http://arxiv.org/abs/1710.05857} {arXiv:1710.05857 [astro-ph.HE]}
  \BibitemShut {NoStop}%
\bibitem [{\citenamefont {{Lazzati}}\ \emph {et~al.}(2018)\citenamefont
  {{Lazzati}}, \citenamefont {{Perna}}, \citenamefont {{Morsony}},
  \citenamefont {{Lopez-Camara}}, \citenamefont {{Cantiello}}, \citenamefont
  {{Ciolfi}}, \citenamefont {{Giacomazzo}},\ and\ \citenamefont
  {{Workman}}}]{Lazzati2018PRL}%
  \BibitemOpen
  \bibfield  {author} {\bibinfo {author} {\bibfnamefont {D.}~\bibnamefont
  {{Lazzati}}}, \bibinfo {author} {\bibfnamefont {R.}~\bibnamefont {{Perna}}},
  \bibinfo {author} {\bibfnamefont {B.~J.}\ \bibnamefont {{Morsony}}}, \bibinfo
  {author} {\bibfnamefont {D.}~\bibnamefont {{Lopez-Camara}}}, \bibinfo
  {author} {\bibfnamefont {M.}~\bibnamefont {{Cantiello}}}, \bibinfo {author}
  {\bibfnamefont {R.}~\bibnamefont {{Ciolfi}}}, \bibinfo {author}
  {\bibfnamefont {B.}~\bibnamefont {{Giacomazzo}}}, \ and\ \bibinfo {author}
  {\bibfnamefont {J.~C.}\ \bibnamefont {{Workman}}},\ }\href {\doibase
  10.1103/PhysRevLett.120.241103} {\bibfield  {journal} {\bibinfo  {journal}
  {\prl}\ }\textbf {\bibinfo {volume} {120}},\ \bibinfo {eid} {241103}
  (\bibinfo {year} {2018})},\ \Eprint {http://arxiv.org/abs/1712.03237}
  {arXiv:1712.03237 [astro-ph.HE]} \BibitemShut {NoStop}%
\bibitem [{\citenamefont {{Lyman}}\ \emph {et~al.}(2018)\citenamefont
  {{Lyman}}, \citenamefont {{Lamb}}, \citenamefont {{Levan}}, \citenamefont
  {{Mandel}}, \citenamefont {{Tanvir}} \emph {et~al.}}]{Lyman2018NatAs}%
  \BibitemOpen
  \bibfield  {author} {\bibinfo {author} {\bibfnamefont {J.~D.}\ \bibnamefont
  {{Lyman}}}, \bibinfo {author} {\bibfnamefont {G.~P.}\ \bibnamefont {{Lamb}}},
  \bibinfo {author} {\bibfnamefont {A.~J.}\ \bibnamefont {{Levan}}}, \bibinfo
  {author} {\bibfnamefont {I.}~\bibnamefont {{Mandel}}}, \bibinfo {author}
  {\bibfnamefont {N.~R.}\ \bibnamefont {{Tanvir}}},  \emph {et~al.},\ }\href
  {\doibase 10.1038/s41550-018-0511-3} {\bibfield  {journal} {\bibinfo
  {journal} {Nature Astronomy}\ }\textbf {\bibinfo {volume} {2}},\ \bibinfo
  {pages} {751} (\bibinfo {year} {2018})},\ \Eprint
  {http://arxiv.org/abs/1801.02669} {arXiv:1801.02669 [astro-ph.HE]}
  \BibitemShut {NoStop}%
\bibitem [{\citenamefont {{Mooley}}\ \emph {et~al.}(2018)\citenamefont
  {{Mooley}}, \citenamefont {{Deller}}, \citenamefont {{Gottlieb}},
  \citenamefont {{Nakar}}, \citenamefont {{Hallinan}}, \citenamefont
  {{Bourke}}, \citenamefont {{Frail}}, \citenamefont {{Horesh}}, \citenamefont
  {{Corsi}},\ and\ \citenamefont {{Hotokezaka}}}]{Mooley2018Nat2}%
  \BibitemOpen
  \bibfield  {author} {\bibinfo {author} {\bibfnamefont {K.~P.}\ \bibnamefont
  {{Mooley}}}, \bibinfo {author} {\bibfnamefont {A.~T.}\ \bibnamefont
  {{Deller}}}, \bibinfo {author} {\bibfnamefont {O.}~\bibnamefont
  {{Gottlieb}}}, \bibinfo {author} {\bibfnamefont {E.}~\bibnamefont {{Nakar}}},
  \bibinfo {author} {\bibfnamefont {G.}~\bibnamefont {{Hallinan}}}, \bibinfo
  {author} {\bibfnamefont {S.}~\bibnamefont {{Bourke}}}, \bibinfo {author}
  {\bibfnamefont {D.~A.}\ \bibnamefont {{Frail}}}, \bibinfo {author}
  {\bibfnamefont {A.}~\bibnamefont {{Horesh}}}, \bibinfo {author}
  {\bibfnamefont {A.}~\bibnamefont {{Corsi}}}, \ and\ \bibinfo {author}
  {\bibfnamefont {K.}~\bibnamefont {{Hotokezaka}}},\ }\href {\doibase
  10.1038/s41586-018-0486-3} {\bibfield  {journal} {\bibinfo  {journal} {\nat}\
  }\textbf {\bibinfo {volume} {561}},\ \bibinfo {pages} {355} (\bibinfo {year}
  {2018})},\ \Eprint {http://arxiv.org/abs/1806.09693} {arXiv:1806.09693
  [astro-ph.HE]} \BibitemShut {NoStop}%
\bibitem [{\citenamefont {{Yue}}\ \emph {et~al.}(2018)\citenamefont {{Yue}},
  \citenamefont {{Hu}}, \citenamefont {{Zhang}}, \citenamefont {{Liang}},
  \citenamefont {{Jin}}, \citenamefont {{Zou}}, \citenamefont {{Fan}},\ and\
  \citenamefont {{Wei}}}]{Yue2018ApJ}%
  \BibitemOpen
  \bibfield  {author} {\bibinfo {author} {\bibfnamefont {C.}~\bibnamefont
  {{Yue}}}, \bibinfo {author} {\bibfnamefont {Q.}~\bibnamefont {{Hu}}},
  \bibinfo {author} {\bibfnamefont {F.-W.}\ \bibnamefont {{Zhang}}}, \bibinfo
  {author} {\bibfnamefont {Y.-F.}\ \bibnamefont {{Liang}}}, \bibinfo {author}
  {\bibfnamefont {Z.-P.}\ \bibnamefont {{Jin}}}, \bibinfo {author}
  {\bibfnamefont {Y.-C.}\ \bibnamefont {{Zou}}}, \bibinfo {author}
  {\bibfnamefont {Y.-Z.}\ \bibnamefont {{Fan}}}, \ and\ \bibinfo {author}
  {\bibfnamefont {D.-M.}\ \bibnamefont {{Wei}}},\ }\href {\doibase
  10.3847/2041-8213/aaa66c} {\bibfield  {journal} {\bibinfo  {journal} {\apjl}\
  }\textbf {\bibinfo {volume} {853}},\ \bibinfo {eid} {L10} (\bibinfo {year}
  {2018})},\ \Eprint {http://arxiv.org/abs/1710.05942} {arXiv:1710.05942
  [astro-ph.HE]} \BibitemShut {NoStop}%
\bibitem [{\citenamefont {{Troja}}\ \emph {et~al.}(2019)\citenamefont
  {{Troja}}, \citenamefont {{van Eerten}}, \citenamefont {{Ryan}},
  \citenamefont {{Ricci}}, \citenamefont {{Burgess}}, \citenamefont
  {{Wieringa}}, \citenamefont {{Piro}}, \citenamefont {{Cenko}},\ and\
  \citenamefont {{Sakamoto}}}]{Troja2019MNRAS}%
  \BibitemOpen
  \bibfield  {author} {\bibinfo {author} {\bibfnamefont {E.}~\bibnamefont
  {{Troja}}}, \bibinfo {author} {\bibfnamefont {H.}~\bibnamefont {{van
  Eerten}}}, \bibinfo {author} {\bibfnamefont {G.}~\bibnamefont {{Ryan}}},
  \bibinfo {author} {\bibfnamefont {R.}~\bibnamefont {{Ricci}}}, \bibinfo
  {author} {\bibfnamefont {J.~M.}\ \bibnamefont {{Burgess}}}, \bibinfo {author}
  {\bibfnamefont {M.~H.}\ \bibnamefont {{Wieringa}}}, \bibinfo {author}
  {\bibfnamefont {L.}~\bibnamefont {{Piro}}}, \bibinfo {author} {\bibfnamefont
  {S.~B.}\ \bibnamefont {{Cenko}}}, \ and\ \bibinfo {author} {\bibfnamefont
  {T.}~\bibnamefont {{Sakamoto}}},\ }\href {\doibase 10.1093/mnras/stz2248}
  {\bibfield  {journal} {\bibinfo  {journal} {\mnras}\ }\textbf {\bibinfo
  {volume} {489}},\ \bibinfo {pages} {1919} (\bibinfo {year} {2019})},\ \Eprint
  {http://arxiv.org/abs/1808.06617} {arXiv:1808.06617 [astro-ph.HE]}
  \BibitemShut {NoStop}%
\bibitem [{\citenamefont {{Ghirlanda}}\ \emph {et~al.}(2019)\citenamefont
  {{Ghirlanda}}, \citenamefont {{Salafia}}, \citenamefont {{Paragi}},
  \citenamefont {{Giroletti}}, \citenamefont {{Yang}} \emph
  {et~al.}}]{Ghirlanda2019Sci}%
  \BibitemOpen
  \bibfield  {author} {\bibinfo {author} {\bibfnamefont {G.}~\bibnamefont
  {{Ghirlanda}}}, \bibinfo {author} {\bibfnamefont {O.~S.}\ \bibnamefont
  {{Salafia}}}, \bibinfo {author} {\bibfnamefont {Z.}~\bibnamefont {{Paragi}}},
  \bibinfo {author} {\bibfnamefont {M.}~\bibnamefont {{Giroletti}}}, \bibinfo
  {author} {\bibfnamefont {J.}~\bibnamefont {{Yang}}},  \emph {et~al.},\ }\href
  {\doibase 10.1126/science.aau8815} {\bibfield  {journal} {\bibinfo  {journal}
  {Science}\ }\textbf {\bibinfo {volume} {363}},\ \bibinfo {pages} {968}
  (\bibinfo {year} {2019})},\ \Eprint {http://arxiv.org/abs/1808.00469}
  {arXiv:1808.00469 [astro-ph.HE]} \BibitemShut {NoStop}%
\bibitem [{\citenamefont {{Wu}}\ and\ \citenamefont
  {{MacFadyen}}(2019{\natexlab{b}})}]{Wu2019APJL}%
  \BibitemOpen
  \bibfield  {author} {\bibinfo {author} {\bibfnamefont {Y.}~\bibnamefont
  {{Wu}}}\ and\ \bibinfo {author} {\bibfnamefont {A.}~\bibnamefont
  {{MacFadyen}}},\ }\href {\doibase 10.3847/2041-8213/ab2fd4} {\bibfield
  {journal} {\bibinfo  {journal} {\apjl}\ }\textbf {\bibinfo {volume} {880}},\
  \bibinfo {eid} {L23} (\bibinfo {year} {2019}{\natexlab{b}})},\ \Eprint
  {http://arxiv.org/abs/1905.02665} {arXiv:1905.02665 [astro-ph.HE]}
  \BibitemShut {NoStop}%
\bibitem [{\citenamefont {{Hajela}}\ \emph {et~al.}(2019)\citenamefont
  {{Hajela}}, \citenamefont {{Margutti}}, \citenamefont {{Alexander}},
  \citenamefont {{Kathirgamaraju}}, \citenamefont {{Baldeschi}}, \citenamefont
  {{Guidorzi}} \emph {et~al.}}]{Hajela2019ApJL}%
  \BibitemOpen
  \bibfield  {author} {\bibinfo {author} {\bibfnamefont {A.}~\bibnamefont
  {{Hajela}}}, \bibinfo {author} {\bibfnamefont {R.}~\bibnamefont
  {{Margutti}}}, \bibinfo {author} {\bibfnamefont {K.~D.}\ \bibnamefont
  {{Alexander}}}, \bibinfo {author} {\bibfnamefont {A.}~\bibnamefont
  {{Kathirgamaraju}}}, \bibinfo {author} {\bibfnamefont {A.}~\bibnamefont
  {{Baldeschi}}}, \bibinfo {author} {\bibfnamefont {C.}~\bibnamefont
  {{Guidorzi}}},  \emph {et~al.},\ }\href {\doibase 10.3847/2041-8213/ab5226}
  {\bibfield  {journal} {\bibinfo  {journal} {\apjl}\ }\textbf {\bibinfo
  {volume} {886}},\ \bibinfo {eid} {L17} (\bibinfo {year} {2019})},\ \Eprint
  {http://arxiv.org/abs/1909.06393} {arXiv:1909.06393 [astro-ph.HE]}
  \BibitemShut {NoStop}%
\bibitem [{\citenamefont {{Takahashi}}\ and\ \citenamefont
  {{Ioka}}(2020)}]{Ioka2020MNRAS}%
  \BibitemOpen
  \bibfield  {author} {\bibinfo {author} {\bibfnamefont {K.}~\bibnamefont
  {{Takahashi}}}\ and\ \bibinfo {author} {\bibfnamefont {K.}~\bibnamefont
  {{Ioka}}},\ }\href {\doibase 10.1093/mnras/staa1984} {\bibfield  {journal}
  {\bibinfo  {journal} {\mnras}\ }\textbf {\bibinfo {volume} {497}},\ \bibinfo
  {pages} {1217} (\bibinfo {year} {2020})},\ \Eprint
  {http://arxiv.org/abs/1912.01871} {arXiv:1912.01871 [astro-ph.HE]}
  \BibitemShut {NoStop}%
\bibitem [{\citenamefont {{Hotokezaka}}\ \emph {et~al.}(2019)\citenamefont
  {{Hotokezaka}}, \citenamefont {{Nakar}}, \citenamefont {{Gottlieb}},
  \citenamefont {{Nissanke}}, \citenamefont {{Masuda}}, \citenamefont
  {{Hallinan}}, \citenamefont {{Mooley}},\ and\ \citenamefont
  {{Deller}}}]{Hotokezaka2019NatAs}%
  \BibitemOpen
  \bibfield  {author} {\bibinfo {author} {\bibfnamefont {K.}~\bibnamefont
  {{Hotokezaka}}}, \bibinfo {author} {\bibfnamefont {E.}~\bibnamefont
  {{Nakar}}}, \bibinfo {author} {\bibfnamefont {O.}~\bibnamefont {{Gottlieb}}},
  \bibinfo {author} {\bibfnamefont {S.}~\bibnamefont {{Nissanke}}}, \bibinfo
  {author} {\bibfnamefont {K.}~\bibnamefont {{Masuda}}}, \bibinfo {author}
  {\bibfnamefont {G.}~\bibnamefont {{Hallinan}}}, \bibinfo {author}
  {\bibfnamefont {K.~P.}\ \bibnamefont {{Mooley}}}, \ and\ \bibinfo {author}
  {\bibfnamefont {A.~T.}\ \bibnamefont {{Deller}}},\ }\href {\doibase
  10.1038/s41550-019-0820-1} {\bibfield  {journal} {\bibinfo  {journal} {Nature
  Astronomy}\ }\textbf {\bibinfo {volume} {3}},\ \bibinfo {pages} {940}
  (\bibinfo {year} {2019})},\ \Eprint {http://arxiv.org/abs/1806.10596}
  {arXiv:1806.10596 [astro-ph.CO]} \BibitemShut {NoStop}%
\bibitem [{\citenamefont {{Coughlin}}\ \emph {et~al.}(2020)\citenamefont
  {{Coughlin}}, \citenamefont {{Antier}}, \citenamefont {{Dietrich}},
  \citenamefont {{Foley}}, \citenamefont {{Heinzel}}, \citenamefont {{Bulla}}
  \emph {et~al.}}]{Coughlin2020NatCo}%
  \BibitemOpen
  \bibfield  {author} {\bibinfo {author} {\bibfnamefont {M.~W.}\ \bibnamefont
  {{Coughlin}}}, \bibinfo {author} {\bibfnamefont {S.}~\bibnamefont
  {{Antier}}}, \bibinfo {author} {\bibfnamefont {T.}~\bibnamefont
  {{Dietrich}}}, \bibinfo {author} {\bibfnamefont {R.~J.}\ \bibnamefont
  {{Foley}}}, \bibinfo {author} {\bibfnamefont {J.}~\bibnamefont {{Heinzel}}},
  \bibinfo {author} {\bibfnamefont {M.}~\bibnamefont {{Bulla}}},  \emph
  {et~al.},\ }\href {\doibase 10.1038/s41467-020-17998-5} {\bibfield  {journal}
  {\bibinfo  {journal} {Nature Communications}\ }\textbf {\bibinfo {volume}
  {11}},\ \bibinfo {eid} {4129} (\bibinfo {year} {2020})},\ \Eprint
  {http://arxiv.org/abs/2008.07420} {arXiv:2008.07420 [astro-ph.HE]}
  \BibitemShut {NoStop}%
\bibitem [{\citenamefont {{Heinzel}}\ \emph {et~al.}(2021)\citenamefont
  {{Heinzel}}, \citenamefont {{Coughlin}}, \citenamefont {{Dietrich}},
  \citenamefont {{Bulla}}, \citenamefont {{Antier}}, \citenamefont
  {{Christensen}}, \citenamefont {{Coulter}}, \citenamefont {{Foley}},
  \citenamefont {{Issa}},\ and\ \citenamefont {{Khetan}}}]{Heinzel2021MNRAS}%
  \BibitemOpen
  \bibfield  {author} {\bibinfo {author} {\bibfnamefont {J.}~\bibnamefont
  {{Heinzel}}}, \bibinfo {author} {\bibfnamefont {M.~W.}\ \bibnamefont
  {{Coughlin}}}, \bibinfo {author} {\bibfnamefont {T.}~\bibnamefont
  {{Dietrich}}}, \bibinfo {author} {\bibfnamefont {M.}~\bibnamefont {{Bulla}}},
  \bibinfo {author} {\bibfnamefont {S.}~\bibnamefont {{Antier}}}, \bibinfo
  {author} {\bibfnamefont {N.}~\bibnamefont {{Christensen}}}, \bibinfo {author}
  {\bibfnamefont {D.~A.}\ \bibnamefont {{Coulter}}}, \bibinfo {author}
  {\bibfnamefont {R.~J.}\ \bibnamefont {{Foley}}}, \bibinfo {author}
  {\bibfnamefont {L.}~\bibnamefont {{Issa}}}, \ and\ \bibinfo {author}
  {\bibfnamefont {N.}~\bibnamefont {{Khetan}}},\ }\href {\doibase
  10.1093/mnras/stab221} {\bibfield  {journal} {\bibinfo  {journal} {\mnras}\
  }\textbf {\bibinfo {volume} {502}},\ \bibinfo {pages} {3057} (\bibinfo {year}
  {2021})},\ \Eprint {http://arxiv.org/abs/2010.10746} {arXiv:2010.10746
  [astro-ph.HE]} \BibitemShut {NoStop}%
\bibitem [{\citenamefont {{Korobkin}}\ \emph {et~al.}(2021)\citenamefont
  {{Korobkin}}, \citenamefont {{Wollaeger}}, \citenamefont {{Fryer}},
  \citenamefont {{Hungerford}}, \citenamefont {{Rosswog}}, \citenamefont
  {{Fontes}} \emph {et~al.}}]{Korobkin2021ApJ}%
  \BibitemOpen
  \bibfield  {author} {\bibinfo {author} {\bibfnamefont {O.}~\bibnamefont
  {{Korobkin}}}, \bibinfo {author} {\bibfnamefont {R.~T.}\ \bibnamefont
  {{Wollaeger}}}, \bibinfo {author} {\bibfnamefont {C.~L.}\ \bibnamefont
  {{Fryer}}}, \bibinfo {author} {\bibfnamefont {A.~L.}\ \bibnamefont
  {{Hungerford}}}, \bibinfo {author} {\bibfnamefont {S.}~\bibnamefont
  {{Rosswog}}}, \bibinfo {author} {\bibfnamefont {C.~J.}\ \bibnamefont
  {{Fontes}}},  \emph {et~al.},\ }\href {\doibase 10.3847/1538-4357/abe1b5}
  {\bibfield  {journal} {\bibinfo  {journal} {\apj}\ }\textbf {\bibinfo
  {volume} {910}},\ \bibinfo {eid} {116} (\bibinfo {year} {2021})},\ \Eprint
  {http://arxiv.org/abs/2004.00102} {arXiv:2004.00102 [astro-ph.HE]}
  \BibitemShut {NoStop}%
\bibitem [{\citenamefont {{Lamb}}\ \emph {et~al.}(2019)\citenamefont {{Lamb}},
  \citenamefont {{Lyman}}, \citenamefont {{Levan}}, \citenamefont {{Tanvir}},
  \citenamefont {{Kangas}}, \citenamefont {{Fruchter}}, \citenamefont
  {{Gompertz}}, \citenamefont {{Hjorth}}, \citenamefont {{Mandel}},
  \citenamefont {{Oates}}, \citenamefont {{Steeghs}},\ and\ \citenamefont
  {{Wiersema}}}]{Lamb2019ApJ}%
  \BibitemOpen
  \bibfield  {author} {\bibinfo {author} {\bibfnamefont {G.~P.}\ \bibnamefont
  {{Lamb}}}, \bibinfo {author} {\bibfnamefont {J.~D.}\ \bibnamefont {{Lyman}}},
  \bibinfo {author} {\bibfnamefont {A.~J.}\ \bibnamefont {{Levan}}}, \bibinfo
  {author} {\bibfnamefont {N.~R.}\ \bibnamefont {{Tanvir}}}, \bibinfo {author}
  {\bibfnamefont {T.}~\bibnamefont {{Kangas}}}, \bibinfo {author}
  {\bibfnamefont {A.~S.}\ \bibnamefont {{Fruchter}}}, \bibinfo {author}
  {\bibfnamefont {B.}~\bibnamefont {{Gompertz}}}, \bibinfo {author}
  {\bibfnamefont {J.}~\bibnamefont {{Hjorth}}}, \bibinfo {author}
  {\bibfnamefont {I.}~\bibnamefont {{Mandel}}}, \bibinfo {author}
  {\bibfnamefont {S.~R.}\ \bibnamefont {{Oates}}}, \bibinfo {author}
  {\bibfnamefont {D.}~\bibnamefont {{Steeghs}}}, \ and\ \bibinfo {author}
  {\bibfnamefont {K.}~\bibnamefont {{Wiersema}}},\ }\href {\doibase
  10.3847/2041-8213/aaf96b} {\bibfield  {journal} {\bibinfo  {journal} {\apjl}\
  }\textbf {\bibinfo {volume} {870}},\ \bibinfo {eid} {L15} (\bibinfo {year}
  {2019})},\ \Eprint {http://arxiv.org/abs/1811.11491} {arXiv:1811.11491
  [astro-ph.HE]} \BibitemShut {NoStop}%
\bibitem [{\citenamefont {{Betoule}}\ \emph {et~al.}(2014)\citenamefont
  {{Betoule}}, \citenamefont {{Kessler}}, \citenamefont {{Guy}}, \citenamefont
  {{Mosher}}, \citenamefont {{Hardin}}, \citenamefont {{Biswas}} \emph
  {et~al.}}]{Betoule2014A&A}%
  \BibitemOpen
  \bibfield  {author} {\bibinfo {author} {\bibfnamefont {M.}~\bibnamefont
  {{Betoule}}}, \bibinfo {author} {\bibfnamefont {R.}~\bibnamefont
  {{Kessler}}}, \bibinfo {author} {\bibfnamefont {J.}~\bibnamefont {{Guy}}},
  \bibinfo {author} {\bibfnamefont {J.}~\bibnamefont {{Mosher}}}, \bibinfo
  {author} {\bibfnamefont {D.}~\bibnamefont {{Hardin}}}, \bibinfo {author}
  {\bibfnamefont {R.}~\bibnamefont {{Biswas}}},  \emph {et~al.},\ }\href
  {\doibase 10.1051/0004-6361/201423413} {\bibfield  {journal} {\bibinfo
  {journal} {\aap}\ }\textbf {\bibinfo {volume} {568}},\ \bibinfo {eid} {A22}
  (\bibinfo {year} {2014})},\ \Eprint {http://arxiv.org/abs/1401.4064}
  {arXiv:1401.4064 [astro-ph.CO]} \BibitemShut {NoStop}%
\bibitem [{\citenamefont {{Scolnic}}\ \emph {et~al.}(2018)\citenamefont
  {{Scolnic}}, \citenamefont {{Jones}}, \citenamefont {{Rest}}, \citenamefont
  {{Pan}}, \citenamefont {{Chornock}}, \citenamefont {{Foley}} \emph
  {et~al.}}]{Scolnic2018ApJ}%
  \BibitemOpen
  \bibfield  {author} {\bibinfo {author} {\bibfnamefont {D.~M.}\ \bibnamefont
  {{Scolnic}}}, \bibinfo {author} {\bibfnamefont {D.~O.}\ \bibnamefont
  {{Jones}}}, \bibinfo {author} {\bibfnamefont {A.}~\bibnamefont {{Rest}}},
  \bibinfo {author} {\bibfnamefont {Y.~C.}\ \bibnamefont {{Pan}}}, \bibinfo
  {author} {\bibfnamefont {R.}~\bibnamefont {{Chornock}}}, \bibinfo {author}
  {\bibfnamefont {R.~J.}\ \bibnamefont {{Foley}}},  \emph {et~al.},\ }\href
  {\doibase 10.3847/1538-4357/aab9bb} {\bibfield  {journal} {\bibinfo
  {journal} {\apj}\ }\textbf {\bibinfo {volume} {859}},\ \bibinfo {eid} {101}
  (\bibinfo {year} {2018})},\ \Eprint {http://arxiv.org/abs/1710.00845}
  {arXiv:1710.00845 [astro-ph.CO]} \BibitemShut {NoStop}%
\bibitem [{\citenamefont {{Riess}}\ \emph {et~al.}(2019)\citenamefont
  {{Riess}}, \citenamefont {{Casertano}}, \citenamefont {{Yuan}}, \citenamefont
  {{Macri}},\ and\ \citenamefont {{Scolnic}}}]{Riess2019ApJ}%
  \BibitemOpen
  \bibfield  {author} {\bibinfo {author} {\bibfnamefont {A.~G.}\ \bibnamefont
  {{Riess}}}, \bibinfo {author} {\bibfnamefont {S.}~\bibnamefont
  {{Casertano}}}, \bibinfo {author} {\bibfnamefont {W.}~\bibnamefont {{Yuan}}},
  \bibinfo {author} {\bibfnamefont {L.~M.}\ \bibnamefont {{Macri}}}, \ and\
  \bibinfo {author} {\bibfnamefont {D.}~\bibnamefont {{Scolnic}}},\ }\href
  {\doibase 10.3847/1538-4357/ab1422} {\bibfield  {journal} {\bibinfo
  {journal} {\apj}\ }\textbf {\bibinfo {volume} {876}},\ \bibinfo {eid} {85}
  (\bibinfo {year} {2019})},\ \Eprint {http://arxiv.org/abs/1903.07603}
  {arXiv:1903.07603 [astro-ph.CO]} \BibitemShut {NoStop}%
\bibitem [{\citenamefont {{Dainotti}}\ \emph {et~al.}(2021)\citenamefont
  {{Dainotti}}, \citenamefont {{De Simone}}, \citenamefont {{Schiavone}},
  \citenamefont {{Montani}}, \citenamefont {{Rinaldi}},\ and\ \citenamefont
  {{Lambiase}}}]{Dainotti2021ApJ}%
  \BibitemOpen
  \bibfield  {author} {\bibinfo {author} {\bibfnamefont {M.~G.}\ \bibnamefont
  {{Dainotti}}}, \bibinfo {author} {\bibfnamefont {B.}~\bibnamefont {{De
  Simone}}}, \bibinfo {author} {\bibfnamefont {T.}~\bibnamefont {{Schiavone}}},
  \bibinfo {author} {\bibfnamefont {G.}~\bibnamefont {{Montani}}}, \bibinfo
  {author} {\bibfnamefont {E.}~\bibnamefont {{Rinaldi}}}, \ and\ \bibinfo
  {author} {\bibfnamefont {G.}~\bibnamefont {{Lambiase}}},\ }\href {\doibase
  10.3847/1538-4357/abeb73} {\bibfield  {journal} {\bibinfo  {journal} {\apj}\
  }\textbf {\bibinfo {volume} {912}},\ \bibinfo {eid} {150} (\bibinfo {year}
  {2021})},\ \Eprint {http://arxiv.org/abs/2103.02117} {arXiv:2103.02117
  [astro-ph.CO]} \BibitemShut {NoStop}%
\bibitem [{\citenamefont {{Riess}}\ \emph {et~al.}(2021)\citenamefont
  {{Riess}}, \citenamefont {{Casertano}}, \citenamefont {{Yuan}}, \citenamefont
  {{Bowers}}, \citenamefont {{Macri}}, \citenamefont {{Zinn}},\ and\
  \citenamefont {{Scolnic}}}]{SHoES_2021}%
  \BibitemOpen
  \bibfield  {author} {\bibinfo {author} {\bibfnamefont {A.~G.}\ \bibnamefont
  {{Riess}}}, \bibinfo {author} {\bibfnamefont {S.}~\bibnamefont
  {{Casertano}}}, \bibinfo {author} {\bibfnamefont {W.}~\bibnamefont {{Yuan}}},
  \bibinfo {author} {\bibfnamefont {J.~B.}\ \bibnamefont {{Bowers}}}, \bibinfo
  {author} {\bibfnamefont {L.}~\bibnamefont {{Macri}}}, \bibinfo {author}
  {\bibfnamefont {J.~C.}\ \bibnamefont {{Zinn}}}, \ and\ \bibinfo {author}
  {\bibfnamefont {D.}~\bibnamefont {{Scolnic}}},\ }\href {\doibase
  10.3847/2041-8213/abdbaf} {\bibfield  {journal} {\bibinfo  {journal} {\apjl}\
  }\textbf {\bibinfo {volume} {908}},\ \bibinfo {eid} {L6} (\bibinfo {year}
  {2021})},\ \Eprint {http://arxiv.org/abs/2012.08534} {arXiv:2012.08534
  [astro-ph.CO]} \BibitemShut {NoStop}%
\bibitem [{\citenamefont {{Abbott}}\ \emph
  {et~al.}(2017{\natexlab{c}})\citenamefont {{Abbott}}, \citenamefont
  {{Abbott}}, \citenamefont {{Abbott}}, \citenamefont {{Acernese}},
  \citenamefont {{Ackley}} \emph {et~al.}}]{Abbott2017Nature}%
  \BibitemOpen
  \bibfield  {author} {\bibinfo {author} {\bibfnamefont {B.~P.}\ \bibnamefont
  {{Abbott}}}, \bibinfo {author} {\bibfnamefont {R.}~\bibnamefont {{Abbott}}},
  \bibinfo {author} {\bibfnamefont {T.~D.}\ \bibnamefont {{Abbott}}}, \bibinfo
  {author} {\bibfnamefont {F.}~\bibnamefont {{Acernese}}}, \bibinfo {author}
  {\bibfnamefont {K.}~\bibnamefont {{Ackley}}},  \emph {et~al.},\ }\href
  {\doibase 10.1038/nature24471} {\bibfield  {journal} {\bibinfo  {journal}
  {\nat}\ }\textbf {\bibinfo {volume} {551}},\ \bibinfo {pages} {425} (\bibinfo
  {year} {2017}{\natexlab{c}})},\ \Eprint {http://arxiv.org/abs/1710.05835}
  {arXiv:1710.05835 [astro-ph.CO]} \BibitemShut {NoStop}%
\bibitem [{\citenamefont {{Chen}}\ \emph {et~al.}(2018)\citenamefont {{Chen}},
  \citenamefont {{Fishbach}},\ and\ \citenamefont {{Holz}}}]{Chen2018Nature}%
  \BibitemOpen
  \bibfield  {author} {\bibinfo {author} {\bibfnamefont {H.-Y.}\ \bibnamefont
  {{Chen}}}, \bibinfo {author} {\bibfnamefont {M.}~\bibnamefont {{Fishbach}}},
  \ and\ \bibinfo {author} {\bibfnamefont {D.~E.}\ \bibnamefont {{Holz}}},\
  }\href {\doibase 10.1038/s41586-018-0606-0} {\bibfield  {journal} {\bibinfo
  {journal} {\nat}\ }\textbf {\bibinfo {volume} {562}},\ \bibinfo {pages} {545}
  (\bibinfo {year} {2018})},\ \Eprint {http://arxiv.org/abs/1712.06531}
  {arXiv:1712.06531 [astro-ph.CO]} \BibitemShut {NoStop}%
\bibitem [{\citenamefont {{Planck Collaboration}}\ \emph
  {et~al.}(2020)\citenamefont {{Planck Collaboration}}, \citenamefont
  {{Aghanim}}, \citenamefont {{Akrami}}, \citenamefont {{Ashdown}},
  \citenamefont {{Aumont}} \emph {et~al.}}]{Planck_2020}%
  \BibitemOpen
  \bibfield  {author} {\bibinfo {author} {\bibnamefont {{Planck
  Collaboration}}}, \bibinfo {author} {\bibfnamefont {N.}~\bibnamefont
  {{Aghanim}}}, \bibinfo {author} {\bibfnamefont {Y.}~\bibnamefont {{Akrami}}},
  \bibinfo {author} {\bibfnamefont {M.}~\bibnamefont {{Ashdown}}}, \bibinfo
  {author} {\bibfnamefont {J.}~\bibnamefont {{Aumont}}},  \emph {et~al.},\
  }\href {\doibase 10.1051/0004-6361/201833910} {\bibfield  {journal} {\bibinfo
   {journal} {\aap}\ }\textbf {\bibinfo {volume} {641}},\ \bibinfo {eid} {A6}
  (\bibinfo {year} {2020})},\ \Eprint {http://arxiv.org/abs/1807.06209}
  {arXiv:1807.06209 [astro-ph.CO]} \BibitemShut {NoStop}%
\bibitem [{\citenamefont {Ryan}\ \emph {et~al.}(2020)\citenamefont {Ryan},
  \citenamefont {Van~Eerten}, \citenamefont {Piro},\ and\ \citenamefont
  {Troja}}]{ryan2020gamma}%
  \BibitemOpen
  \bibfield  {author} {\bibinfo {author} {\bibfnamefont {G.}~\bibnamefont
  {Ryan}}, \bibinfo {author} {\bibfnamefont {H.}~\bibnamefont {Van~Eerten}},
  \bibinfo {author} {\bibfnamefont {L.}~\bibnamefont {Piro}}, \ and\ \bibinfo
  {author} {\bibfnamefont {E.}~\bibnamefont {Troja}},\ }\href@noop {}
  {\bibfield  {journal} {\bibinfo  {journal} {The Astrophysical Journal}\
  }\textbf {\bibinfo {volume} {896}},\ \bibinfo {pages} {166} (\bibinfo {year}
  {2020})}\BibitemShut {NoStop}%
\bibitem [{\citenamefont {{Duan}}\ \emph {et~al.}(2019)\citenamefont {{Duan}},
  \citenamefont {{Jin}}, \citenamefont {{Zhang}}, \citenamefont {{Zhu}},
  \citenamefont {{Li}}, \citenamefont {Yi-Zhong},\ and\ \citenamefont
  {{Wei}}}]{Duan2019ApJL}%
  \BibitemOpen
  \bibfield  {author} {\bibinfo {author} {\bibfnamefont {K.-K.}\ \bibnamefont
  {{Duan}}}, \bibinfo {author} {\bibfnamefont {Z.-P.}\ \bibnamefont {{Jin}}},
  \bibinfo {author} {\bibfnamefont {F.-W.}\ \bibnamefont {{Zhang}}}, \bibinfo
  {author} {\bibfnamefont {Y.-M.}\ \bibnamefont {{Zhu}}}, \bibinfo {author}
  {\bibfnamefont {X.}~\bibnamefont {{Li}}}, \bibinfo {author} {\bibnamefont
  {Yi-Zhong}}, \ and\ \bibinfo {author} {\bibfnamefont {D.-M.}\ \bibnamefont
  {{Wei}}},\ }\href {\doibase 10.3847/2041-8213/ab1c64} {\bibfield  {journal}
  {\bibinfo  {journal} {\apjl}\ }\textbf {\bibinfo {volume} {876}},\ \bibinfo
  {eid} {L28} (\bibinfo {year} {2019})},\ \Eprint
  {http://arxiv.org/abs/1901.01521} {arXiv:1901.01521 [astro-ph.HE]}
  \BibitemShut {NoStop}%
\bibitem [{\citenamefont {Finn}\ and\ \citenamefont
  {Chernoff}(1993)}]{Finn_PRD_1993}%
  \BibitemOpen
  \bibfield  {author} {\bibinfo {author} {\bibfnamefont {L.~S.}\ \bibnamefont
  {Finn}}\ and\ \bibinfo {author} {\bibfnamefont {D.~F.}\ \bibnamefont
  {Chernoff}},\ }\href {\doibase 10.1103/PhysRevD.47.2198} {\bibfield
  {journal} {\bibinfo  {journal} {Phys. Rev. D}\ }\textbf {\bibinfo {volume}
  {47}},\ \bibinfo {pages} {2198} (\bibinfo {year} {1993})}\BibitemShut
  {NoStop}%
\bibitem [{\citenamefont {Cutler}\ and\ \citenamefont
  {Flanagan}(1994)}]{Cutler_PRD_1994}%
  \BibitemOpen
  \bibfield  {author} {\bibinfo {author} {\bibfnamefont {C.}~\bibnamefont
  {Cutler}}\ and\ \bibinfo {author} {\bibfnamefont {E.~E.}\ \bibnamefont
  {Flanagan}},\ }\href {\doibase 10.1103/PhysRevD.49.2658} {\bibfield
  {journal} {\bibinfo  {journal} {Phys. Rev. D}\ }\textbf {\bibinfo {volume}
  {49}},\ \bibinfo {pages} {2658} (\bibinfo {year} {1994})}\BibitemShut
  {NoStop}%
\bibitem [{\citenamefont {Chassande-Mottin}\ \emph {et~al.}(2019)\citenamefont
  {Chassande-Mottin}, \citenamefont {Leyde}, \citenamefont {Mastrogiovanni},\
  and\ \citenamefont {Steer}}]{sigma_epsilon_PRD_2019}%
  \BibitemOpen
  \bibfield  {author} {\bibinfo {author} {\bibfnamefont {E.}~\bibnamefont
  {Chassande-Mottin}}, \bibinfo {author} {\bibfnamefont {K.}~\bibnamefont
  {Leyde}}, \bibinfo {author} {\bibfnamefont {S.}~\bibnamefont
  {Mastrogiovanni}}, \ and\ \bibinfo {author} {\bibfnamefont {D.~A.}\
  \bibnamefont {Steer}},\ }\href {\doibase 10.1103/PhysRevD.100.083514}
  {\bibfield  {journal} {\bibinfo  {journal} {Phys. Rev. D}\ }\textbf {\bibinfo
  {volume} {100}},\ \bibinfo {pages} {083514} (\bibinfo {year}
  {2019})}\BibitemShut {NoStop}%
\bibitem [{\citenamefont {Allen}\ \emph {et~al.}(2012)\citenamefont {Allen},
  \citenamefont {Anderson}, \citenamefont {Brady}, \citenamefont {Brown},\ and\
  \citenamefont {Creighton}}]{Allen:2005fk}%
  \BibitemOpen
  \bibfield  {author} {\bibinfo {author} {\bibfnamefont {B.}~\bibnamefont
  {Allen}}, \bibinfo {author} {\bibfnamefont {W.~G.}\ \bibnamefont {Anderson}},
  \bibinfo {author} {\bibfnamefont {P.~R.}\ \bibnamefont {Brady}}, \bibinfo
  {author} {\bibfnamefont {D.~A.}\ \bibnamefont {Brown}}, \ and\ \bibinfo
  {author} {\bibfnamefont {J.~D.~E.}\ \bibnamefont {Creighton}},\ }\href
  {\doibase 10.1103/PhysRevD.85.122006} {\bibfield  {journal} {\bibinfo
  {journal} {Phys. Rev. D}\ }\textbf {\bibinfo {volume} {85}},\ \bibinfo
  {pages} {122006} (\bibinfo {year} {2012})},\ \Eprint
  {http://arxiv.org/abs/gr-qc/0509116} {arXiv:gr-qc/0509116} \BibitemShut
  {NoStop}%
\bibitem [{\citenamefont {Veitch}\ \emph {et~al.}(2015)\citenamefont {Veitch}
  \emph {et~al.}}]{Veitch:2014wba}%
  \BibitemOpen
  \bibfield  {author} {\bibinfo {author} {\bibfnamefont {J.}~\bibnamefont
  {Veitch}} \emph {et~al.},\ }\href {\doibase 10.1103/PhysRevD.91.042003}
  {\bibfield  {journal} {\bibinfo  {journal} {Phys. Rev. D}\ }\textbf {\bibinfo
  {volume} {91}},\ \bibinfo {pages} {042003} (\bibinfo {year} {2015})},\
  \Eprint {http://arxiv.org/abs/1409.7215} {arXiv:1409.7215 [gr-qc]}
  \BibitemShut {NoStop}%
\bibitem [{\citenamefont {Biwer}\ \emph {et~al.}(2019)\citenamefont {Biwer},
  \citenamefont {Capano}, \citenamefont {De}, \citenamefont {Cabero},
  \citenamefont {Brown}, \citenamefont {Nitz},\ and\ \citenamefont
  {Raymond}}]{Biwer:2018osg}%
  \BibitemOpen
  \bibfield  {author} {\bibinfo {author} {\bibfnamefont {C.~M.}\ \bibnamefont
  {Biwer}}, \bibinfo {author} {\bibfnamefont {C.~D.}\ \bibnamefont {Capano}},
  \bibinfo {author} {\bibfnamefont {S.}~\bibnamefont {De}}, \bibinfo {author}
  {\bibfnamefont {M.}~\bibnamefont {Cabero}}, \bibinfo {author} {\bibfnamefont
  {D.~A.}\ \bibnamefont {Brown}}, \bibinfo {author} {\bibfnamefont {A.~H.}\
  \bibnamefont {Nitz}}, \ and\ \bibinfo {author} {\bibfnamefont
  {V.}~\bibnamefont {Raymond}},\ }\href {\doibase 10.1088/1538-3873/aaef0b}
  {\bibfield  {journal} {\bibinfo  {journal} {Publ. Astron. Soc. Pac.}\
  }\textbf {\bibinfo {volume} {131}},\ \bibinfo {pages} {024503} (\bibinfo
  {year} {2019})},\ \Eprint {http://arxiv.org/abs/1807.10312} {arXiv:1807.10312
  [astro-ph.IM]} \BibitemShut {NoStop}%
\bibitem [{\citenamefont {Ashton}\ \emph {et~al.}(2019)\citenamefont {Ashton}
  \emph {et~al.}}]{Ashton:2018jfp}%
  \BibitemOpen
  \bibfield  {author} {\bibinfo {author} {\bibfnamefont {G.}~\bibnamefont
  {Ashton}} \emph {et~al.},\ }\href {\doibase 10.3847/1538-4365/ab06fc}
  {\bibfield  {journal} {\bibinfo  {journal} {Astrophys. J. Suppl.}\ }\textbf
  {\bibinfo {volume} {241}},\ \bibinfo {pages} {27} (\bibinfo {year} {2019})},\
  \Eprint {http://arxiv.org/abs/1811.02042} {arXiv:1811.02042 [astro-ph.IM]}
  \BibitemShut {NoStop}%
\bibitem [{\citenamefont {{Zackay}}\ \emph {et~al.}(2018)\citenamefont
  {{Zackay}}, \citenamefont {{Dai}},\ and\ \citenamefont
  {{Venumadhav}}}]{fll_Zackay_2018}%
  \BibitemOpen
  \bibfield  {author} {\bibinfo {author} {\bibfnamefont {B.}~\bibnamefont
  {{Zackay}}}, \bibinfo {author} {\bibfnamefont {L.}~\bibnamefont {{Dai}}}, \
  and\ \bibinfo {author} {\bibfnamefont {T.}~\bibnamefont {{Venumadhav}}},\
  }\href@noop {} {\bibfield  {journal} {\bibinfo  {journal} {arXiv e-prints}\
  ,\ \bibinfo {eid} {arXiv:1806.08792}} (\bibinfo {year} {2018})},\ \Eprint
  {http://arxiv.org/abs/1806.08792} {arXiv:1806.08792 [astro-ph.IM]}
  \BibitemShut {NoStop}%
\bibitem [{\citenamefont {{Abbott}}\ \emph {et~al.}(2018)\citenamefont
  {{Abbott}}, \citenamefont {{Abbott}}, \citenamefont {{Abbott}}, \citenamefont
  {{Abernathy}}, \citenamefont {{Acernese}}, \citenamefont {others},\ and\
  \citenamefont {{VIRGO Collaboration}}}]{Abbott2018LRR}%
  \BibitemOpen
  \bibfield  {author} {\bibinfo {author} {\bibfnamefont {B.~P.}\ \bibnamefont
  {{Abbott}}}, \bibinfo {author} {\bibfnamefont {R.}~\bibnamefont {{Abbott}}},
  \bibinfo {author} {\bibfnamefont {T.~D.}\ \bibnamefont {{Abbott}}}, \bibinfo
  {author} {\bibfnamefont {M.~R.}\ \bibnamefont {{Abernathy}}}, \bibinfo
  {author} {\bibfnamefont {F.}~\bibnamefont {{Acernese}}}, \bibinfo {author}
  {\bibfnamefont {L.~S.~C.}\ \bibnamefont {others}}, \ and\ \bibinfo {author}
  {\bibnamefont {{VIRGO Collaboration}}},\ }\href {\doibase
  10.1007/s41114-018-0012-9} {\bibfield  {journal} {\bibinfo  {journal} {Living
  Reviews in Relativity}\ }\textbf {\bibinfo {volume} {21}},\ \bibinfo {eid}
  {3} (\bibinfo {year} {2018})},\ \Eprint {http://arxiv.org/abs/1304.0670}
  {arXiv:1304.0670 [gr-qc]} \BibitemShut {NoStop}%
\bibitem [{\citenamefont {{Abbott}}\ \emph {et~al.}(2016)\citenamefont
  {{Abbott}}, \citenamefont {{Abbott}}, \citenamefont {{Abbott}}, \citenamefont
  {{Abernathy}}, \citenamefont {{Acernese}} \emph {et~al.}}]{Abbott2016LRR}%
  \BibitemOpen
  \bibfield  {author} {\bibinfo {author} {\bibfnamefont {B.~P.}\ \bibnamefont
  {{Abbott}}}, \bibinfo {author} {\bibfnamefont {R.}~\bibnamefont {{Abbott}}},
  \bibinfo {author} {\bibfnamefont {T.~D.}\ \bibnamefont {{Abbott}}}, \bibinfo
  {author} {\bibfnamefont {M.~R.}\ \bibnamefont {{Abernathy}}}, \bibinfo
  {author} {\bibfnamefont {F.}~\bibnamefont {{Acernese}}},  \emph {et~al.},\
  }\href {\doibase 10.1007/lrr-2016-1} {\bibfield  {journal} {\bibinfo
  {journal} {Living Reviews in Relativity}\ }\textbf {\bibinfo {volume} {19}},\
  \bibinfo {eid} {1} (\bibinfo {year} {2016})}\BibitemShut {NoStop}%
\bibitem [{\citenamefont {{Martynov}}\ \emph {et~al.}(2016)\citenamefont
  {{Martynov}}, \citenamefont {{Hall}}, \citenamefont {{Abbott}}, \citenamefont
  {{Abbott}}, \citenamefont {{Abbott}}, \citenamefont {{Adams}} \emph
  {et~al.}}]{Martynov2016PRD}%
  \BibitemOpen
  \bibfield  {author} {\bibinfo {author} {\bibfnamefont {D.~V.}\ \bibnamefont
  {{Martynov}}}, \bibinfo {author} {\bibfnamefont {E.~D.}\ \bibnamefont
  {{Hall}}}, \bibinfo {author} {\bibfnamefont {B.~P.}\ \bibnamefont
  {{Abbott}}}, \bibinfo {author} {\bibfnamefont {R.}~\bibnamefont {{Abbott}}},
  \bibinfo {author} {\bibfnamefont {T.~D.}\ \bibnamefont {{Abbott}}}, \bibinfo
  {author} {\bibfnamefont {C.}~\bibnamefont {{Adams}}},  \emph {et~al.},\
  }\href {\doibase 10.1103/PhysRevD.93.112004} {\bibfield  {journal} {\bibinfo
  {journal} {\prd}\ }\textbf {\bibinfo {volume} {93}},\ \bibinfo {eid} {112004}
  (\bibinfo {year} {2016})},\ \Eprint {http://arxiv.org/abs/1604.00439}
  {arXiv:1604.00439 [astro-ph.IM]} \BibitemShut {NoStop}%
\bibitem [{\citenamefont {{Acernese}}\ \emph {et~al.}(2015)\citenamefont
  {{Acernese}}, \citenamefont {{Agathos}}, \citenamefont {{Agatsuma}},
  \citenamefont {{Aisa}}, \citenamefont {{Allemandou}} \emph
  {et~al.}}]{Acernese2015}%
  \BibitemOpen
  \bibfield  {author} {\bibinfo {author} {\bibfnamefont {F.}~\bibnamefont
  {{Acernese}}}, \bibinfo {author} {\bibfnamefont {M.}~\bibnamefont
  {{Agathos}}}, \bibinfo {author} {\bibfnamefont {K.}~\bibnamefont
  {{Agatsuma}}}, \bibinfo {author} {\bibfnamefont {D.}~\bibnamefont {{Aisa}}},
  \bibinfo {author} {\bibfnamefont {N.}~\bibnamefont {{Allemandou}}},  \emph
  {et~al.},\ }\href {\doibase 10.1088/0264-9381/32/2/024001} {\bibfield
  {journal} {\bibinfo  {journal} {Classical and Quantum Gravity}\ }\textbf
  {\bibinfo {volume} {32}},\ \bibinfo {eid} {024001} (\bibinfo {year}
  {2015})},\ \Eprint {http://arxiv.org/abs/1408.3978} {arXiv:1408.3978 [gr-qc]}
  \BibitemShut {NoStop}%
\bibitem [{\citenamefont {{Kagra Collaboration}}\ \emph
  {et~al.}(2019)\citenamefont {{Kagra Collaboration}}, \citenamefont
  {{Akutsu}}, \citenamefont {{Ando}}, \citenamefont {{Arai}}, \citenamefont
  {{Arai}} \emph {et~al.}}]{Kagra2019NatAs}%
  \BibitemOpen
  \bibfield  {author} {\bibinfo {author} {\bibnamefont {{Kagra
  Collaboration}}}, \bibinfo {author} {\bibfnamefont {T.}~\bibnamefont
  {{Akutsu}}}, \bibinfo {author} {\bibfnamefont {M.}~\bibnamefont {{Ando}}},
  \bibinfo {author} {\bibfnamefont {K.}~\bibnamefont {{Arai}}}, \bibinfo
  {author} {\bibfnamefont {Y.}~\bibnamefont {{Arai}}},  \emph {et~al.},\ }\href
  {\doibase 10.1038/s41550-018-0658-y} {\bibfield  {journal} {\bibinfo
  {journal} {Nature Astronomy}\ }\textbf {\bibinfo {volume} {3}},\ \bibinfo
  {pages} {35} (\bibinfo {year} {2019})},\ \Eprint
  {http://arxiv.org/abs/1811.08079} {arXiv:1811.08079 [gr-qc]} \BibitemShut
  {NoStop}%
\bibitem [{\citenamefont {{Saleem}}\ \emph {et~al.}(2021)\citenamefont
  {{Saleem}}, \citenamefont {{Rana}}, \citenamefont {{Gayathri}}, \citenamefont
  {{Vijaykumar}}, \citenamefont {{Goyal}}, \citenamefont {{Sachdev}},
  \citenamefont {{Suresh}}, \citenamefont {{Sudhagar}}, \citenamefont
  {{Mukherjee}}, \citenamefont {{Gaur}}, \citenamefont {{Sathyaprakash}},
  \citenamefont {{Pai}}, \citenamefont {{Adhikari}}, \citenamefont {{Ajith}},\
  and\ \citenamefont {{Bose}}}]{Saleem2021}%
  \BibitemOpen
  \bibfield  {author} {\bibinfo {author} {\bibfnamefont {M.}~\bibnamefont
  {{Saleem}}}, \bibinfo {author} {\bibfnamefont {J.}~\bibnamefont {{Rana}}},
  \bibinfo {author} {\bibfnamefont {V.}~\bibnamefont {{Gayathri}}}, \bibinfo
  {author} {\bibfnamefont {A.}~\bibnamefont {{Vijaykumar}}}, \bibinfo {author}
  {\bibfnamefont {S.}~\bibnamefont {{Goyal}}}, \bibinfo {author} {\bibfnamefont
  {S.}~\bibnamefont {{Sachdev}}}, \bibinfo {author} {\bibfnamefont
  {J.}~\bibnamefont {{Suresh}}}, \bibinfo {author} {\bibfnamefont
  {S.}~\bibnamefont {{Sudhagar}}}, \bibinfo {author} {\bibfnamefont
  {A.}~\bibnamefont {{Mukherjee}}}, \bibinfo {author} {\bibfnamefont
  {G.}~\bibnamefont {{Gaur}}}, \bibinfo {author} {\bibfnamefont
  {B.}~\bibnamefont {{Sathyaprakash}}}, \bibinfo {author} {\bibfnamefont
  {A.}~\bibnamefont {{Pai}}}, \bibinfo {author} {\bibfnamefont {R.~X.}\
  \bibnamefont {{Adhikari}}}, \bibinfo {author} {\bibfnamefont
  {P.}~\bibnamefont {{Ajith}}}, \ and\ \bibinfo {author} {\bibfnamefont
  {S.}~\bibnamefont {{Bose}}},\ }\href@noop {} {\bibfield  {journal} {\bibinfo
  {journal} {arXiv e-prints}\ ,\ \bibinfo {eid} {arXiv:2105.01716}} (\bibinfo
  {year} {2021})},\ \Eprint {http://arxiv.org/abs/2105.01716} {arXiv:2105.01716
  [gr-qc]} \BibitemShut {NoStop}%
\bibitem [{\citenamefont {{Ma}}\ and\ \citenamefont
  {{Pan}}(2014)}]{Ma_Pan2014MNRAS}%
  \BibitemOpen
  \bibfield  {author} {\bibinfo {author} {\bibfnamefont {Y.-Z.}\ \bibnamefont
  {{Ma}}}\ and\ \bibinfo {author} {\bibfnamefont {J.}~\bibnamefont {{Pan}}},\
  }\href {\doibase 10.1093/mnras/stt2038} {\bibfield  {journal} {\bibinfo
  {journal} {\mnras}\ }\textbf {\bibinfo {volume} {437}},\ \bibinfo {pages}
  {1996} (\bibinfo {year} {2014})},\ \Eprint {http://arxiv.org/abs/1311.6888}
  {arXiv:1311.6888 [astro-ph.CO]} \BibitemShut {NoStop}%
\bibitem [{\citenamefont {{Graziani}}\ \emph {et~al.}(2019)\citenamefont
  {{Graziani}}, \citenamefont {{Courtois}}, \citenamefont {{Lavaux}},
  \citenamefont {{Hoffman}}, \citenamefont {{Tully}}, \citenamefont {{Copin}},\
  and\ \citenamefont {{Pomar{\`e}de}}}]{Graziani2019MNRAS}%
  \BibitemOpen
  \bibfield  {author} {\bibinfo {author} {\bibfnamefont {R.}~\bibnamefont
  {{Graziani}}}, \bibinfo {author} {\bibfnamefont {H.~M.}\ \bibnamefont
  {{Courtois}}}, \bibinfo {author} {\bibfnamefont {G.}~\bibnamefont
  {{Lavaux}}}, \bibinfo {author} {\bibfnamefont {Y.}~\bibnamefont {{Hoffman}}},
  \bibinfo {author} {\bibfnamefont {R.~B.}\ \bibnamefont {{Tully}}}, \bibinfo
  {author} {\bibfnamefont {Y.}~\bibnamefont {{Copin}}}, \ and\ \bibinfo
  {author} {\bibfnamefont {D.}~\bibnamefont {{Pomar{\`e}de}}},\ }\href
  {\doibase 10.1093/mnras/stz078} {\bibfield  {journal} {\bibinfo  {journal}
  {\mnras}\ }\textbf {\bibinfo {volume} {488}},\ \bibinfo {pages} {5438}
  (\bibinfo {year} {2019})},\ \Eprint {http://arxiv.org/abs/1901.01818}
  {arXiv:1901.01818 [astro-ph.CO]} \BibitemShut {NoStop}%
\bibitem [{\citenamefont {Buonanno}\ \emph {et~al.}(2009)\citenamefont
  {Buonanno}, \citenamefont {Iyer}, \citenamefont {Ochsner}, \citenamefont
  {Pan},\ and\ \citenamefont {Sathyaprakash}}]{TaylorF2_PRD_2009}%
  \BibitemOpen
  \bibfield  {author} {\bibinfo {author} {\bibfnamefont {A.}~\bibnamefont
  {Buonanno}}, \bibinfo {author} {\bibfnamefont {B.~R.}\ \bibnamefont {Iyer}},
  \bibinfo {author} {\bibfnamefont {E.}~\bibnamefont {Ochsner}}, \bibinfo
  {author} {\bibfnamefont {Y.}~\bibnamefont {Pan}}, \ and\ \bibinfo {author}
  {\bibfnamefont {B.~S.}\ \bibnamefont {Sathyaprakash}},\ }\href {\doibase
  10.1103/PhysRevD.80.084043} {\bibfield  {journal} {\bibinfo  {journal} {Phys.
  Rev. D}\ }\textbf {\bibinfo {volume} {80}},\ \bibinfo {pages} {084043}
  (\bibinfo {year} {2009})}\BibitemShut {NoStop}%
\bibitem [{\citenamefont {{Ashton}}\ \emph {et~al.}(2019)\citenamefont
  {{Ashton}}, \citenamefont {{H{\"u}bner}}, \citenamefont {{Lasky}},
  \citenamefont {{Talbot}}, \citenamefont {{Ackley}} \emph
  {et~al.}}]{bilby2019ApJS}%
  \BibitemOpen
  \bibfield  {author} {\bibinfo {author} {\bibfnamefont {G.}~\bibnamefont
  {{Ashton}}}, \bibinfo {author} {\bibfnamefont {M.}~\bibnamefont
  {{H{\"u}bner}}}, \bibinfo {author} {\bibfnamefont {P.~D.}\ \bibnamefont
  {{Lasky}}}, \bibinfo {author} {\bibfnamefont {C.}~\bibnamefont {{Talbot}}},
  \bibinfo {author} {\bibfnamefont {K.}~\bibnamefont {{Ackley}}},  \emph
  {et~al.},\ }\href {\doibase 10.3847/1538-4365/ab06fc} {\bibfield  {journal}
  {\bibinfo  {journal} {\apjs}\ }\textbf {\bibinfo {volume} {241}},\ \bibinfo
  {eid} {27} (\bibinfo {year} {2019})},\ \Eprint
  {http://arxiv.org/abs/1811.02042} {arXiv:1811.02042 [astro-ph.IM]}
  \BibitemShut {NoStop}%
\bibitem [{\citenamefont {{Romero-Shaw}}\ \emph {et~al.}(2020)\citenamefont
  {{Romero-Shaw}}, \citenamefont {{Talbot}}, \citenamefont {{Biscoveanu}},
  \citenamefont {{D'Emilio}}, \citenamefont {{Ashton}}, \citenamefont {{Berry}}
  \emph {et~al.}}]{bilby2020MNRAS}%
  \BibitemOpen
  \bibfield  {author} {\bibinfo {author} {\bibfnamefont {I.~M.}\ \bibnamefont
  {{Romero-Shaw}}}, \bibinfo {author} {\bibfnamefont {C.}~\bibnamefont
  {{Talbot}}}, \bibinfo {author} {\bibfnamefont {S.}~\bibnamefont
  {{Biscoveanu}}}, \bibinfo {author} {\bibfnamefont {V.}~\bibnamefont
  {{D'Emilio}}}, \bibinfo {author} {\bibfnamefont {G.}~\bibnamefont
  {{Ashton}}}, \bibinfo {author} {\bibfnamefont {C.~P.~L.}\ \bibnamefont
  {{Berry}}},  \emph {et~al.},\ }\href {\doibase 10.1093/mnras/staa2850}
  {\bibfield  {journal} {\bibinfo  {journal} {\mnras}\ }\textbf {\bibinfo
  {volume} {499}},\ \bibinfo {pages} {3295} (\bibinfo {year} {2020})},\ \Eprint
  {http://arxiv.org/abs/2006.00714} {arXiv:2006.00714 [astro-ph.IM]}
  \BibitemShut {NoStop}%
\bibitem [{\citenamefont {{Biwer}}\ \emph {et~al.}(2019)\citenamefont
  {{Biwer}}, \citenamefont {{Capano}}, \citenamefont {{De}}, \citenamefont
  {{Cabero}}, \citenamefont {{Brown}}, \citenamefont {{Nitz}},\ and\
  \citenamefont {{Raymond}}}]{pycbc2019PASP}%
  \BibitemOpen
  \bibfield  {author} {\bibinfo {author} {\bibfnamefont {C.~M.}\ \bibnamefont
  {{Biwer}}}, \bibinfo {author} {\bibfnamefont {C.~D.}\ \bibnamefont
  {{Capano}}}, \bibinfo {author} {\bibfnamefont {S.}~\bibnamefont {{De}}},
  \bibinfo {author} {\bibfnamefont {M.}~\bibnamefont {{Cabero}}}, \bibinfo
  {author} {\bibfnamefont {D.~A.}\ \bibnamefont {{Brown}}}, \bibinfo {author}
  {\bibfnamefont {A.~H.}\ \bibnamefont {{Nitz}}}, \ and\ \bibinfo {author}
  {\bibfnamefont {V.}~\bibnamefont {{Raymond}}},\ }\href {\doibase
  10.1088/1538-3873/aaef0b} {\bibfield  {journal} {\bibinfo  {journal} {\pasp}\
  }\textbf {\bibinfo {volume} {131}},\ \bibinfo {pages} {024503} (\bibinfo
  {year} {2019})},\ \Eprint {http://arxiv.org/abs/1807.10312} {arXiv:1807.10312
  [astro-ph.IM]} \BibitemShut {NoStop}%
\bibitem [{\citenamefont {{Speagle}}(2020)}]{dynesty2020MNRAS}%
  \BibitemOpen
  \bibfield  {author} {\bibinfo {author} {\bibfnamefont {J.~S.}\ \bibnamefont
  {{Speagle}}},\ }\href {\doibase 10.1093/mnras/staa278} {\bibfield  {journal}
  {\bibinfo  {journal} {\mnras}\ }\textbf {\bibinfo {volume} {493}},\ \bibinfo
  {pages} {3132} (\bibinfo {year} {2020})},\ \Eprint
  {http://arxiv.org/abs/1904.02180} {arXiv:1904.02180 [astro-ph.IM]}
  \BibitemShut {NoStop}%
\bibitem [{\citenamefont {Seto}(2007)}]{Seto_iota_psi_PRD_2007}%
  \BibitemOpen
  \bibfield  {author} {\bibinfo {author} {\bibfnamefont {N.}~\bibnamefont
  {Seto}},\ }\href {\doibase 10.1103/PhysRevD.75.024016} {\bibfield  {journal}
  {\bibinfo  {journal} {Phys. Rev. D}\ }\textbf {\bibinfo {volume} {75}},\
  \bibinfo {pages} {024016} (\bibinfo {year} {2007})}\BibitemShut {NoStop}%
\bibitem [{\citenamefont {Dietrich}\ \emph {et~al.}(2017)\citenamefont
  {Dietrich}, \citenamefont {Bernuzzi},\ and\ \citenamefont
  {Tichy}}]{IMRPhenomDNRTidal-2_PRD_2017}%
  \BibitemOpen
  \bibfield  {author} {\bibinfo {author} {\bibfnamefont {T.}~\bibnamefont
  {Dietrich}}, \bibinfo {author} {\bibfnamefont {S.}~\bibnamefont {Bernuzzi}},
  \ and\ \bibinfo {author} {\bibfnamefont {W.}~\bibnamefont {Tichy}},\ }\href
  {\doibase 10.1103/PhysRevD.96.121501} {\bibfield  {journal} {\bibinfo
  {journal} {Phys. Rev. D}\ }\textbf {\bibinfo {volume} {96}},\ \bibinfo
  {pages} {121501} (\bibinfo {year} {2017})}\BibitemShut {NoStop}%
\bibitem [{\citenamefont {Dietrich}\ \emph {et~al.}(2019)\citenamefont
  {Dietrich}, \citenamefont {Khan}, \citenamefont {Dudi}, \citenamefont
  {Kapadia}, \citenamefont {Kumar},  \emph
  {et~al.}}]{IMRPhenomDNRTidal-1_PRD_2019}%
  \BibitemOpen
  \bibfield  {author} {\bibinfo {author} {\bibfnamefont {T.}~\bibnamefont
  {Dietrich}}, \bibinfo {author} {\bibfnamefont {S.}~\bibnamefont {Khan}},
  \bibinfo {author} {\bibfnamefont {R.}~\bibnamefont {Dudi}}, \bibinfo {author}
  {\bibfnamefont {S.~J.}\ \bibnamefont {Kapadia}}, \bibinfo {author}
  {\bibfnamefont {P.}~\bibnamefont {Kumar}}, ,  \emph {et~al.},\ }\href
  {\doibase 10.1103/PhysRevD.99.024029} {\bibfield  {journal} {\bibinfo
  {journal} {Phys. Rev. D}\ }\textbf {\bibinfo {volume} {99}},\ \bibinfo
  {pages} {024029} (\bibinfo {year} {2019})}\BibitemShut {NoStop}%
\bibitem [{\citenamefont {Finstad}\ and\ \citenamefont
  {Brown}(2020)}]{fll_Finstad_2020}%
  \BibitemOpen
  \bibfield  {author} {\bibinfo {author} {\bibfnamefont {D.}~\bibnamefont
  {Finstad}}\ and\ \bibinfo {author} {\bibfnamefont {D.~A.}\ \bibnamefont
  {Brown}},\ }\href {\doibase 10.3847/2041-8213/abca9e} {\bibfield  {journal}
  {\bibinfo  {journal} {The Astrophysical Journal}\ }\textbf {\bibinfo {volume}
  {905}},\ \bibinfo {pages} {L9} (\bibinfo {year} {2020})}\BibitemShut
  {NoStop}%
\bibitem [{\citenamefont {{Chen}}(2020)}]{Chen2020PRL}%
  \BibitemOpen
  \bibfield  {author} {\bibinfo {author} {\bibfnamefont {H.-Y.}\ \bibnamefont
  {{Chen}}},\ }\href {\doibase 10.1103/PhysRevLett.125.201301} {\bibfield
  {journal} {\bibinfo  {journal} {\prl}\ }\textbf {\bibinfo {volume} {125}},\
  \bibinfo {eid} {201301} (\bibinfo {year} {2020})},\ \Eprint
  {http://arxiv.org/abs/2006.02779} {arXiv:2006.02779 [astro-ph.HE]}
  \BibitemShut {NoStop}%
\bibitem [{\citenamefont {{Jin}}\ \emph {et~al.}(2018)\citenamefont {{Jin}},
  \citenamefont {{Li}}, \citenamefont {{Wang}}, \citenamefont {{Wang}},
  \citenamefont {{He}}, \citenamefont {{Yuan}}, \citenamefont {{Zhang}},
  \citenamefont {{Zou}}, \citenamefont {{Fan}},\ and\ \citenamefont
  {{Wei}}}]{Jin2018ApJ}%
  \BibitemOpen
  \bibfield  {author} {\bibinfo {author} {\bibfnamefont {Z.-P.}\ \bibnamefont
  {{Jin}}}, \bibinfo {author} {\bibfnamefont {X.}~\bibnamefont {{Li}}},
  \bibinfo {author} {\bibfnamefont {H.}~\bibnamefont {{Wang}}}, \bibinfo
  {author} {\bibfnamefont {Y.-Z.}\ \bibnamefont {{Wang}}}, \bibinfo {author}
  {\bibfnamefont {H.-N.}\ \bibnamefont {{He}}}, \bibinfo {author}
  {\bibfnamefont {Q.}~\bibnamefont {{Yuan}}}, \bibinfo {author} {\bibfnamefont
  {F.-W.}\ \bibnamefont {{Zhang}}}, \bibinfo {author} {\bibfnamefont {Y.-C.}\
  \bibnamefont {{Zou}}}, \bibinfo {author} {\bibfnamefont {Y.-Z.}\ \bibnamefont
  {{Fan}}}, \ and\ \bibinfo {author} {\bibfnamefont {D.-M.}\ \bibnamefont
  {{Wei}}},\ }\href {\doibase 10.3847/1538-4357/aab76d} {\bibfield  {journal}
  {\bibinfo  {journal} {\apj}\ }\textbf {\bibinfo {volume} {857}},\ \bibinfo
  {eid} {128} (\bibinfo {year} {2018})},\ \Eprint
  {http://arxiv.org/abs/1708.07008} {arXiv:1708.07008 [astro-ph.HE]}
  \BibitemShut {NoStop}%
\bibitem [{\citenamefont {{Mastrogiovanni}}\ \emph {et~al.}(2021)\citenamefont
  {{Mastrogiovanni}}, \citenamefont {{Duque}}, \citenamefont
  {{Chassande-Mottin}}, \citenamefont {{Daigne}},\ and\ \citenamefont
  {{Mochkovitch}}}]{Mastrogiovanni2021AA}%
  \BibitemOpen
  \bibfield  {author} {\bibinfo {author} {\bibfnamefont {S.}~\bibnamefont
  {{Mastrogiovanni}}}, \bibinfo {author} {\bibfnamefont {R.}~\bibnamefont
  {{Duque}}}, \bibinfo {author} {\bibfnamefont {E.}~\bibnamefont
  {{Chassande-Mottin}}}, \bibinfo {author} {\bibfnamefont {F.}~\bibnamefont
  {{Daigne}}}, \ and\ \bibinfo {author} {\bibfnamefont {R.}~\bibnamefont
  {{Mochkovitch}}},\ }\href {\doibase 10.1051/0004-6361/202040229} {\bibfield
  {journal} {\bibinfo  {journal} {\aap}\ }\textbf {\bibinfo {volume} {652}},\
  \bibinfo {eid} {A1} (\bibinfo {year} {2021})},\ \Eprint
  {http://arxiv.org/abs/2012.12836} {arXiv:2012.12836 [astro-ph.HE]}
  \BibitemShut {NoStop}%
\bibitem [{\citenamefont {{Ryan}}\ \emph {et~al.}(2015)\citenamefont {{Ryan}},
  \citenamefont {{van Eerten}}, \citenamefont {{MacFadyen}},\ and\
  \citenamefont {{Zhang}}}]{Ryan2015ApJ}%
  \BibitemOpen
  \bibfield  {author} {\bibinfo {author} {\bibfnamefont {G.}~\bibnamefont
  {{Ryan}}}, \bibinfo {author} {\bibfnamefont {H.}~\bibnamefont {{van
  Eerten}}}, \bibinfo {author} {\bibfnamefont {A.}~\bibnamefont {{MacFadyen}}},
  \ and\ \bibinfo {author} {\bibfnamefont {B.-B.}\ \bibnamefont {{Zhang}}},\
  }\href {\doibase 10.1088/0004-637X/799/1/3} {\bibfield  {journal} {\bibinfo
  {journal} {\apj}\ }\textbf {\bibinfo {volume} {799}},\ \bibinfo {eid} {3}
  (\bibinfo {year} {2015})},\ \Eprint {http://arxiv.org/abs/1405.5516}
  {arXiv:1405.5516 [astro-ph.HE]} \BibitemShut {NoStop}%
\bibitem [{\citenamefont {{Patricelli}}\ \emph {et~al.}(2022)\citenamefont
  {{Patricelli}}, \citenamefont {{Bernardini}}, \citenamefont {{Mapelli}},
  \citenamefont {{D'Avanzo}}, \citenamefont {{Santoliquido}}, \citenamefont
  {{Cella}}, \citenamefont {{Razzano}},\ and\ \citenamefont
  {{Cuoco}}}]{2022PatricelliMNRAS}%
  \BibitemOpen
  \bibfield  {author} {\bibinfo {author} {\bibfnamefont {B.}~\bibnamefont
  {{Patricelli}}}, \bibinfo {author} {\bibfnamefont {M.~G.}\ \bibnamefont
  {{Bernardini}}}, \bibinfo {author} {\bibfnamefont {M.}~\bibnamefont
  {{Mapelli}}}, \bibinfo {author} {\bibfnamefont {P.}~\bibnamefont
  {{D'Avanzo}}}, \bibinfo {author} {\bibfnamefont {F.}~\bibnamefont
  {{Santoliquido}}}, \bibinfo {author} {\bibfnamefont {G.}~\bibnamefont
  {{Cella}}}, \bibinfo {author} {\bibfnamefont {M.}~\bibnamefont {{Razzano}}},
  \ and\ \bibinfo {author} {\bibfnamefont {E.}~\bibnamefont {{Cuoco}}},\ }\href
  {\doibase 10.1093/mnras/stac1167} {\bibfield  {journal} {\bibinfo  {journal}
  {\mnras}\ } (\bibinfo {year} {2022}),\ 10.1093/mnras/stac1167},\ \Eprint
  {http://arxiv.org/abs/2204.12504} {arXiv:2204.12504 [astro-ph.HE]}
  \BibitemShut {NoStop}%
\bibitem [{\citenamefont {{La Mura}}\ \emph {et~al.}(2021)\citenamefont {{La
  Mura}}, \citenamefont {{Barres de Almeida}}, \citenamefont
  {{Concei{\c{c}}{\~a}o}}, \citenamefont {{De Angelis}}, \citenamefont
  {{Longo}}, \citenamefont {{Pimenta}}, \citenamefont {{Prandini}},
  \citenamefont {{Ruiz-Velasco}},\ and\ \citenamefont
  {{Tom{\'e}}}}]{Mura2021MNRAS}%
  \BibitemOpen
  \bibfield  {author} {\bibinfo {author} {\bibfnamefont {G.}~\bibnamefont {{La
  Mura}}}, \bibinfo {author} {\bibfnamefont {U.}~\bibnamefont {{Barres de
  Almeida}}}, \bibinfo {author} {\bibfnamefont {R.}~\bibnamefont
  {{Concei{\c{c}}{\~a}o}}}, \bibinfo {author} {\bibfnamefont {A.}~\bibnamefont
  {{De Angelis}}}, \bibinfo {author} {\bibfnamefont {F.}~\bibnamefont
  {{Longo}}}, \bibinfo {author} {\bibfnamefont {M.}~\bibnamefont {{Pimenta}}},
  \bibinfo {author} {\bibfnamefont {E.}~\bibnamefont {{Prandini}}}, \bibinfo
  {author} {\bibfnamefont {E.}~\bibnamefont {{Ruiz-Velasco}}}, \ and\ \bibinfo
  {author} {\bibfnamefont {B.}~\bibnamefont {{Tom{\'e}}}},\ }\href {\doibase
  10.1093/mnras/stab2544} {\bibfield  {journal} {\bibinfo  {journal} {\mnras}\
  }\textbf {\bibinfo {volume} {508}},\ \bibinfo {pages} {671} (\bibinfo {year}
  {2021})},\ \Eprint {http://arxiv.org/abs/2109.06676} {arXiv:2109.06676
  [astro-ph.IM]} \BibitemShut {NoStop}%
\bibitem [{\citenamefont {{Ronchini}}\ \emph {et~al.}(2022)\citenamefont
  {{Ronchini}}, \citenamefont {{Branchesi}}, \citenamefont {{Oganesyan}},
  \citenamefont {{Banerjee}}, \citenamefont {{Dupletsa}}, \citenamefont
  {{Ghirlanda}}, \citenamefont {{Harms}}, \citenamefont {{Mapelli}},\ and\
  \citenamefont {{Santoliquido}}}]{2022Ronchini}%
  \BibitemOpen
  \bibfield  {author} {\bibinfo {author} {\bibfnamefont {S.}~\bibnamefont
  {{Ronchini}}}, \bibinfo {author} {\bibfnamefont {M.}~\bibnamefont
  {{Branchesi}}}, \bibinfo {author} {\bibfnamefont {G.}~\bibnamefont
  {{Oganesyan}}}, \bibinfo {author} {\bibfnamefont {B.}~\bibnamefont
  {{Banerjee}}}, \bibinfo {author} {\bibfnamefont {U.}~\bibnamefont
  {{Dupletsa}}}, \bibinfo {author} {\bibfnamefont {G.}~\bibnamefont
  {{Ghirlanda}}}, \bibinfo {author} {\bibfnamefont {J.}~\bibnamefont
  {{Harms}}}, \bibinfo {author} {\bibfnamefont {M.}~\bibnamefont {{Mapelli}}},
  \ and\ \bibinfo {author} {\bibfnamefont {F.}~\bibnamefont {{Santoliquido}}},\
  }\href@noop {} {\bibfield  {journal} {\bibinfo  {journal} {arXiv e-prints}\
  ,\ \bibinfo {eid} {arXiv:2204.01746}} (\bibinfo {year} {2022})},\ \Eprint
  {http://arxiv.org/abs/2204.01746} {arXiv:2204.01746 [astro-ph.HE]}
  \BibitemShut {NoStop}%
\bibitem [{\citenamefont {{Zhang}}\ \emph {et~al.}(2019)\citenamefont
  {{Zhang}}, \citenamefont {{Li}}, \citenamefont {{Xiong}}, \citenamefont
  {{Li}}, \citenamefont {{Sun}}, \citenamefont {{An}}, \citenamefont {{Xu}},
  \citenamefont {{Zhu}}, \citenamefont {{Peng}}, \citenamefont {{Wang}},\ and\
  \citenamefont {{Zhang}}}]{GeCAM}%
  \BibitemOpen
  \bibfield  {author} {\bibinfo {author} {\bibfnamefont {D.}~\bibnamefont
  {{Zhang}}}, \bibinfo {author} {\bibfnamefont {X.}~\bibnamefont {{Li}}},
  \bibinfo {author} {\bibfnamefont {S.}~\bibnamefont {{Xiong}}}, \bibinfo
  {author} {\bibfnamefont {Y.}~\bibnamefont {{Li}}}, \bibinfo {author}
  {\bibfnamefont {X.}~\bibnamefont {{Sun}}}, \bibinfo {author} {\bibfnamefont
  {Z.}~\bibnamefont {{An}}}, \bibinfo {author} {\bibfnamefont {Y.}~\bibnamefont
  {{Xu}}}, \bibinfo {author} {\bibfnamefont {Y.}~\bibnamefont {{Zhu}}},
  \bibinfo {author} {\bibfnamefont {W.}~\bibnamefont {{Peng}}}, \bibinfo
  {author} {\bibfnamefont {H.}~\bibnamefont {{Wang}}}, \ and\ \bibinfo {author}
  {\bibfnamefont {F.}~\bibnamefont {{Zhang}}},\ }\href {\doibase
  10.1016/j.nima.2018.12.032} {\bibfield  {journal} {\bibinfo  {journal}
  {Nuclear Instruments and Methods in Physics Research A}\ }\textbf {\bibinfo
  {volume} {921}},\ \bibinfo {pages} {8} (\bibinfo {year} {2019})},\ \Eprint
  {http://arxiv.org/abs/1804.04499} {arXiv:1804.04499 [physics.ins-det]}
  \BibitemShut {NoStop}%
\bibitem [{\citenamefont {{McEnery}}\ \emph {et~al.}(2019)\citenamefont
  {{McEnery}}, \citenamefont {{van der Horst}}, \citenamefont {{Dominguez}},
  \citenamefont {{Moiseev}}, \citenamefont {{Marcowith}}, \citenamefont
  {{Harding}} \emph {et~al.}}]{AMEGO}%
  \BibitemOpen
  \bibfield  {author} {\bibinfo {author} {\bibfnamefont {J.}~\bibnamefont
  {{McEnery}}}, \bibinfo {author} {\bibfnamefont {A.}~\bibnamefont {{van der
  Horst}}}, \bibinfo {author} {\bibfnamefont {A.}~\bibnamefont {{Dominguez}}},
  \bibinfo {author} {\bibfnamefont {A.}~\bibnamefont {{Moiseev}}}, \bibinfo
  {author} {\bibfnamefont {A.}~\bibnamefont {{Marcowith}}}, \bibinfo {author}
  {\bibfnamefont {A.}~\bibnamefont {{Harding}}},  \emph {et~al.},\ }in\
  \href@noop {} {\emph {\bibinfo {booktitle} {Bulletin of the American
  Astronomical Society}}},\ Vol.~\bibinfo {volume} {51}\ (\bibinfo {year}
  {2019})\ p.\ \bibinfo {pages} {245},\ \Eprint
  {http://arxiv.org/abs/1907.07558} {arXiv:1907.07558 [astro-ph.IM]}
  \BibitemShut {NoStop}%
\bibitem [{\citenamefont {{Wei}}\ \emph {et~al.}(2016)\citenamefont {{Wei}},
  \citenamefont {{Cordier}}, \citenamefont {{Antier}}, \citenamefont
  {{Antilogus}}, \citenamefont {{Atteia}}, \citenamefont {{Bajat}} \emph
  {et~al.}}]{SVOM}%
  \BibitemOpen
  \bibfield  {author} {\bibinfo {author} {\bibfnamefont {J.}~\bibnamefont
  {{Wei}}}, \bibinfo {author} {\bibfnamefont {B.}~\bibnamefont {{Cordier}}},
  \bibinfo {author} {\bibfnamefont {S.}~\bibnamefont {{Antier}}}, \bibinfo
  {author} {\bibfnamefont {P.}~\bibnamefont {{Antilogus}}}, \bibinfo {author}
  {\bibfnamefont {J.~L.}\ \bibnamefont {{Atteia}}}, \bibinfo {author}
  {\bibfnamefont {A.}~\bibnamefont {{Bajat}}},  \emph {et~al.},\ }\href@noop {}
  {\bibfield  {journal} {\bibinfo  {journal} {arXiv e-prints}\ ,\ \bibinfo
  {eid} {arXiv:1610.06892}} (\bibinfo {year} {2016})},\ \Eprint
  {http://arxiv.org/abs/1610.06892} {arXiv:1610.06892 [astro-ph.IM]}
  \BibitemShut {NoStop}%
\bibitem [{\citenamefont {{Albert}}\ \emph {et~al.}(2019)\citenamefont
  {{Albert}}, \citenamefont {{Alfaro}}, \citenamefont {{Ashkar}}, \citenamefont
  {{Alvarez}}, \citenamefont {{{\'A}lvarez}}, \citenamefont
  {{Arteaga-Vel{\'a}zquez}} \emph {et~al.}}]{SGWO}%
  \BibitemOpen
  \bibfield  {author} {\bibinfo {author} {\bibfnamefont {A.}~\bibnamefont
  {{Albert}}}, \bibinfo {author} {\bibfnamefont {R.}~\bibnamefont {{Alfaro}}},
  \bibinfo {author} {\bibfnamefont {H.}~\bibnamefont {{Ashkar}}}, \bibinfo
  {author} {\bibfnamefont {C.}~\bibnamefont {{Alvarez}}}, \bibinfo {author}
  {\bibfnamefont {J.}~\bibnamefont {{{\'A}lvarez}}}, \bibinfo {author}
  {\bibfnamefont {J.~C.}\ \bibnamefont {{Arteaga-Vel{\'a}zquez}}},  \emph
  {et~al.},\ }\href@noop {} {\bibfield  {journal} {\bibinfo  {journal} {arXiv
  e-prints}\ ,\ \bibinfo {eid} {arXiv:1902.08429}} (\bibinfo {year} {2019})},\
  \Eprint {http://arxiv.org/abs/1902.08429} {arXiv:1902.08429 [astro-ph.HE]}
  \BibitemShut {NoStop}%
\bibitem [{\citenamefont {{Huentemeyer}}\ \emph {et~al.}(2019)\citenamefont
  {{Huentemeyer}}, \citenamefont {{BenZvi}}, \citenamefont {{Dingus}},
  \citenamefont {{Fleischhack}}, \citenamefont {{Schoorlemmer}},\ and\
  \citenamefont {{Weisgarber}}}]{SGWO2}%
  \BibitemOpen
  \bibfield  {author} {\bibinfo {author} {\bibfnamefont {P.}~\bibnamefont
  {{Huentemeyer}}}, \bibinfo {author} {\bibfnamefont {S.}~\bibnamefont
  {{BenZvi}}}, \bibinfo {author} {\bibfnamefont {B.}~\bibnamefont {{Dingus}}},
  \bibinfo {author} {\bibfnamefont {H.}~\bibnamefont {{Fleischhack}}}, \bibinfo
  {author} {\bibfnamefont {H.}~\bibnamefont {{Schoorlemmer}}}, \ and\ \bibinfo
  {author} {\bibfnamefont {T.}~\bibnamefont {{Weisgarber}}},\ }in\ \href@noop
  {} {\emph {\bibinfo {booktitle} {Bulletin of the American Astronomical
  Society}}},\ Vol.~\bibinfo {volume} {51}\ (\bibinfo {year} {2019})\ p.\
  \bibinfo {pages} {109},\ \Eprint {http://arxiv.org/abs/1907.07737}
  {arXiv:1907.07737 [astro-ph.IM]} \BibitemShut {NoStop}%
\bibitem [{\citenamefont {{Fiore}}\ \emph {et~al.}(2020)\citenamefont
  {{Fiore}}, \citenamefont {{Burderi}}, \citenamefont {{Lavagna}},
  \citenamefont {{Bertacin}}, \citenamefont {{Evangelista}}, \citenamefont
  {{Campana}} \emph {et~al.}}]{HERMES}%
  \BibitemOpen
  \bibfield  {author} {\bibinfo {author} {\bibfnamefont {F.}~\bibnamefont
  {{Fiore}}}, \bibinfo {author} {\bibfnamefont {L.}~\bibnamefont {{Burderi}}},
  \bibinfo {author} {\bibfnamefont {M.}~\bibnamefont {{Lavagna}}}, \bibinfo
  {author} {\bibfnamefont {R.}~\bibnamefont {{Bertacin}}}, \bibinfo {author}
  {\bibfnamefont {Y.}~\bibnamefont {{Evangelista}}}, \bibinfo {author}
  {\bibfnamefont {R.}~\bibnamefont {{Campana}}},  \emph {et~al.}\ }(\bibinfo
  {year} {2020})\ p.\ \bibinfo {pages} {114441R},\ \Eprint
  {http://arxiv.org/abs/2101.03078} {arXiv:2101.03078 [astro-ph.HE]}
  \BibitemShut {NoStop}%
\bibitem [{\citenamefont {{Steeghs}}\ \emph {et~al.}(2022)\citenamefont
  {{Steeghs}}, \citenamefont {{Galloway}}, \citenamefont {{Ackley}},
  \citenamefont {{Dyer}}, \citenamefont {{Lyman}}, \citenamefont {{Ulaczyk}}
  \emph {et~al.}}]{GOTO}%
  \BibitemOpen
  \bibfield  {author} {\bibinfo {author} {\bibfnamefont {D.}~\bibnamefont
  {{Steeghs}}}, \bibinfo {author} {\bibfnamefont {D.~K.}\ \bibnamefont
  {{Galloway}}}, \bibinfo {author} {\bibfnamefont {K.}~\bibnamefont
  {{Ackley}}}, \bibinfo {author} {\bibfnamefont {M.~J.}\ \bibnamefont
  {{Dyer}}}, \bibinfo {author} {\bibfnamefont {J.}~\bibnamefont {{Lyman}}},
  \bibinfo {author} {\bibfnamefont {K.}~\bibnamefont {{Ulaczyk}}},  \emph
  {et~al.},\ }\href {\doibase 10.1093/mnras/stac013} {\bibfield  {journal}
  {\bibinfo  {journal} {\mnras}\ }\textbf {\bibinfo {volume} {511}},\ \bibinfo
  {pages} {2405} (\bibinfo {year} {2022})},\ \Eprint
  {http://arxiv.org/abs/2110.05539} {arXiv:2110.05539 [astro-ph.IM]}
  \BibitemShut {NoStop}%
\bibitem [{\citenamefont {{Camp}}\ and\ \citenamefont {{TAP
  Team}}(2019)}]{TAP}%
  \BibitemOpen
  \bibfield  {author} {\bibinfo {author} {\bibfnamefont {J.}~\bibnamefont
  {{Camp}}}\ and\ \bibinfo {author} {\bibnamefont {{TAP Team}}}\ }(\bibinfo
  {year} {2019})\ p.\ \bibinfo {pages} {5027}\BibitemShut {NoStop}%
\bibitem [{\citenamefont {{Fong}}\ \emph {et~al.}(2015)\citenamefont {{Fong}},
  \citenamefont {{Berger}}, \citenamefont {{Margutti}},\ and\ \citenamefont
  {{Zauderer}}}]{Fong2015ApJ}%
  \BibitemOpen
  \bibfield  {author} {\bibinfo {author} {\bibfnamefont {W.}~\bibnamefont
  {{Fong}}}, \bibinfo {author} {\bibfnamefont {E.}~\bibnamefont {{Berger}}},
  \bibinfo {author} {\bibfnamefont {R.}~\bibnamefont {{Margutti}}}, \ and\
  \bibinfo {author} {\bibfnamefont {B.~A.}\ \bibnamefont {{Zauderer}}},\ }\href
  {\doibase 10.1088/0004-637X/815/2/102} {\bibfield  {journal} {\bibinfo
  {journal} {\apj}\ }\textbf {\bibinfo {volume} {815}},\ \bibinfo {eid} {102}
  (\bibinfo {year} {2015})},\ \Eprint {http://arxiv.org/abs/1509.02922}
  {arXiv:1509.02922 [astro-ph.HE]} \BibitemShut {NoStop}%
\bibitem [{\citenamefont {{Abbott}}\ \emph {et~al.}(2021)\citenamefont
  {{Abbott}}, \citenamefont {{Abbott}}, \citenamefont {{Abraham}},
  \citenamefont {{Acernese}}, \citenamefont {{Ackley}}, \citenamefont
  {{Adams}}, \citenamefont {{Adams}}, \citenamefont {{Adhikari}} \emph
  {et~al.}}]{Abbott2021ApJL}%
  \BibitemOpen
  \bibfield  {author} {\bibinfo {author} {\bibfnamefont {R.}~\bibnamefont
  {{Abbott}}}, \bibinfo {author} {\bibfnamefont {T.~D.}\ \bibnamefont
  {{Abbott}}}, \bibinfo {author} {\bibfnamefont {S.}~\bibnamefont {{Abraham}}},
  \bibinfo {author} {\bibfnamefont {F.}~\bibnamefont {{Acernese}}}, \bibinfo
  {author} {\bibfnamefont {K.}~\bibnamefont {{Ackley}}}, \bibinfo {author}
  {\bibfnamefont {A.}~\bibnamefont {{Adams}}}, \bibinfo {author} {\bibfnamefont
  {C.}~\bibnamefont {{Adams}}}, \bibinfo {author} {\bibfnamefont {R.~X.}\
  \bibnamefont {{Adhikari}}},  \emph {et~al.},\ }\href {\doibase
  10.3847/2041-8213/abe949} {\bibfield  {journal} {\bibinfo  {journal} {\apjl}\
  }\textbf {\bibinfo {volume} {913}},\ \bibinfo {eid} {L7} (\bibinfo {year}
  {2021})},\ \Eprint {http://arxiv.org/abs/2010.14533} {arXiv:2010.14533
  [astro-ph.HE]} \BibitemShut {NoStop}%
\bibitem [{\citenamefont {{Gardner}}\ \emph {et~al.}(2006)\citenamefont
  {{Gardner}}, \citenamefont {{Mather}}, \citenamefont {{Clampin}},
  \citenamefont {{Doyon}}, \citenamefont {{Greenhouse}}, \citenamefont
  {{Hammel}} \emph {et~al.}}]{JWST}%
  \BibitemOpen
  \bibfield  {author} {\bibinfo {author} {\bibfnamefont {J.~P.}\ \bibnamefont
  {{Gardner}}}, \bibinfo {author} {\bibfnamefont {J.~C.}\ \bibnamefont
  {{Mather}}}, \bibinfo {author} {\bibfnamefont {M.}~\bibnamefont {{Clampin}}},
  \bibinfo {author} {\bibfnamefont {R.}~\bibnamefont {{Doyon}}}, \bibinfo
  {author} {\bibfnamefont {M.~A.}\ \bibnamefont {{Greenhouse}}}, \bibinfo
  {author} {\bibfnamefont {H.~B.}\ \bibnamefont {{Hammel}}},  \emph {et~al.},\
  }\href {\doibase 10.1007/s11214-006-8315-7} {\bibfield  {journal} {\bibinfo
  {journal} {ssr}\ }\textbf {\bibinfo {volume} {123}},\ \bibinfo {pages} {485}
  (\bibinfo {year} {2006})},\ \Eprint {http://arxiv.org/abs/astro-ph/0606175}
  {arXiv:astro-ph/0606175 [astro-ph]} \BibitemShut {NoStop}%
\bibitem [{\citenamefont {{Piro}}\ \emph {et~al.}(2021)\citenamefont {{Piro}},
  \citenamefont {{Ahlers}}, \citenamefont {{Coleiro}}, \citenamefont {{Colpi}},
  \citenamefont {{de O{\~n}a Wilhelmi}}, \citenamefont {{Guainazzi}} \emph
  {et~al.}}]{Athena}%
  \BibitemOpen
  \bibfield  {author} {\bibinfo {author} {\bibfnamefont {L.}~\bibnamefont
  {{Piro}}}, \bibinfo {author} {\bibfnamefont {M.}~\bibnamefont {{Ahlers}}},
  \bibinfo {author} {\bibfnamefont {A.}~\bibnamefont {{Coleiro}}}, \bibinfo
  {author} {\bibfnamefont {M.}~\bibnamefont {{Colpi}}}, \bibinfo {author}
  {\bibfnamefont {E.}~\bibnamefont {{de O{\~n}a Wilhelmi}}}, \bibinfo {author}
  {\bibfnamefont {M.}~\bibnamefont {{Guainazzi}}},  \emph {et~al.},\
  }\href@noop {} {\bibfield  {journal} {\bibinfo  {journal} {arXiv e-prints}\
  ,\ \bibinfo {eid} {arXiv:2110.15677}} (\bibinfo {year} {2021})},\ \Eprint
  {http://arxiv.org/abs/2110.15677} {arXiv:2110.15677 [astro-ph.HE]}
  \BibitemShut {NoStop}%
\bibitem [{\citenamefont {{Nissanke}}\ \emph {et~al.}(2010)\citenamefont
  {{Nissanke}}, \citenamefont {{Holz}}, \citenamefont {{Hughes}}, \citenamefont
  {{Dalal}},\ and\ \citenamefont {{Sievers}}}]{Nissanke2010ApJ}%
  \BibitemOpen
  \bibfield  {author} {\bibinfo {author} {\bibfnamefont {S.}~\bibnamefont
  {{Nissanke}}}, \bibinfo {author} {\bibfnamefont {D.~E.}\ \bibnamefont
  {{Holz}}}, \bibinfo {author} {\bibfnamefont {S.~A.}\ \bibnamefont
  {{Hughes}}}, \bibinfo {author} {\bibfnamefont {N.}~\bibnamefont {{Dalal}}}, \
  and\ \bibinfo {author} {\bibfnamefont {J.~L.}\ \bibnamefont {{Sievers}}},\
  }\href {\doibase 10.1088/0004-637X/725/1/496} {\bibfield  {journal} {\bibinfo
   {journal} {\apj}\ }\textbf {\bibinfo {volume} {725}},\ \bibinfo {pages}
  {496} (\bibinfo {year} {2010})},\ \Eprint {http://arxiv.org/abs/0904.1017}
  {arXiv:0904.1017 [astro-ph.CO]} \BibitemShut {NoStop}%
\bibitem [{\citenamefont {Abbott}\ \emph {et~al.}(2021)\citenamefont {Abbott}
  \emph {et~al.}}]{LIGOScientific:2020ibl}%
  \BibitemOpen
  \bibfield  {author} {\bibinfo {author} {\bibfnamefont {R.}~\bibnamefont
  {Abbott}} \emph {et~al.} (\bibinfo {collaboration} {LIGO Scientific,
  Virgo}),\ }\href {\doibase 10.1103/PhysRevX.11.021053} {\bibfield  {journal}
  {\bibinfo  {journal} {Phys. Rev. X}\ }\textbf {\bibinfo {volume} {11}},\
  \bibinfo {pages} {021053} (\bibinfo {year} {2021})},\ \Eprint
  {http://arxiv.org/abs/2010.14527} {arXiv:2010.14527 [gr-qc]} \BibitemShut
  {NoStop}%
\bibitem [{\citenamefont {{Jin}}\ \emph {et~al.}(2021)\citenamefont {{Jin}},
  \citenamefont {{Zhou}}, \citenamefont {{Covino}}, \citenamefont {{Liao}},
  \citenamefont {{Li}}, \citenamefont {{Lei}}, \citenamefont {{D'Avanzo}},
  \citenamefont {{Fan}},\ and\ \citenamefont {{Wei}}}]{JinZP2021}%
  \BibitemOpen
  \bibfield  {author} {\bibinfo {author} {\bibfnamefont {Z.-P.}\ \bibnamefont
  {{Jin}}}, \bibinfo {author} {\bibfnamefont {H.}~\bibnamefont {{Zhou}}},
  \bibinfo {author} {\bibfnamefont {S.}~\bibnamefont {{Covino}}}, \bibinfo
  {author} {\bibfnamefont {N.-H.}\ \bibnamefont {{Liao}}}, \bibinfo {author}
  {\bibfnamefont {X.}~\bibnamefont {{Li}}}, \bibinfo {author} {\bibfnamefont
  {L.}~\bibnamefont {{Lei}}}, \bibinfo {author} {\bibfnamefont
  {P.}~\bibnamefont {{D'Avanzo}}}, \bibinfo {author} {\bibfnamefont {Y.-Z.}\
  \bibnamefont {{Fan}}}, \ and\ \bibinfo {author} {\bibfnamefont {D.-M.}\
  \bibnamefont {{Wei}}},\ }\href@noop {} {\bibfield  {journal} {\bibinfo
  {journal} {arXiv e-prints}\ ,\ \bibinfo {eid} {arXiv:2109.07694}} (\bibinfo
  {year} {2021})},\ \Eprint {http://arxiv.org/abs/2109.07694} {arXiv:2109.07694
  [astro-ph.HE]} \BibitemShut {NoStop}%
\bibitem [{\citenamefont {{Gehrels}}\ \emph {et~al.}(2004)\citenamefont
  {{Gehrels}}, \citenamefont {{Chincarini}}, \citenamefont {{Giommi}},
  \citenamefont {{Mason}}, \citenamefont {{Nousek}},  \emph
  {et~al.}}]{2004ApJ...611.1005G}%
  \BibitemOpen
  \bibfield  {author} {\bibinfo {author} {\bibfnamefont {N.}~\bibnamefont
  {{Gehrels}}}, \bibinfo {author} {\bibfnamefont {G.}~\bibnamefont
  {{Chincarini}}}, \bibinfo {author} {\bibfnamefont {P.}~\bibnamefont
  {{Giommi}}}, \bibinfo {author} {\bibfnamefont {K.~O.}\ \bibnamefont
  {{Mason}}}, \bibinfo {author} {\bibfnamefont {J.~A.}\ \bibnamefont
  {{Nousek}}}, ,  \emph {et~al.},\ }\href {\doibase 10.1086/422091} {\bibfield
  {journal} {\bibinfo  {journal} {\apj}\ }\textbf {\bibinfo {volume} {611}},\
  \bibinfo {pages} {1005} (\bibinfo {year} {2004})},\ \Eprint
  {http://arxiv.org/abs/astro-ph/0405233} {arXiv:astro-ph/0405233 [astro-ph]}
  \BibitemShut {NoStop}%
\end{thebibliography}%
\bibliographystyle{apsrev4-1}

\end{document}